\documentclass[prd,groupedaddress,twocolumn,nofootinbib,showpacs,preprintnumbers,floatfix]{revtex4}

\usepackage{mathrsfs}
\usepackage{amsfonts,amssymb,amsmath}
\usepackage{feynmp}
\usepackage{graphicx}
\usepackage{multirow}
\usepackage{slashed}
\usepackage{fancybox}
\usepackage{subdepth}
\usepackage{float}
\usepackage{fancyvrb}
\usepackage{accents}
\usepackage{stmaryrd}
\usepackage{wasysym}
\usepackage{comment}
\usepackage[T1]{fontenc}
\usepackage{etoolbox}
\usepackage{bm}
\usepackage{url}
\usepackage{hyperref}
\usepackage{breakurl}
\usepackage{ragged2e}
\usepackage{lipsum}
\usepackage{array,booktabs}
\usepackage[nodisplayskipstretch]{setspace}

\declareslashed{}{/}{0}{-0.05}{E}
\declareslashed{}{/}{0}{-0.05}{P}
\declareslashed{}{/}{0}{-0.10}{\Widehat{P}}
\cornersize*{15pt}
\newfloat{card}{htb}{lop}
\floatname{card}{Card}
\makeatletter \let\@currsize\normalsize \makeatother
\makeatletter
\renewcommand\floatc@plain[2]{\setbox\@tempboxa\hbox{{\footnotesize {\bf {\@fs@cfont #1:}} #2}}%
\ifdim\wd\@tempboxa>\hsize {\begin{minipage}[t]{\linewidth}\justifying \small {\raggedleft {\@fs@cfont #1:} #2}\end{minipage}}\par
\else\hbox to\hsize{\hfil\box\@tempboxa\hfil}\fi}
\@namedef{thecard}{\Alph{card}}
\makeatother

\DeclareMathSymbol{\widehatsym}{\mathord}{largesymbols}{"62}
\newcommand\lowerwidehatsym{\text{\smash{\raisebox{-1.3ex}{$\widehatsym$}}}}
\newcommand\Widehat[1]{\accentset{\textstyle\lowerwidehatsym}{#1}}

\DeclareMathSymbol{\widetildesym}{\mathord}{largesymbols}{"65}
\newcommand\lowerwidetildesym{\text{\smash{\raisebox{-1.3ex}{$\widetildesym$}}}}
\newcommand\Widetilde[1]{\accentset{\textstyle\lowerwidetildesym}{#1}}

\begin{document}

\title{A Complete Solution Classification and Unified Algorithmic Treatment for the\\
	One- and Two-Step Asymmetric S-Transverse Mass (\texorpdfstring{\smash{$\boldsymbol{{\Widetilde{M}}_{\rm T2}}$}}{MT2}) Event Scale Statistic}

\author{Joel W. Walker}

\affiliation{
	\mbox{Department of Physics, Sam Houston State University, Huntsville, TX 77341-2267, USA}\\
	\mbox{Kavli Institute for Theoretical Physics, University of California, Santa Barbara, CA 93106-4030, USA}\\
	{\tt jwalker@shsu.edu} \hspace{9pt} {\tt www.joelwalker.net}}


\begin{abstract}
The $M_{\rm T2}$, or ``s-transverse mass'', statistic was developed to
associate a parent mass scale to a missing transverse energy signature, given that
escaping particles are generally expected in pairs, while collider experiments are sensitive
to just a single transverse momentum vector sum.
This document focuses on the generalized \smash{${\Widetilde{M}}_{\rm T2}$} extension of that statistic to
asymmetric one- and two-step decay chains, with arbitrary child particle masses and upstream missing transverse momentum.
It provides a unified theoretical formulation, complete solution classification,
taxonomy of critical points, and technical algorithmic prescription
for treatment of the \smash{${\Widetilde{M}}_{\rm T2}$} event scale.
An implementation of the described algorithm is available for download,
and is also a deployable component of the author's selection cut software
package \mbox{\sc AEACuS} ({\sc A}lgorithmic {\sc E}vent {\sc A}rbiter and {\sc Cu}t {\sc S}elector).
Appendices address combinatoric event assembly, algorithm validation, and a complete pseudocode.
\end{abstract}



\preprint{NSF-KITP-13-169}

\maketitle


\makeatletter
\close@column@grid
\onecolumngrid
\vspace{-8pt}
\begin{center}
\includegraphics[width=0.8\textwidth]{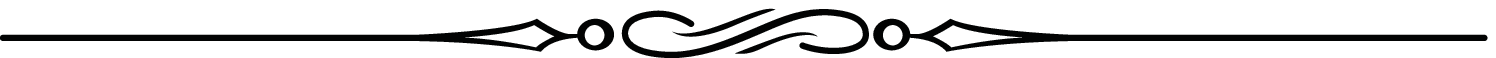}
\end{center}
\vspace{8pt}
\close@column@grid
\twocolumngrid
\makeatother


\begin{center}
\bf{\small Introduction}
\end{center}

Transverse kinematic variables are ubiquitous to the description of collider
events, due to the simple reality that physical detectors cannot be sensitive
to particle flow down the beamline.  Moreover, since the initial portion of
longitudinal momentum carried by the interacting ``parton'' constituents in a hadron collision
cannot be ascertained, an imbalance in the detected momentum sum can only be established in the beam-transverse plane,
where the net conserved momentum may be safely approximated as zero.  This projection bestows an intrinsic kinematic
incompleteness on the transverse variables, by dint of which they will generically tend to indicate new physics via
a termination point rather than a peak.

The most basic definitions of this sort are the 
transverse energy and momentum of a single particle, which feature invariance under the
subset of Lorentz transformations corresponding to longitudinal boosts.
\begin{eqnarray}
{\vec{P}}_{\rm T} &\equiv& P_x \,\hat{x} + P_y \,\hat{y}
\nonumber \\
E_{\rm T} &\equiv& \sqrt{ M^2 + {\vec{P}}_{\rm T} \cdot {\vec{P}}_{\rm T} } \,\,=\, \sqrt{E^2 - P_z^2}
\nonumber \\
{\displaystyle \lim_{M = 0}} &\Rightarrow& \vert {\vec{P}}_{\rm T}\vert \,\,=\, \sqrt{P_x^2 + P_y^2}
\label{eq:et}
\end{eqnarray}

A second statistic of great traditional importance to collider studies is the mutual ``transverse mass'',
which for any two objects $A$ and $B$, may be defined as follows~\cite{kinematics},
referencing the transverse energy $E_{\rm T}$ of each constituent object in a recursively consistent manner.
\begin{eqnarray}
M^{A,B}_{\rm T} &\equiv& \sqrt{ \left(E^{A}_{\rm T} + E^{B}_{\rm T} \right)^2 - \left(\vert {\vec{P}}^{\raisebox{-2.5pt}{$\scriptstyle A$}}_{\rm T} +
	{\vec{P}}^{\raisebox{-2.5pt}{$\scriptstyle B$}}_{\rm T} \vert\right)^2}
\nonumber \\
&=& \sqrt{ M_{A}^2 + M_{B}^2 + 2\left( E^{A}_{\rm T} E^{B}_{\rm T} - {\vec{P}}^{\raisebox{-2.5pt}{$\scriptstyle A$}}_{\rm T} \cdot
	{\vec{P}}^{\raisebox{-2.5pt}{$\scriptstyle B$}}_{\rm T} \right)}
\nonumber \\
{\displaystyle \lim_{M_{A} = M_{B} = 0}} &\Rightarrow& \sqrt{ 2\, \vert {\vec{P}}^{\raisebox{-2.5pt}{$\scriptstyle A$}}_{\rm T}\vert
	\vert {\vec{P}}^{\raisebox{-2.5pt}{$\scriptstyle B$}}_{\rm T}\vert
\left( 1 - \cos {\Delta \phi}^{\,B,A} \right) }
\label{eq:mt}
\end{eqnarray}

The $M_{\rm T}^{A,B}$ transverse mass distribution exhibits an endpoint corresponding to the mass of a presumed common parent to 
particle species $A$ and $B$, as enforced by the inequality \mbox{\{ $M_{\rm T}^{A,B} \,\le\, M^{A,B} \equiv \surd[\,(P^\mu_A + P^\mu_{B})^2\,] \,\}$}.
This property is commonly employed as a selection cut against leptonic decay of the $W$~boson, where the associated neutrino is hypothetically
linked to an observed imbalance in the vector sum over transverse momentum.  Beyond application as a discovery statistic,
$M_{\rm T}^{A,B}$ has potential to facilitate future detailed mass measurement of new particles
predicted by Supersymmetry (SUSY).

However, association of a custodial mass scale to missing transverse energy signatures may be a physically subtle task, complicated 
by the practicality that escaping particles are commonly expected in models of new physics to be produced in pairs.  These pairs may
become kinematically decoupled by intervening steps in the decay cascade, implying that a substantial portion of the unseen tracks
may internally cancel in the observable missing momentum vector sum.  This dilemma has historically
motivated the consideration of various increasingly sophisticated discovery and measurement variables. 

One such variable is a generalization of the previously described transverse mass statistic referred to as $M_{\rm T2}^{A,B}$ or the
``s-transverse mass''.  First introduced in Ref.~\cite{Lester:1999tx},
the s-transverse mass construction is based upon the observation that each such visible shower may be convolved with its affiliated contribution to
the missing momentum to yield a transverse mass, as defined in Eq.~(\ref{eq:mt}), that is bounded above by the mass of the system's parent particle.
Since only the unified missing transverse momentum vector \smash{${\vec{{\slashed{P}}}} {}_{\rm T}$} is experimentally available,
this analysis requires the specialization to a suitably optimized location in the distribution of this sum between the pair of
missing momentum candidates.  The $M_{\rm T2}$ prescription is to assume a symmetric structure for each decay chain, so that the larger of the two
transverse masses may be used imply a lower bound on a common parent species, subsequent to minimization of that maximum value with respect to all
consistent \smash{${\vec{{\slashed{P}}}} {}_{\rm T}$} partitions.  If event pollution by upstream decays of the initial state prior to the targeted pair production
can be neglected, the minimization may actually be performed analytically~\cite{Cho:2007dh}.  If the mass of the escaping particle species
is additionally set equal to zero, the following simple expression is realized.
\begin{eqnarray}
\Sigma_{\rm T} &\equiv& E^{A}_{\rm T} E^{B}_{\rm T} + {\vec{P}}^{\raisebox{-2.5pt}{$\scriptstyle A$}}_{\rm T} \cdot {\vec{P}}^{\raisebox{-2.5pt}{$\scriptstyle B$}}_{\rm T}
\vphantom{\sqrt{\left( {\Delta}^{\,B,A} \right) }}
\nonumber \\
M^{A,B}_{\rm T2} &\equiv& \sqrt{ \Sigma_{\rm T} + \sqrt{ ( \Sigma_{\rm T} )^2 - M_{A}^2 M_{B}^2 }}
\nonumber \\
{\displaystyle \lim_{M_{A} = M_{B} = 0}} &\Rightarrow& \sqrt{ 2\, \vert {\vec{P}}^{\raisebox{-2.5pt}{$\scriptstyle A$}}_{\rm T}\vert \vert {\vec{P}}^{\raisebox{-2.5pt}{$\scriptstyle B$}}_{\rm T}\vert
\left( 1 + \cos {\Delta \phi}^{\,B,A} \right) }
\label{eq:mt2}
\end{eqnarray}
Relative to Eq.~(\ref{eq:mt}), one immediately notices the curious emergence of a Euclidean signature for the inner product.
Although originally introduced as tool for deducing the mass scale of decaying SUSY particles, the s-transverse mass
(like the transverse mass itself) is currently enjoying service as a discovery statistic, with variously tuned
cutoff thresholds at the Large Hadron Collider (LHC) ranging from around $100$~GeV up to about half a TeV~\cite{PAS-SUS-12-002}.

However, this event hypothesis is much too restrictive for the handling of many physically
interesting scenarios, which may variously require the freedom to model
\mbox{({\it i}\hspace{1.6pt})} decay chains that proceed asymmetrically from the initial pair production,
\mbox{({\it ii}\hspace{1.6pt})} arbitrary masses for child particle species,
\mbox{({\it iii}\hspace{1.6pt})} specification of multiple independently visible
four-momentum tracks associated with each event leg, and ({\it iv}\hspace{1.2pt}) an inclusive ``upstream'' determination
of the missing transverse momentum including imbalances from event constituents beyond
those explicitly hypothesized to descend from the targeted pair production.
In response, the generalized one- and two-step asymmetric s-transverse mass event scale statistic
\smash{${\Widetilde{M}}_{\rm T2}$} was developed by various
authors~\cite{Cheng:2008hk,Barr:2009jv,Konar:2009qr,Bai:2012gs,Bai:2013ema},
and has seen duty in the ongoing LHC SUSY search~\cite{ATLAS-CONF-2013-037}.
The full weight of this added complexity necessitates the shift to a numerical framework for analysis.
Modern treatments of this variety have been driven by the insight~\cite{Cheng:2008hk} that the original $M_{\rm T2}$
statistic may be equivalently recast as the minimal parent particle species mass compatible with all kinematic
constraints on the hypothesized event topology.

This document provides a 
\mbox{({\it i}\hspace{1.6pt})} unified theoretical formulation,
\mbox{({\it ii}\hspace{1.6pt})} complete solution classification,
\mbox{({\it iii}\hspace{1.6pt})} taxonomy of critical points,
and \mbox{({\it iv}\hspace{1.2pt})} technical algorithmic prescription 
for treatment of the generalized one- and two-step asymmetric \smash{${\Widetilde{M}}_{\rm T2}$} event scale.
In many regards, this presentation follows, or builds upon, the excellent pedagogical summary of Ref.~\cite{mincompatiblemasses}.
Potentially novel elements of this presentation include
\mbox{({\it i}\hspace{1.6pt})} a unified symbolic and computational environment encompassing arbitrary event configurations,
\mbox{({\it ii}\hspace{1.6pt})} a graphically enhanced intuition for the organization and interrelation of solution classes, 
\mbox{({\it iii}\hspace{1.6pt})} a heightened sensitivity in the identification and handling of exception cases,
\mbox{({\it iv}\hspace{1.2pt})} an optimization of the mass search bounds without resort to an arbitrary hard upper scale,
and \mbox{({\it v}\hspace{1.2pt})} implementation of an existence test for solutions interior to some scale bounds without sampling the enclosed bulk.
A triad of appendices address
\mbox{({\it a}\hspace{1.2pt})} combinatoric assembly of event objects, 
\mbox{({\it b}\hspace{1.2pt})} validation by public codes against Monte Carlo events, 
and \mbox{({\it c}\hspace{1.2pt})} the algorithm pseudocode and logical flow diagram.

A standalone implementation of the described \smash{${\Widetilde{M}}_{\rm T2}$} algorithm in the {\sc Perl}
programming language is available for download from the author's personal website~\cite{amt2}, and
is also included as an ancillary file \mbox{``{\tt anc/amt2.pl}''} with
this document's electronic source at the {\it ar{\raisebox{1.8pt}{$\chi$}}{\hspace{-0.8pt}}iv.org} repository.
A broader context for the practical usage of this algorithm is provided
by its inclusion in the author's fully-featured selection cut software
package \mbox{\sc AEACuS}
--- in Greek myth, the judges Minos, Rhadamanthus and Aeacus, sons of Zeus,
weigh the fate of souls departing the mortal sphere ---
({\sc A}lgorithmic {\sc E}vent {\sc A}rbiter and {\sc Cu}t {\sc S}elector),
which was formerly developed under the name \mbox{\sc CutLHCO}~\cite{Walker:2012vf,aeacus}.
All described programs have been freely released into the public domain under the terms of the GNU General Public License~\cite{gnugpl}.


\section{\protecting{Theoretical Formulation of \texorpdfstring{\smash{$\boldsymbol{{\Widetilde{M}}_{\rm T2}}$}}{MT2}}\label{sct:theory}}

\begin{figure}[tp]
  \centering
  \unitlength = 1mm
  \begin{fmffile}{twostep}
    \begin{fmfgraph*}(60,25)
      \fmfbottom{b1,b2,b3}
      \fmf{fermion,label=$Y$}{b1,v1}
      \fmf{photon,label=$X$}{v1,v2}
      \fmf{dashes,label=$H$}{v2,b3}
      \fmftop{t1,t2,t3}
      \fmf{phantom,tension=0.6}{t1,v3}
      \fmf{phantom,tension=1.0}{v3,v4}
      \fmf{phantom,tension=3.0}{v4,t3}
      \fmf{fermion,tension=0,label=$V$}{v1,v3}
      \fmf{fermion,tension=0,label=$S$}{v2,v4}
    \end{fmfgraph*}
  \end{fmffile}
  \caption{Two-step decay topology of a single \protecting{\smash{${\Widetilde{M}}_{\rm T2}$}} event leg, including
    a parent particle species $Y$, a detected visible product $V$, an on-shell internal mediator
    species $X$, a detected secondary product $S$, and an undetected hidden product $H$.}
  \label{fig:twostep}
\end{figure}
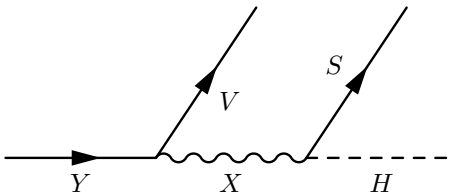

The event topologies addressed by the s-transverse mass scale
statistic family \smash{${\Widetilde{M}}_{\rm T2}$} begin
with pair production of a massive parent particle species $Y$,
possibly accompanied by some component of upstream transverse momentum.
Both legs of this initial production event are permitted to decay
independently, by either a one-step ({\it cf.} FIG.~\ref{fig:onestep})
or two-step ({\it cf.} FIG.~\ref{fig:twostep}) process into distinct
(possibly massive) daughter particle species, and distinct assumptions about
the invariant mass of an escaping hidden product $H$ (or detection anomalies,
{\it e.g.} a lost lepton), may be imposed.

\begin{figure}[tp]
  \centering
  \unitlength = 1mm
  \begin{fmffile}{onestep}
    \begin{fmfgraph*}(60,25)
      \fmfbottom{b1,b2,b3}
      \fmf{fermion,label=$Y$}{b1,v1}
      \fmf{dashes,label=$H$}{v1,b3}
      \fmftop{t1,t2,t3}
      \fmf{phantom,tension=0.5}{t1,v3}
      \fmf{phantom,tension=1.3}{v3,t3}
      \fmf{fermion,tension=0,label=$V$}{v1,v3}
    \end{fmfgraph*}
  \end{fmffile}
  \caption{One-step decay topology of a single \protecting{\smash{${\Widetilde{M}}_{\rm T2}$}} event leg, including
    a parent particle species $Y$, a detected visible product $V$, and an undetected hidden product $H$.}
  \label{fig:onestep}
\end{figure}
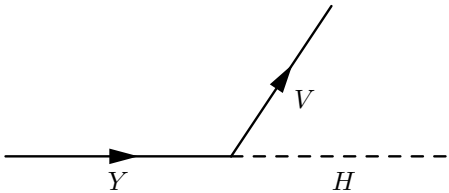

In the one-step topology the complete four-momentum of a single
visible decay product $V$ (or an effective product summed over
independently observed leptons and jets) is specified.  Superior 
event discrimination may be possible in the two-step topology, where a
secondarily resolved visible decay product $S$ (or, again, effective product)
is hypothesized to branch away from $V$ via an on-shell internal
mediator species $X$ of specified mass.

The \smash{${\Widetilde{M}}_{\rm T2}$} statistic may be succinctly summarized as the minimal
parent particle species mass $M_Y$ that is compatible with all kinematic constraints
on the hypothesized event topology.  For a one-step event leg, the decay must respect
the following energy-momentum conservation rule and mass-shell identities.
\begin{subequations}
\label{eq:onestepepm}
\begin{eqnarray}
{(P_V^\mu + P_H^\mu)}^2 = {(P_Y^\mu)}^2 &\equiv& M_Y^2 \label{eq:onestepepm_a} \\
{(P_H^\mu)}^2 &\equiv& M_H^2 \label{eq:onestepepm_b}
\end{eqnarray}
\end{subequations}

The corresponding formulae for a two-step event leg are provided subsequently.
\begin{subequations}
\label{eq:twostepepm}
\begin{eqnarray}
{(P_V^\mu + P_S^\mu + P_H^\mu)}^2 = {(P_Y^\mu)}^2 &\equiv& M_Y^2 \label{eq:twostepepm_a} \\
{(P_S^\mu + P_H^\mu)}^2 = {(P_X^\mu)}^2 &\equiv& M_X^2 \label{eq:twostepepm_b} \\
{(P_H^\mu)}^2 &\equiv& M_H^2 \label{eq:twostepepm_c}
\end{eqnarray}
\end{subequations}

In the one-step scenario of Eqs.~(\ref{eq:onestepepm}), taking
$P_V^\mu$ as observational input and $M_H$ as a definition of the event
hypothesis, the four-momentum of the hidden decay product $P_H^\mu$ and
the mass of the parent particle species $M_Y$ comprise five residual
unknown parameters.  The mass-shell condition on $M_H$ may be considered to
eliminate the time-like component of $P_H^\mu$ in favor of the
three undetermined spatial components.  The longitudinal component of $P_H^\mu$
is experimentally inaccessible, but obviously still bounded to positive magnitude-square
\smash{${(P_H^z)}^2 \ge 0$}.  This inequality may then be projected back onto the transverse
components $(P_H^{x,y})$ of the hidden momentum vector, defining a bounded
consistency region in that coordinate plane for each trial value of the
parent mass $M_Y$ that admits real solutions.  The perimeter of this region
corresponds to $P_H^z = 0$, and all locations interior to this perimeter are
scanned by increasing the magnitude of $P_H^z$ until the phase space
afforded by the trial mass $M_Y$ saturates and the contracted geometry becomes singular.

In the two-step scenario of Eqs.~(\ref{eq:twostepepm}), taking both $P_V^\mu$
and $P_S^\mu$ as observational input, and both $M_H$ and $M_X$ as definitions
of the event hypothesis, the same five unknowns, namely $P_H^\mu$ and $M_Y$,
persist.  However, there are now two supplementary mass-shell conditions, on
$M_H$ and $M_X$, which may be considered to eliminate both the time-like and
longitudinal components of $P_H^\mu$, such that there is no need to invoke a
softer inequality condition on the magnitude of $P_H^z$.  The expectation in
this case is that of a continuous thin contour of consistent solutions in the
$(P_H^x,P_H^y)$ coordinate plane for each admissible parent particle
species trial mass $M_Y$.

The resulting expressions of constraint may be somewhat formally simplified by boosting
into the longitudinally resting frame of the primary visible decay product $V$;
implementing this protocol, the notation $E_V$ will always actually represent
the transverse energy $E_V^{\rm T}$.  The only other four-momentum sensitive to this
choice of reference frame will be that of the secondary visible product $S$ in the
two-step decay topology.  Whenever referenced, $P_S^z$ will thus represent
the residual longitudinal momentum component of $S$, subsequent to that transformation.
Interestingly, it is legitimate to perform such a boost
independently for each of the two event legs, given that observables will only be
cross-correlated between legs in the invariant transverse directions.

The consistent $(P_H^x,P_H^y)$ two-dimensional area and one-dimensional curve
subspaces are embodied in the following relationships.
\begin{eqnarray}
& a\, {(\Widehat{P}_H^x)}^2 + 2\, b\, \Widehat{P}_H^x \Widehat{P}_H^y + c\, {(\Widehat{P}_H^y)}^2 + 2\, d\, \Widehat{P}_H^x + 2\, f\, \Widehat{P}_H^y + g & \nonumber \\
& \left\{ \begin{array}{ll}
     \le 0 & \text{1-Step} \\
       = 0 & \text{2-Step}
	\end{array} \right.
&
\label{eq:ellipseph}
\end{eqnarray}

Here, and throughout, application of the circumflex accent implies a dimensionless ratio of
the given kinematic construct against the leading visible event scale $E_V$.
\begin{equation}
\Widehat{P}_\mu \equiv \left( \frac{P_\mu}{E_V} \right) \quad;\quad
\Widehat{M} \equiv \left( \frac{M}{E_V} \right)
\quad;\quad \text{etc.}
\label{eq:circumflex}
\end{equation}

Solutions for the coefficients employed in Eqs.~(\ref{eq:ellipseph}) are as follows.
A global (possibly signed and/or dimensionful) factor, {\it e.g.} $E_V^2$, may be
freely multiplied through according to taste or convention.
\begin{eqnarray}
a &=& \Lambda_x^2 \,+\, \Omega \left\{ 1 - {( \Widehat{P}_V^x )}^2 \right\} \vphantom{{\Big(\Big)}^2\}} \nonumber \\
b &=& \Lambda_x\, \Lambda_y \,-\, \Omega\, \Widehat{P}_V^x\, \Widehat{P}_V^y \vphantom{\left\{{(A)}^2\right\}} \vphantom{{\Big(\Big)}^2\}} \nonumber \\
c &=& \Lambda_y^2 \,+\, \Omega \left\{ 1 - {( \Widehat{P}_V^y )}^2 \right\} \vphantom{{\Big(\Big)}^2\}} \nonumber \\
d &=& \Lambda_x \Big( \Pi\, \Gamma - \Delta \Big) \,-\, \Omega\, \Gamma\, \Widehat{P}_V^x \vphantom{{\Big(\Big)}^2\}} \nonumber \\
f &=& \Lambda_y \Big( \Pi\, \Gamma - \Delta \Big) \,-\, \Omega\, \Gamma\, \Widehat{P}_V^y \vphantom{{\Big(\Big)}^2\}} \nonumber \\
g &=& {\Big( \Pi\, \Gamma - \Delta \Big)}^2 + \Omega \left( \Widehat{M}_H^2 - \Gamma^2 \right)
\label{eq:mt2ell}
\end{eqnarray}

Definitions for the dimensionless consolidating factors referenced in Eqs.~(\ref{eq:mt2ell}) are as follows.
In particular, all dependencies on the $(P_S^\mu,M_X)$ factors unique to the two-step topology
are sequestered here.
\begin{alignat}{3}
&\Gamma&\, &\equiv\,\, \frac{M_Y^2 - M_X^2 - M_V^2 - 2\, P_\mu^{\raisebox{-1pt}{$\scriptstyle V$}}  P_S^\mu }{2\, E_V^2}&& \nonumber \\
&&\, &\Rightarrow\,\, \frac{M_Y^2 - M_H^2 - M_V^2}{2\, E_V^2} && \!\! \left\{ \lim_{\text{1-Step}} \right\} \nonumber \\
&\Delta&\, &\equiv\,\, \frac{M_X^2 - M_H^2 - M_S^2}{2\, E_V E_S} \vphantom{\frac{M_H^2}{E_V^2}} && \Rightarrow\, 0 \nonumber \\
&\Lambda_x&\, &\equiv\,\, \frac{ E_S P_V^x - E_V P_S^x }{E_V E_S} \vphantom{\frac{M_H^2}{E_V^2}} && \Rightarrow\, 0 \nonumber \\
&\Lambda_y&\, &\equiv\,\, \frac{ E_S P_V^y - E_V P_S^y }{E_V E_S} \vphantom{\frac{M_H^2}{E_V^2}} && \Rightarrow\, 0 \nonumber \\
&\Omega&\, &\equiv\,\, {\left( \frac{P_S^z}{E_S} \right) }^2 \vphantom{\frac{M_H^2}{E_V^2}} && \Rightarrow\, 1 \nonumber \\
&\Pi&\, &\equiv\,\, 1 \vphantom{\frac{M_H^2}{E_V^2}} && \Rightarrow\, 0
\label{eq:mt2del}
\end{alignat}

Heuristically, the FIG.~\ref{fig:onestep} one-step topology is recovered from the FIG.~\ref{fig:twostep}
two-step topology as $(P_S^\mu \Rightarrow 0)$, and $(M_X \Rightarrow M_H)$.  In practice, this limit
is rather delicate, but its purpose is effected under the rules provided in Eqs.~(\ref{eq:mt2del}), in
conjunction with the discrimination of boundary versus bulk solutions made in Eqs.~(\ref{eq:ellipseph}).
Idempotency $(\Pi^2 \equiv \Pi)$ of the logical switch $\Pi$ will be freely leveraged in propagation to descendent expressions.

The two (or three) conditions of Eqs.~(\ref{eq:onestepepm} or \ref{eq:twostepepm})
have been condensed in Eq.~(\ref{eq:ellipseph}) down to only a single logical expression;
the orthogonal informatics, again, may be considered to take on the role of determining the
physically inaccessible hidden energy $E_H$ (and the longitudinal momentum $P_H^z$).
Insofar as this procedure may be performed consistently in principle, there is no tangible need to actually
implement it in practice.  For all of the one-step event topologies, and also the majority of interesting
two-step event topologies, this does prove to be the case; however, there are two-step edge cases
for which consistency requires a measure of supplemental ``shepherding'' from the remaining kinematic
constraints.  To this purpose, it is noted that the logical subset of Eqs.~(\ref{eq:twostepepm_b},\ref{eq:twostepepm_c})
is structurally equivalent to an effective one-step subsystem, as described by Eqs.~(\ref{eq:onestepepm_a},\ref{eq:onestepepm_b}),
modulo the mappings $(P_V^\mu \Rightarrow P_S^\mu)$ and $(M_Y \Rightarrow M_X)$.
Since all constraints are posed in a Lorentz-invariant fashion, this sub-analysis may be independently
boosted into the longitudinal rest frame of the secondary decay product $S$
to induce a symmetry in the formal rendering of any resulting component expressions;
consistent with this understanding, the substitution $(\Gamma \Rightarrow \Delta)$ may also
be considered in effect.

In order to extract a unified minimal compatible parent species mass $M_Y$ from the joint consideration
of both pair-production event legs $A$ and $B$, it is necessary to project the kinematic constraints on
each leg into a common space for comparison.  This is accomplished by associating the net missing
transverse momentum vector for the event with the vector sum on transverse momentum of the hidden decay
products $H$ from each event leg.
\begin{equation}
{\slashed{P}} {}_{\rm T}^{x,y} \equiv {(P_H^{x,y})}_A + {(P_H^{x,y})}_B
\label{eq:metab}
\end{equation}

An alternate (primed) set of coefficients may then be extracted for use in Eqs.~(\ref{eq:ellipseph}),
which perform a mapping of either event leg's kinematically consistent solution space into the conjugate
leg's $(\Widehat{P}_H^x,\Widehat{P}_H^y)$ hidden momentum plane.  These coefficients are provided subsequently, in terms of
the referenced leg's native coefficient set from Eqs.~(\ref{eq:mt2ell}), and the joined event
missing transverse momentum \smash{${\vec{{\slashed{P}}}} {}_{\rm T}$}.
The energy scale divided out of \smash{${\vec{{\slashed{P}}}} {}_{\rm T}$} according to the manner of Eqs.~(\ref{eq:circumflex})
is $E_V$, that native to the projected ellipse, rather than $E'_V$ of the target host ellipse.
\begin{eqnarray}
a' &=& a \quad;\quad b' \;=\; b \quad;\quad c' \;=\; c \vphantom{{\bigg(\bigg)}^2\bigg\}} \nonumber \\
d' &=& -\, \bigg\{ a\, {\slashed{\Widehat{P}}} {}_{\rm T}^{x} + b\, {\slashed{\Widehat{P}}} {}_{\rm T}^{y} + d \bigg\} \times {\left( \frac{E_V}{E'_V} \right)} \vphantom{{\bigg(\bigg)}^2\bigg\}} \nonumber \\
f' &=& -\, \bigg\{ c\, {\slashed{\Widehat{P}}} {}_{\rm T}^{y} + b\, {\slashed{\Widehat{P}}} {}_{\rm T}^{x} + f \bigg\} \times {\left( \frac{E_V}{E'_V} \right)} \vphantom{{\bigg(\bigg)}^2\bigg\}} \nonumber \\
g' &=& \bigg\{ a\, {\big( {\slashed{\Widehat{P}}} {}_{\rm T}^{x} \big)}^2 + 2\, b\, {\slashed{\Widehat{P}}} {}_{\rm T}^{x} {\slashed{\Widehat{P}}} {}_{\rm T}^{y} + c\, {\big( {\slashed{\Widehat{P}}} {}_{\rm T}^{y} \big)}^2
	\vphantom{{\bigg(\bigg)}^2\bigg\}} \nonumber \\
&+& 2\, d\, {\slashed{\Widehat{P}}} {}_{\rm T}^{x} + 2\, f\, {\slashed{\Widehat{P}}} {}_{\rm T}^{y} + g \bigg\} \times {\bigg( \frac{E_V}{E'_V} \bigg)}^2
\label{eq:mt2ellp}
\end{eqnarray}

Finally, the \smash{${\Widetilde{M}}_{\rm T2}$} statistic is recognized as the smallest trial value of $M_Y$,
if any, that is capable of producing an intersection between the overlaid regions of kinematic
consistency for the two event legs. 


\section{\protecting{Classification of \texorpdfstring{\smash{$\boldsymbol{{\Widetilde{M}}_{\rm T2}}$}}{MT2} Solutions}\label{sct:classification}}

The consistency conditions from Eqs.~(\ref{eq:ellipseph}) on the perimeter circumscribed
by the transverse momentum components $(P_H^{x,y})$ of the hidden particle $H$ exhibit the form of a general
conic section in some pair of coordinates $(x,y)$, as follows.
\begin{equation}
a\, x^2 + 2\, b\, x\, y + c\, y^2 + 2\, d\, x + 2\, f\, y + g = 0
\label{eq:ellipse}
\end{equation}

The conic discriminant $(a\,c - b^2)$ may be computed in terms of the explicit coefficient
substitutions provided in Eqs.~(\ref{eq:mt2ell}), as shown subsequently.
\begin{eqnarray}
a\,c - b^2 \,=\, \Omega &\times& \bigg\{\,
{\Big( \Lambda_x\, \Widehat{P}_V^x + \Lambda_y\, \Widehat{P}_V^y \Big)}^2
\nonumber \\
&+& \Widehat{M}_V^2 \Big( \Lambda_x^2 + \Lambda_y^2 + \Omega \Big) \bigg\}
\label{eq:conicdisc}
\end{eqnarray}

Referencing Eqs.~(\ref{eq:mt2del}), and assuming physical, non-singular
\smash{$(E_V > 0 \;\&\; E_S > 0)$} input kinematics, this expression is
well defined and manifestly positive semi-definite.  For strictly positive values of the conic discriminant,
the specified geometry is that of an ellipse.  The conic discriminant vanishes for the one-step
topologies if and only if $(M_V \Rightarrow 0)$, or for the two-step topologies,
if and only if $(\Omega \Rightarrow 0)$, {\it i.e.} if the two visible decay products $V$ and $S$
share a common longitudinal rest frame.  In this case, the corresponding geometry is that of a parabola.
More specifically, as $(\Omega \Rightarrow 0)$, the conic prescription given in Eqs.~(\ref{eq:ellipseph})
reduces to the perfect square in Eq.~(\ref{eq:2stepdegen}), encoding the geometry of a line,
{\it i.e.} a degenerately compacted (infinitely sharply folded) parabola.
\begin{equation}
{\left\{ \Lambda_x\, \Widehat{P}_H^x \,+\, \Lambda_y\, \Widehat{P}_H^y \,+\, ( \Gamma - \Delta ) \right\}}^2 \Rightarrow \,0\,
\left\{ \lim_{\text{2-Step}}^{\Omega \Rightarrow 0} \right\}
\label{eq:2stepdegen}
\end{equation}

\begin{figure}[tp]
	\centering
	\includegraphics[width=0.48\textwidth]{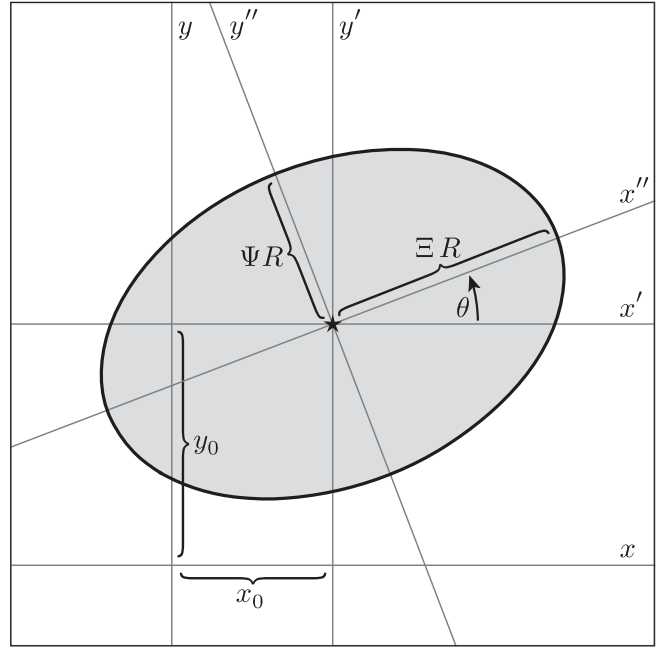}
	\caption{A generic ellipse classification, in terms of translation
	$(x_0,y_0)$, rotation $(\theta)$, and scaling $(\Xi\,R, \Psi R)$ parameters.}
	\label{fig:ellipse}
\end{figure}

Observing, however, that the main solution trunk is elliptical, it is beneficial to review the structure
of that geometric class explicitly; due consideration will be given in turn to the parabolic solution branch,
to which a majority of the results developed here will likewise apply by common inheritance.
As depicted in FIG.~\ref{fig:ellipse}, the generic ellipse may be characterized by six
parameters, including two offsets of the center by $(x_0,y_0)$ from the coordinate origin,
one rotation angle $\theta$, two relative scaling factors $(\Xi,\Psi)$ for displacements along
the translated, rotated axes, and an overall scale parameter $R$. Only five of these
parameters are actually independent, as there is a degeneracy in the definitions of $(\Xi,\Psi,R)$
that preserves as invariants only the combinations $\Xi\, R$ and $\Psi R$, with a combined dimensionality
equivalent to that of the coordinates; this is in precise correspondence with the freedom to extract an
overall scale from the six coefficients of Eq.~(\ref{eq:ellipse}).
The equations governing the generic elliptical structure are as follows.
\begin{gather}
x' \equiv x - x_0 \quad;\quad y' \equiv y - y_0 \nonumber \\
x'' \equiv x' \cos \theta + y' \sin \theta \nonumber \\
y'' \equiv y' \cos \theta - x' \sin \theta \nonumber \\
{\left( \frac{\hphantom{''}x''}{\Xi} \right)}^2 +\, {\left( \frac{\hphantom{''}y''}{\Psi} \right)}^2 = R^2
\label{eq:elldef}
\end{gather}

The corresponding ellipse coefficients may be read off by sequential application of the prior coordinate
redefinitions, with recursive references used to simplify
the form, as shown next.  The sign convention $(a \ge 0 \;\&\; c \ge 0)$ is in effect here, in agreement
with Eqs.~(\ref{eq:mt2ell}).
\begin{eqnarray}
a &=& \frac{\cos^2 \theta}{\Xi^2} + \frac{\sin^2 \theta}{\Psi^2} \nonumber \\
b &=& \sin \theta \cos \theta \left( \frac{1}{\Xi^2} - \frac{1}{\Psi^2} \right) \nonumber \\
c &=& \frac{\sin^2 \theta}{\Xi^2} + \frac{\cos^2 \theta}{\Psi^2} \nonumber \\
d &=& - \left( a\, x_0 + b\, y_0 \right) \vphantom{\frac{\sin^2 \theta}{\Psi^2}} \nonumber \\
f &=& - \left( c\, y_0 + b\, x_0 \right) \vphantom{\frac{\sin^2 \theta}{\Psi^2}} \nonumber \\
g &=& \frac{{( x_0 \cos \theta + y_0 \sin \theta )}^2}{\Xi^2} \nonumber \\
  &+& \frac{{( y_0 \cos \theta - x_0 \sin \theta )}^2}{\Psi^2} - R^2
\label{eq:genellcoeff}
\end{eqnarray}

The content of Eqs.~(\ref{eq:genellcoeff}) may also be inverted in favor of the physical parameter
set, as follows.
\begin{eqnarray}
x_0 &=& \frac{ b\,f - c\,d }{ a\,c - b^2 } \vphantom{\frac{{(a^2)}^2}{{(b^2)}^2}} \nonumber \\
y_0 &=& \frac{ b\,d - a\,f }{ a\,c - b^2 } \vphantom{\frac{{(a^2)}^2}{{(b^2)}^2}} \nonumber \\
\tan \theta &=& \frac{ (c - a) - \surd[\, {( c - a )}^2 + 4\,b^2 \,] }{2\,b} \vphantom{\frac{{(a^2)}^2}{{(b^2)}^2}} \nonumber \\
{\left\{\, \Xi^2, \Psi^2 \right\}} &=& \frac{2}{ (a + c) \mp \surd[\, {( c - a )}^2 + 4\,b^2 \,]} \vphantom{\frac{{(a^2)}^2}{{(b^2)}^2}} \nonumber \\
R^2 &=& \frac{ a\,f^2 + c\,d^2 - 2\, b\,d\,f}{ a\, c - b^2 } - g \vphantom{\frac{{(a^2)}^2}{{(b^2)}^2}}
\label{eq:physellcoeff}
\end{eqnarray}

Rotation by $\pi$ radians is a symmetry of the elliptical geometry, as reflected by the $\pi$-wide principal
range of the inverse tangent function.  A further potential ambiguity
of $\pi/2$ radians associated with exchange of the axial scaling factors $(\Xi \Leftrightarrow \Psi)$ has
been alleviated by choosing signs for root splittings to enforce the conventions of FIG.~\ref{fig:ellipse},
wherein $\Xi$ gauges extension of the major axis, and $\theta$ is the standard angular orientation of that axis relative
to the positive $(x)$ coordinate direction.  The ellipse eccentricity $\varepsilon$, as defined subsequently, and the angle $\theta$,
depend only upon the parameter subset $(a,b,c)$, which are static in Eqs.~(\ref{eq:mt2ell},\ref{eq:mt2del},\ref{eq:mt2ellp}) 
under variation of the trial parent particle species mass $M_Y$, and also under projection onto the conjugate event leg hidden momentum plane.
By contrast, the ellipse coordinate center $(x_0,y_0)$ and squared scale factor $R^2$ do vary with respect to $M_Y$, via
dependence upon the parameter subset $(d,f,g)$, which are in turn dependent upon the factor $\Gamma$.
\begin{equation}
\varepsilon \,\equiv\, \sqrt{ 1 - \Psi^2 /\, \Xi^2}
\label{eq:eccentricity}
\end{equation}

Reality constraints $(\Xi^2,\Psi^2) \ge 0$ on the axial scaling factors are satisfied trivially,
given kinematic positivity of the conic discriminant $(a\,c - b^2)$; the larger of this pair
$(\Xi^2)$ will diverge on the parabolic branch, as is intuitively consistent
with the infinite displacement of an elliptical focal point as the geometry ``opens up'' on one side.
Conversely, enforcing the matching reality condition $(R^2 \ge 0)$ on the unified ellipse
scale provides a vital event consistency classification in terms of kinematic boundaries on $M_Y$.
The search for transitions into or out of compliance with this requirement may be posed as a quadratic equation
\smash{$(R^2 \propto \alpha_\odot \Gamma^2_\odot + \beta_\odot \Gamma_\odot + \gamma_\odot = 0)$} in the factor $\Gamma$
from Eqs.~(\ref{eq:mt2del}), with coefficients as defined in Eqs.~(\ref{eq:mt2min}).
A global positive dimensionless factor $\Omega^2 /\, (a\,c\, - b^2)$ has been divided out;
incidentally, the proportionality $(R^2 \propto \Omega)$ explains 
the previously observed geometric degeneracy in the $(\Omega~\Rightarrow~0)$ limit.
Corresponding real $(M_Y^2 \ge 0)$ parent mass scales may then be extracted from the
quadratic inversions \smash{$\{\, \Gamma_{\odot\pm} \equiv ( - \beta_\odot \pm \surd[\, \beta_\odot^2 - 4\, \alpha_\odot \gamma_\odot \,] \,) /\, 2\,\alpha_\odot \,\}$},
which are linearly dependent upon $M_Y^2$, with a positive slope $+(1 / 2\, E_V^2)$, associating the positive root with $M_Y$ itself.
\begin{widetext}
\vspace{-12pt}
\begin{eqnarray}
\alpha_\odot &\equiv&
\Omega\, -\, \Pi\,
+\, {\left\{ \Lambda_x - \Pi\, \Widehat{P}_V^x \right\}}^2
+\, {\left\{ \Lambda_y - \Pi\, \Widehat{P}_V^y \right\}}^2
\vphantom{\left[{\left\{\Widehat{P}_V^x\right\}}^2\right]} \nonumber \\
\beta_\odot &\equiv& 2\, \Delta \left\{ \Pi\, \Widehat{M}_V^2 \,+\, \Lambda_x\, \Widehat{P}_V^x \,+\, \Lambda_y\, \Widehat{P}_V^y \right\}
	\vphantom{\left[{\left\{\Widehat{P}_V^x\right\}}^2\right]} \nonumber \\
\gamma_\odot &\equiv& \Widehat{M}_H^2 \left[ {\left\{ \Lambda_x\, \Widehat{P}_V^y - \Lambda_y\, \Widehat{P}_V^x \right\}}^2
	-\, \Lambda_x^2\, -\, \Lambda_y^2\, -\, \Omega\, \Widehat{M}_V^2 \right] - \Delta^2\, \Widehat{M}_V^2
\label{eq:mt2min}
\end{eqnarray}
\end{widetext}

For the one-step event topology, applying the limits furnished in Eqs.~(\ref{eq:mt2del}), the
$R^2$ parabolic dependence on $\Gamma$ develops upward concavity as
\smash{$(\alpha_\odot \Rightarrow +1 > 0)$}.  The solutions \smash{$(\Gamma_{\odot\pm} \Rightarrow \pm \Widehat{M}_H \Widehat{M}_V )$}
generically admit two real, positive roots $\left| M_H \pm M_V \right|$ for $M_Y$, between which $(R^2 < 0)$.
The physical $(+)$ root provides a lower limit in the search for \smash{${\Widetilde{M}}_{\rm T2}$},
corresponding to the parent particle species mass at which the kinematically consistent elliptical
region initially materializes as a singular point \smash{$(P_H^{x,y})_{\odot+}$}, and beyond which it may expand in scale without bound.
This root \smash{$(M_Y^{\odot+} \Rightarrow M_H + M_V)$} is
readily interpreted ({\it cf.} FIG.~\ref{fig:onestep}) as the threshold 
rest mass necessary for decay into the $(H,V)$ particle species.

For the two-step event topology, applying factors from Eqs.~(\ref{eq:mt2del}),
\smash{$(\alpha_\odot \Rightarrow - {\{M_S / E_S \}}^2 \le 0)$}, which pushes the $R^2$
parabola into a regime of downward concavity (for $M_S > 0$) and reverses the ordering in $\Gamma$ of the
$(\pm)$ signed $(R^2 = 0)$ quadratic roots.  The quadratic discriminant
\smash{$(\beta_\odot^2 - 4\,\alpha_\odot \gamma_\odot)$}
may be computed explicitly, as follows.
\begin{eqnarray}
\beta_\odot^2 - 4\,\alpha_\odot \gamma_\odot
&=&
\left\{ M_X^2 - ( M_H + M_S)^2  \right\}
\nonumber \\
&\times&
\left\{ M_X^2 - ( M_H - M_S)^2  \right\}
\nonumber \\
&\times&
\left\{\, ( P_\mu^{\raisebox{-1pt}{$\scriptstyle V$}}  P_S^\mu )^2 - M_V^2 M_S^2 \,\right\}
\nonumber \\
&\div&
(E_V E_S)^4 \vphantom{ \left\{(P_\mu^{\raisebox{-1pt}{$\scriptstyle V$}})^2\right\} }
\label{eq:gammadisc}
\end{eqnarray}

Real roots for \smash{$\Gamma_\odot$} will thus be yielded if the production threshold $(M_X \ge M_H + M_S)$ is met.  Curiously, real
roots are also developed in the kinematically off-shell region $M_X \le \left| M_H - M_S \right|$,
whereas imaginary roots do appear in the mass gap separating the two regimes.
Restricting consideration to the scenario where genuine, real solutions for \smash{$\Gamma_\odot$} do exist, energy-momentum
conservation at each vertex ensures that a consistent, minimally on-shell parent particle species mass
limit may be extracted in each case.  The role of the $(+)$ root persists from the
one-step topology, as \smash{$M_Y^{\odot+}$} continues to represent the threshold at which the
corresponding elliptical region kinematically onsets.  However, the $(-)$ root is now the heavier
of the two, and it attains physical significance here as well; growth of the hidden momentum
ellipse scale is now bounded, and \smash{$M_Y^{\odot-}$} represents a conjugate parent mass threshold
at which the squared radial scale factor $R^2$ recollapses to zero following successive phases of expansion
and contraction.  The loci of these singular points \smash{$(P_H^{x,y})_{\odot\pm}$} in the hidden momentum plane
may be disjoint, such that the related ellipse transits across the solution space
as the parent particle species mass $M_Y$ is scanned; this is in contrast to the one-step decay topology,
where the origination point \smash{$(P_H^{x,y})_{\odot+}$} is persistently encompassed by the ellipse at all scale sizes.

Intuitively, the appearance of an upper $M_Y$ bound reflects the greater degree of kinematic constraint
that is in force for the two-step event topology.  In the one-step
decay topology ({\it cf.} FIG.~\ref{fig:onestep}) inherent kinematic indeterminacy of the escaping hidden
product $H$ implies that elevation of the parent particle species mass $M_Y$ may simply be shunted onto
a compensating relativistic boost of $P_H^\mu$.  By contrast, in the two-step decay topology
({\it cf.} FIG.~\ref{fig:twostep}) the on-shell internal mediator species is kinematically constrained at
vertices on both ends, rendering it too slender a conduit for the evacuation of indefinite amounts of energy and momentum.

Quantitative insight into this effect is garnered by consideration of the two extremal scenarios
in which the quadratic discriminant for \smash{$\Gamma_\odot$} vanishes, cutting off the mass splitting between
\smash{$M_Y^{\odot+}$} and \smash{$M_Y^{\odot-}$}.  The first such scenario places $(M_X \Rightarrow M_H + M_S)$
directly at its minimal production threshold, which is equivalent to the statement
\smash{$(P_\mu^{\raisebox{-1pt}{$\scriptstyle H$}} P_S^\mu \Rightarrow M_H M_S)$}, 
implying that $(P_H^\mu \propto P_S^\mu)$ develop co-moving rest frames.
The measurement of $P_S^\mu$ thus suggests also (given the mass $M_H$) a full knowledge
of $P_X^\mu$, which may be combined in turn with the observable four-vector $P_V^\mu$ to exactly calculate $M_Y$;
if $M_H$ is zero, the four-vector proportionality $(M_H/M_S)$ becomes indeterminate and is replaced by $(E_H/E_S)$,
which is unspecified and unmeasured, and may thus assume any positive semi-definite value.
The second such scenario directly pushes the invariant four-vector contraction
\smash{$(P_\mu^{\raisebox{-1pt}{$\scriptstyle V$}}  P_S^\mu \Rightarrow M_V M_S)$}
to its lower kinematic limit, corresponding again to the onset of co-moving rest frames, now for the visible
decay products $(P_V^\mu \propto P_S^\mu)$; boosting into this rest frame, and (hypothetically) rotating into alignment with
the (unknown) three-velocity ${\vec{v}}_Y$ of the parent species $Y$, the full transfer of momentum and
partial transfer of energy (modulo the dropping out of $M_V$ and $M_S$) from particle $Y$ to particle
$X$ to particle $H$ (the latter two having specified masses) enforce four algebraic conditions on
four unknowns $(M_Y,v_Y,v_X,v_H)$, such that $M_Y$ is exactly calculable; note that
$(\Omega \Rightarrow 0)$ is an implied superclass of $(P_V^\mu \propto P_S^\mu)$.
Relaxing away from either of these kinematic proportionalities,
a finite splitting between the allowed boundaries on $M_Y$ is recovered.

The imposition of a two-step decay topology on either event leg yields a natural upper and lower
boundary on the viable search space for a minimal kinematically consistent parent particle species mass.
However, it is also potentially useful to systematically establish an upper trial mass boundary for the scenario where a one-step
decay topology is imposed on both event legs.  This task may be equivalently reframed as the establishment
of a single value of $M_Y$ where the two ellipses are positively known to possess an overlap.  Given that
elliptical regions associated with one-step decay topologies must necessarily grow in all directions without
bound as a function of increasing $M_Y$, such a mass may always be established.  One suitable strategy
consists of solving for the scale $M_Y^\oplus$ at which a given ellipse's kinematic perimeter eclipses the hidden
momentum origination coordinates \smash{$( {\Widehat{\mathcal{X}}}_{\odot+}, {\Widehat{\mathcal{Y}}}_{\odot+} )$} associated with its partner.
For each event leg, the expression of this onset point \smash{$(\Widehat{P}_H^{x,y})_{\odot+}$} in its self-hosted dimensionless
coordinates may be identified with the ellipse center $(x_0,y_0)$ from Eqs.~(\ref{eq:physellcoeff}),
inserting the Eqs.~(\ref{eq:mt2ell}) coefficients, with the factors (specifically $\Gamma$) from
Eqs.~(\ref{eq:mt2del}) evaluated at the \smash{$M_Y^{\odot+}$} kinematic threshold mass;
the result may be redimensioned by multiplying with $E_V$.  For the one-step event topology, this
location reduces to $\{M_H/M_V\}\times(P_V^{x,y})$.  When projected onto the conjugate event leg's
hidden momentum plane via Eq.~(\ref{eq:metab}), this, or any, pair of coordinates shift, as shown following.
\begin{equation}
( {\Widehat{\mathcal{X}}}, {\Widehat{\mathcal{Y}}} ) \;\equiv\;
\big[\, ({\slashed{\Widehat{P}}} {}_{\rm T}^{x,y}) - {(\Widehat{P}_H^{x,y}}) \,\big] \times \left( \frac{E_V}{E'_V} \right)
\label{eq:projcoords}
\end{equation}

In the prior, the energy scale $E_V$ represents that native to the projected ellipse,
whereas $E'_V$ is that of the target host ellipse.
In lieu of subtraction, Eqs.~(\ref{eq:mt2ellp},\ref{eq:physellcoeff}) may be used to equivalently compute
the projected ellipse's modified origination coordinates directly in the target system.
Associating some \smash{$( {\Widehat{\mathcal{X}}}, {\Widehat{\mathcal{Y}}} )$}
with points $(x,y)$ on the host ellipse's perimeter, as expressed in Eq.~(\ref{eq:ellipse}),
employing again the content of Eqs.~(\ref{eq:mt2ell},\ref{eq:mt2del}), a quadratic intersection criterion
\smash{$(\alpha_\oplus \Gamma_\oplus^2 + \beta_\oplus \Gamma_\oplus + \gamma_\oplus = 0)$} in the factor $\Gamma$
from Eqs.~(\ref{eq:mt2del}) emerges, with coefficients as defined in Eqs.~(\ref{eq:mt2max}).
\begin{widetext}
\vspace{-12pt}
\begin{eqnarray}
\alpha_\oplus &\equiv& \Omega\, -\, \Pi \vphantom{{\left\{ {\Widehat{\mathcal{X}}} \right\}}^2} \nonumber \\
\beta_\oplus &\equiv& 2\,\Omega\, {\left\{  \Widehat{P}_V^x\, {\Widehat{\mathcal{X}}} +  \Widehat{P}_V^y\, {\Widehat{\mathcal{Y}}} \right\}}
	\,+\, 2\, \Pi\, \left\{ \Delta - \Lambda_x\, {\Widehat{\mathcal{X}}} - \Lambda_y\, {\Widehat{\mathcal{Y}}} \right\}
	\vphantom{{\left\{ {\Widehat{\mathcal{X}}} \right\}}^2} \nonumber \\
\gamma_\oplus &\equiv& \Omega\, {\left\{ \Widehat{P}_V^x\, {\Widehat{\mathcal{X}}} +  \Widehat{P}_V^y\, {\Widehat{\mathcal{Y}}} \right\}}^2
	\,-\, \Omega\, {\left\{ \Widehat{M}_H^2 + {\Widehat{\mathcal{X}}}^2 + {\Widehat{\mathcal{Y}}}^2 \right\}}
	\,-\, {\left\{ \Delta - \Lambda_x\, {\Widehat{\mathcal{X}}} - \Lambda_y\, {\Widehat{\mathcal{Y}}} \right\}}^2
\label{eq:mt2max}
\end{eqnarray}
\end{widetext}

Although Eqs.~(\ref{eq:mt2max}) are written in a form generic to both the one- and two-step decay
topologies, applicability to the latter scenarios is limited, due to the related facts that the
ellipse scale does not grow indefinitely with $M_Y$ and persistent elliptical containment of the
origination coordinate is not guaranteed.
In the former scenario, applying the limits furnished in Eqs.~(\ref{eq:mt2del}), the 
parabolic dependence on $\Gamma$ develops upward concavity
as \smash{$(\alpha_\oplus \Rightarrow + 1 > 0)$}.
The quadratic inversions 
\smash{$\{\, \Gamma_{\oplus\pm} \equiv ( - \beta_\oplus \pm \surd[\, \beta_\oplus^2 - 4\, \alpha_\oplus \gamma_\oplus \,] \,) /\, 2\,\alpha_\oplus \,\}$},
are generically real, and the physical $(+)$ root exhibited in Eq. (\ref{eq:gmmaxplus}) also necessarily admits a real,
positive root for $M_Y$, providing an upper limit in the search for \smash{${\Widetilde{M}}_{\rm T2}$}.
This scale \smash{$\{\, M_Y^{\oplus+} \ge ( M_Y^{\odot+} \Rightarrow M_H + M_V)\,\}$} is at least as large as the
mass threshold at which the kinematically consistent elliptical region initially materializes as a singular
point, and beyond which it may expand to intersect any specified position in the hidden momentum plane.
\begin{equation}
\Gamma_{\oplus+} \Rightarrow\,
\surd[\, \Widehat{M}_H^2 + {\Widehat{\mathcal{X}}}^2 + {\Widehat{\mathcal{Y}}}^2 \,]
- \Widehat{P}_V^x\, {\Widehat{\mathcal{X}}} -  \Widehat{P}_V^y\, {\Widehat{\mathcal{Y}}}
\,\, \left\{ \lim_{\text{1-Step}} \right\} 
\label{eq:gmmaxplus}
\end{equation}

It will be conventional to adopt hidden momentum coordinates hosted by the
heavier event leg, commencing primary analysis at the associated scale \smash{${(M_Y^{\odot+})}$},
with first annexation of the plane occupied in waiting by the lighter projected event leg.
In practice, however, it can be beneficial to compute \smash{$M_Y^{\oplus+}$} from both
directions, interchanging the respective host and projection roles.
An interesting inversion case may be identified whenever the intersection mass
\smash{${M_Y^{\oplus+}}$} of a lighter host leg with the origination coordinates
\smash{$( {\Widehat{\mathcal{X}}}_{\odot+}, {\Widehat{\mathcal{Y}}}_{\odot+} )$} of
a heavier projected leg is less than the turn-on mass \smash{${M_Y^{\odot+}}$} of the
projected leg. This implies that the host ellipse will already have passed over the projected
origination coordinate at the scale of its initial materialization.
Since the one-step elliptical perimeter of Eqs.~(\ref{eq:ellipseph}) inclusively bounds a
closed two-dimensional area, the minimal consistent kinematic overlap between the two
legs is satisfied immediately at the scale \smash{${M_Y^{\odot+}}$},
defining \smash{${\Widetilde{M}}_{\rm T2}$} in prior of any actual intersection.
This ``unbalanced'' solution, to be called ``type~I\nobreak\hspace{0.6pt}\nobreak'',
may also be relevant to mixed one- and two-step event topologies,
if the lighter ellipse is of the one-step variety, despite the fact that \smash{$M_Y^{\oplus+}$}
is not then suitable (or necessary) for setting an upper bound on \smash{${\Widetilde{M}}_{\rm T2}$} 
if the ellipse perimeters are ``balanced'', {\it i.e.} non-overlapping
at the turn-on mass of the heavier event leg.  A case study exhibiting trivial unbalanced
kinematics in the dual one-step decay topology context is presented in a footnote\footnote{
	This note provides the kinematic blueprint for an example event with dual one-step decay topology
	that satisfies minimal parent mass $M_Y$ constraints in an unbalanced fashion. 
	The first (heavier, host) ellipse is selected with $P_V^\mu = (125, +75, -50, +75)$~GeV and $M_H = 80.4$~GeV.
	The second (lighter, projected) ellipse is selected with $P_V^\mu = (75, +50, -25, -25)$~GeV and $M_H = 0.0$~GeV.
	The event missing energy components are selected as \smash{${\slashed{P}} {}_{\rm T}^{x,y} = (+125.0,-75.0)$}~GeV.
	The projected ellipse materializes at the threshold mass \smash{$M_Y^{\odot+} = 43.3$}~GeV, at the position \smash{$(P_H^{x,y})_{\odot+} = (+125.0,-75.0)$}~GeV.
	The host ellipse materializes on the interior of the closed projected ellipse, at the threshold mass \smash{$M_Y^{\odot+} = 123.7$}~GeV,
	at the position \smash{$(P_H^{x,y})_{\odot+} = (+139.3, -92.8)$}~GeV,
	immediately defining a type I unbalanced solution for the \smash{${\Widetilde{M}}_{\rm T2}$} event scale.}.

If no such trivial solution exists, then \smash{${\Widetilde{M}}_{\rm T2}$}
must be established by scanning in $M_Y$ for the first intersection (if any)
to occur above the heavier of the two leg's minimal mass scales \smash{$M_Y^{\odot+}$} and below
the lighter of the two leg's maximal mass scales, either of the \smash{$M_Y^{\odot-}$} kinematic
turn-off variety (for two-step event legs) or of the \smash{$M_Y^{\oplus+}$} origination coordinate
intersection variety (for dual one-step event legs).  If the minimal and maximal search
boundaries are disjoint, or if the scan reveals that intersection is precluded (this cannot occur with
dual one-step event legs), then no physical asymmetric \smash{${\Widetilde{M}}_{\rm T2}$} event scale may
be defined.  If a numeric report is still desired in this case, the value of \smash{${\Widetilde{M}}_{\rm T2}$} 
may revert to either zero or some arbitrarily large cap, depending on the application; the former
choice is typical when attempting to fit the mass scale of a signal for new physics and the latter choice
is typical when attempting to suppress background from known physics for a specific event topology hypothesis.

\begin{figure}[tp]
	\centering
	\includegraphics[width=0.48\textwidth]{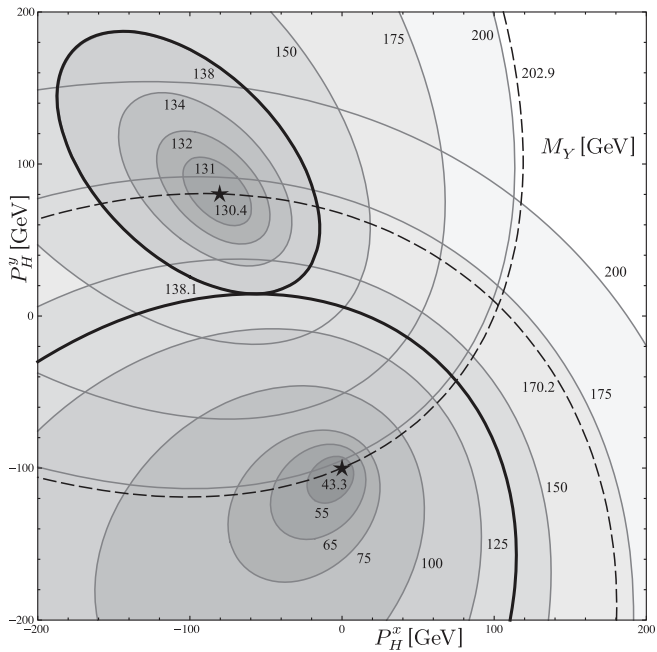}
	\caption{Intersection of elliptical kinematic consistency regions
	for an example event with dual one-step decay topology.
	The upper (heavier, host) ellipse is selected with $P_V^\mu = (100, -50, +50, -50)$~GeV and $M_H = 80.4$~GeV.
	The lower (lighter, projected) ellipse is selected with $P_V^\mu = (75, +25, +25, +50)$~GeV and $M_H = 0.0$~GeV.
	The event missing energy components are selected as \smash{${\slashed{P}} {}_{\rm T}^{x,y} = (0.0,-100.0)$}~GeV.
	The projected ellipse materializes at the threshold mass \smash{$M_Y^{\odot+} = 43.3$}~GeV, at the position \smash{$(P_H^{x,y})_{\odot+} = (0.0,-100.0)$}~GeV.
	The host ellipse materializes on the exterior of the closed projected ellipse, at the threshold mass \smash{$M_Y^{\odot+} = 130.4$}~GeV,
	at the position $(P_H^{x,y})_{\odot+} = (-80.4,+80.4)$~GeV.
	The projected and host ellipses intersect their conjugate leg's origination coordinates at mass scales of \smash{$M_Y^{\oplus+} = (170.2,202.9)$}~GeV, respectively.
	The ellipses first intersect at $M_Y = 138.1$~GeV, defining the \protecting{\smash{${\Widetilde{M}}_{\rm T2}$}} event scale.\vspace{0.5pt}}
	\label{fig:intersection_1}
\end{figure}

\begin{figure}[tp]
	\centering
	\includegraphics[width=0.48\textwidth]{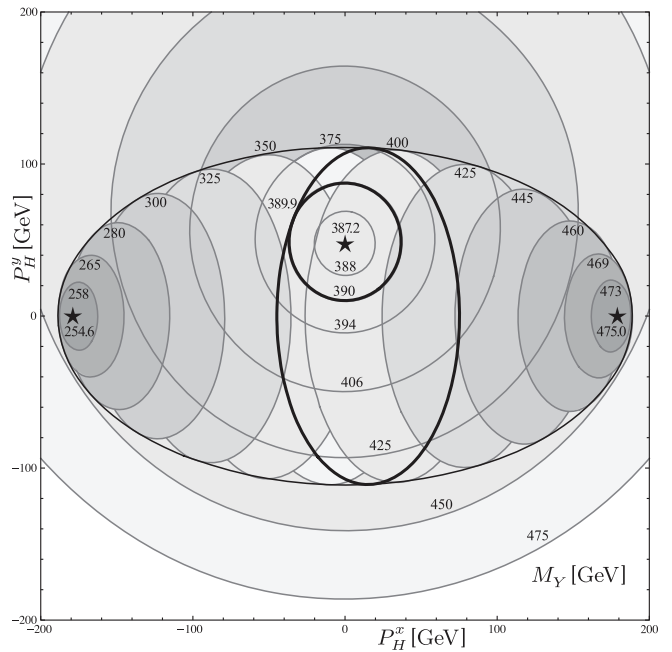}
	\caption{Intersection of elliptical kinematic consistency regions
	for an example event with mixed one- and two-step decay topology.
	The central (heavier, host, one-step) ellipse is selected with $P_V^\mu = (250, 0, +75, +25)$~GeV and $M_H = 150.0$~GeV.
	The left-to-right transiting (lighter, projected, two-step) ellipse is selected with $P_V^\mu = (150, +50, 0, -25)$~GeV,
	$P_S^\mu = (50, -35, 0, +25)$~GeV, $M_H = 55.0$~GeV and $M_X = 100.0$~GeV;
	the dynamic trajectory of this geometry is statically bounded by a one-step elliptical shepherd.
	The event missing energy components are selected as \smash{${\slashed{P}} {}_{\rm T}^{x,y} = (-170.3,0.0)$}~GeV.
	The projected ellipse materializes at the threshold mass \smash{$M_Y^{\odot+} = 254.6$}~GeV, at the position \smash{$(P_H^{x,y})_{\odot+} = (-178.9,0.0)$}~GeV,
	and recollapses at the threshold mass \smash{$M_Y^{\odot-} = 475.0$}~GeV, at the position \smash{$(P_H^{x,y})_{\odot-} = (+178.9,0.0)$}~GeV.
	The host ellipse materializes on the interior of the open projected ellipse, at the threshold mass \smash{$M_Y^{\odot+} = 387.2$}~GeV,
	at the position \smash{$(P_H^{x,y})_{\odot+} = (0.0,+47.4)$}~GeV.
	The ellipses first intersect at $M_Y = 389.9$~GeV, defining the \protecting{\smash{${\Widetilde{M}}_{\rm T2}$}} event scale.\vspace{6.6pt}}
	\label{fig:intersection_2}
\end{figure}

\interfootnotelinepenalty=10000
The establishment of \smash{${\Widetilde{M}}_{\rm T2}$} by the method of intersecting
kinematically consistent perimeters in the $(P_H^{x,y})$ transverse hidden momentum plane
is depicted for a pair of canonical examples in FIG.~\ref{fig:intersection_1}
(dual one-step decay topology) and and FIG.~\ref{fig:intersection_2} (mixed one- and two-step decay topology),
with overlaid elliptical contours sampled at various trial masses $M_Y$ for the parent particle species $Y$.
The shepherding role of the subordinate one-step ellipse is clearly exhibited in FIG.~\ref{fig:intersection_2},
as a static perimeter that encompasses and perfectly silhouettes the inception, translation, and dissolution
of its two-step analog.  A case study exhibiting the malady of spurious compatibility with reality constraints, which
may afflict events with a two-step decay topology,
is presented in a footnote\footnote{
	This note provides the kinematic blueprint for an example event that inauthentically satisfies
	mass-shell reality constraints with a mixed one- and two-step decay topology.
	The one-step ellipse is selected with $P_V^\mu = (50, 0, +25, -25)$~GeV and $M_H = 25.0$~GeV.
	The two-step ellipse is selected with $P_V^\mu = (75, -50, -25, +50)$~GeV,
	$P_S^\mu = (130, 0, 0, +50)$~GeV, $M_H = 30.0$~GeV and $M_X = 30.0$~GeV.
	The event missing energy components are selected as \smash{${\slashed{P}} {}_{\rm T}^{x,y} = (+50.0,+25.0)$}~GeV.
	The mass threshold for production of the secondary visible product $S$ and escaping hidden product $H$ ({\it cf.} FIG.~\ref{fig:twostep})
	is $(M_X \ge M_S + M_H = 150)$~GeV, which is manifestly unsatisfied.
	However, since the hierarchy $(M_X \le \left| M_S - M_H \right| = 90)$~GeV
	is satisfied, the two-step ellipse specification may remain real.
	Codes that do not actively filter against this scenario will
	spuriously report a first intersection of the two ellipses at $M_Y = 74.2$~GeV, although no
	genuine asymmetric \smash{${\Widetilde{M}}_{\rm T2}$} event scale may be defined in this case.}.
Incidentally, the one-step shepherd also constitutes an effective filter against this spurious two-step root.
\interfootnotelinepenalty=100

\section{\protecting{Critical Point Behavior of \texorpdfstring{\smash{$\boldsymbol{{\Widetilde{M}}_{\rm T2}}$}}{MT2}}\label{sct:limits}}

A complete taxonomy of possible \smash{${\Widetilde{M}}_{\rm T2}$} solutions must
supplement the preceding description of canonical event scenarios with an analysis
of the geometric reaction to approaching various kinematic critical points.
Certain of these solution classes are physically essential, while others are
of primarily academic interest, and unlikely to be represented in realistic data
samples.  Nevertheless, comparative intuition and computational surety will benefit
from the study, for completeness' sake, of transitions toward and among all
demonstrable edge cases.

\begin{figure}[tp]
	\centering
	\includegraphics[width=0.48\textwidth]{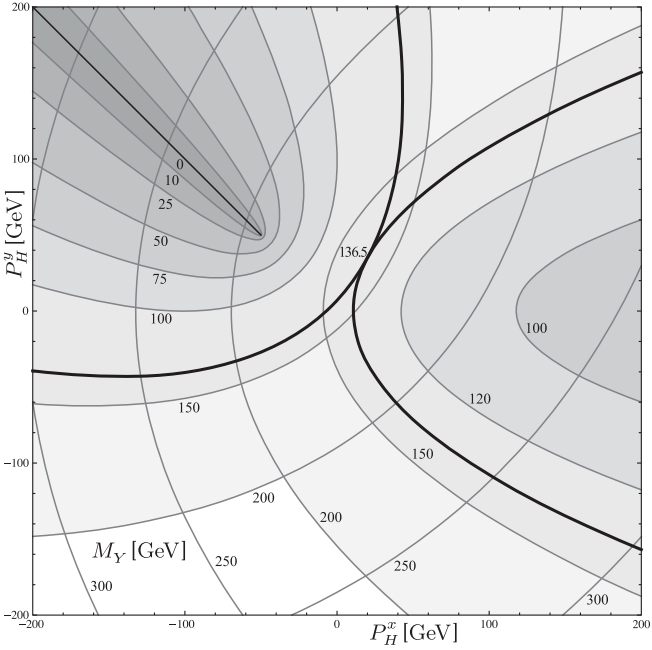}
	\caption{Intersection of parabolic kinematic consistency regions
	for an example event with dual one-step decay topology, where
	the visible decay product of each leg is massless $(M_V \Rightarrow 0)$.
	The central-right (heavier, host) parabola is selected with $P_V^\mu = (100, 100, 0, 0)$~GeV and $M_H = 75.0$~GeV.
	The upper-left (lighter, projected) parabola is selected with $P_V^\mu = (75, +50, -50, 25)$~GeV and $M_H = 0.0$~GeV.
	The event missing energy components are selected as \smash{${\slashed{P}} {}_{\rm T}^{x,y} = (-50.0,+50.0)$}~GeV.
	The projected parabola materializes at the threshold mass \smash{$M_Y^{\odot+} = 0.0$}~GeV, with vertex at the position ${(P_H^{x,y})}_{\odot+} = (-50.0,+50.0)$~GeV.
	The host parabola materializes on the exterior of the closed projected parabola, at the threshold mass \smash{$M_Y^{\odot+} = 75.0$}~GeV,
	with vertex infinitely displaced from the coordinate origin.
	The parabolas first intersect at $M_Y = 136.5$~GeV, defining the \protecting{\smash{${\Widetilde{M}}_{\rm T2}$}} event scale.\vspace{2.3pt}}
	\label{fig:intersection_3}
\end{figure}

Beginning with the one-step event topology, the previously referenced parabolic
branch is accessed in the limit $(M_V \Rightarrow 0)$.  The coordinate center
$(x_0,y_0)$, as specified in Eqs.~(\ref{eq:physellcoeff}), becomes undefined 
due to vanishing of the conic discriminant and severe associated
distention of the underlying geometry.  Correspondingly, the parabola does
not kinematically onset as a point, but rather as a line segment.
A suitable proxy for the elliptical coordinate center is the location
of the parabolic vertex.  The associated transverse hidden momentum $(P_H^{x,y})$ coordinate
pair may be determined by implicitly differentiating the Eq.~(\ref{eq:ellipse}) specification of the
conic perimeter to establish a slope function $(dy/dx)$, forcing that slope to take a value
\smash{$\{ -1\,/\tan \theta \Rightarrow (b/a) \equiv -1\times(\Widehat{V}_x/\Widehat{V}_y)\,\}$}
oriented normal to the inclination of the parabolic symmetry axis ---
the form of which carries over intact from Eqs.~(\ref{eq:physellcoeff}),
and which has been further simplified by applying the degeneracy relation $(b^2 \Rightarrow a\,c)$ and the
corresponding kinematic limit $(M_V \Rightarrow 0)$ --- and subsequently convolving the deduced 
constraint on $(x)$ and $(y)$ back into Eq.~(\ref{eq:ellipse}).
The result of this procedure, reprojected onto physical kinematic parameters, is as follows.
\begin{equation}
{(\Widehat{P}_H^{x,y})}_{\odot}
\,\,\Rightarrow\,\,
(\Widehat{V}_x,\Widehat{V}_y) \times
\left(
\frac{\Widehat{M}_H^2-\Gamma^2}{2\,\Gamma}
\right)
\,\, \left\{ \lim_{\text{1-Step}}^{M_V \hspace{0.4pt}\Rightarrow\, 0} \right\}
\label{eq:orgparproxy}
\end{equation}

If the limit $(M_H \Rightarrow 0)$ is also in effect, the vertex of the parabola
will onset at the coordinate origin of its self-hosted \smash{$(P_H^{x,y})$}
transverse hidden momentum plane; otherwise, the vertex will onset with infinite
displacement from the origin, but rapidly proceed toward it.  In both cases,
as the mass scale $M_Y$ increases, the parabola unfolds, and it's vertex advances 
directly away from the point of focus, along the axis of reflection symmetry.
This geometry represents a smooth transition away from the one-step ellipse,
which grows increasingly narrow and elongated as $(M_V \Rightarrow 0)$ from
positive values.  Like the one-step ellipse, the one-step parabola bounds
a planar continuum of bulk solutions to its interior, and the solution perimeter 
expands indefinitely with $M_Y$, without yielding subjugated territory,
encompassing the complete coordinate plane as $(M_Y \Rightarrow \infty)$.
A depiction of two intersecting one-step parabolas is provided in FIG.~\ref{fig:intersection_3}.

\begin{figure}[tp]
	\centering
	\includegraphics[width=0.48\textwidth]{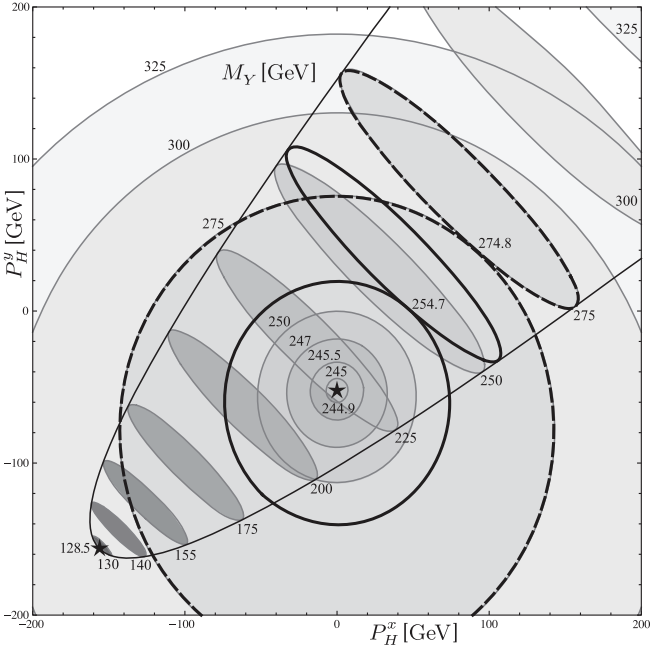}
	\caption{Intersection of elliptical kinematic consistency regions
	for an example event with mixed one- and two-step decay topology,
	where the secondary visible decay product of the latter leg is massless $(M_S \Rightarrow 0)$.
	The central (heavier, host, one-step) ellipse is selected with $P_V^\mu = (150, 0, -50, +75)$~GeV and $M_H = 125.0$~GeV.
	The lower-left-to-upper-right transiting (lighter, projected, two-step) ellipse is selected with $P_V^\mu = (50, +25, +25, 0)$~GeV,
	$P_S^\mu = (75, -50, -50, +25)$~GeV, $M_H = 0.0$~GeV and $M_X = 50.0$~GeV;
	the dynamic trajectory of this geometry is statically bounded by a one-step parabolic shepherd.
	The event missing energy components are selected as \smash{${\slashed{P}} {}_{\rm T}^{x,y} = (-150.0,-150.0)$}~GeV.
	The projected ellipse materializes at the threshold mass \smash{$M_Y^{\odot+} = 128.5$}~GeV, at the position \smash{$(P_H^{x,y})_{\odot+} = (-156.0,-156.0)$}~GeV,
	and does not recollapse.
	The host ellipse materializes on the exterior of the open projected ellipse, at the threshold mass \smash{$M_Y^{\odot+} = 244.9$}~GeV,
	at the position \smash{$(P_H^{x,y})_{\odot+} = (0.0,-52.1)$}~GeV.
	The host ellipse initially expands rapidly, catching up to the projected ellipse for a first intersection at
	$M_Y = 254.7$~GeV, defining the \protecting{\smash{${\Widetilde{M}}_{\rm T2}$}} event scale.
	Subsequently, expansion of the host ellipse slows, and the transiting projected ellipse pulls away, terminating their intersection at $M_Y = 274.8$~GeV.\vspace{3.7pt}}
	\label{fig:intersection_4}
\end{figure}

Segueing to the two-step event topology, the first critical transition to
elaborate is that associated with $(M_S \Rightarrow 0)$.  The crucial
observation here is that the leading quadratic coefficient
in the search for $R^2 = 0$ roots, namely $\alpha_\odot$ from
Eqs.~(\ref{eq:mt2min}), vanishes, implying that the root relation is linear
(at most), with no more than a single solution.  The slope parameter,
\smash{$\beta_\odot$} from Eqs.~(\ref{eq:mt2min}), is positive semi-definite,
vanishing if and only if $(P_H^\mu \propto P_S^\mu)$ or $(P_V^\mu \propto P_S^\mu)$,
{\it i.e.} precisely the two conditions that trigger degeneracy
\smash{$(\beta_\odot^2 - 4\,\alpha_\odot \gamma_\odot \Rightarrow 0)$}
in the associated quadratic discriminant; incidentally, either of these circumstances
will simultaneously nullify the zeroth order coefficient \smash{$\gamma_\odot$}, rendering 
Eqs.~(\ref{eq:mt2min}) logically moot.  Avoiding these two scenarios, there will
be an isolated real root \smash{$M_Y^{\odot+}$} at which the elliptical kinematic
consistency region materializes out of a point.  Proceeding upward from this scale,
the ellipse transits across the $(P_H^{x,y})$ transverse hidden momentum plane,
enlarging along with the unbounded \smash{${\Widetilde{M}}_{\rm T2}$} search range.
The subordinate one-step shepherd geometry, with $P_S^\mu$ assuming the traditional
role of $P_V^\mu$, simultaneously accesses its parabolic branch, as is consistent with
the absence of a recontraction phase for the two-step geometry.
The intersection of a mixed one- and two-step decay topology, where the secondary
visible decay product of the latter event leg is massless $(M_S \Rightarrow 0)$,
is depicted in FIG.~\ref{fig:intersection_4}.

The next two-step boundary case to be described will be that associated with $(\Omega \Rightarrow 0)$,
which corresponds to vanishing of the conic discriminant and a related transition onto the
parabolic branch.  The elliptical scale factor $(R^2 \propto \Omega)$ defined in Eqs.~(\ref{eq:physellcoeff})
is identically null; this implies that the primary geometry is automatically degenerate,
in the mode of a line (compacted parabola), as specified previously in Eq.~(\ref{eq:2stepdegen}).
The role of the subordinate one-step shepherd becomes particularly acute here, as it must literally
truncate into a line segment the spurious infinite extent displayed by the two-step line.
Despite these facts, roots for the mass thresholds \smash{$M_Y^{\odot\pm}$} deriving from
Eqs.~(\ref{eq:mt2min}), from which the impact of $(\Omega \Rightarrow 0)$ was divided out, remain valid.
The coordinate center $(x_0,y_0)$ from Eqs.~(\ref{eq:physellcoeff}) is presently undefined,
in keeping with generic expectation for the conic discriminant going to zero. Nevertheless,
definite loci \smash{$(P_H^{x,y})_{\odot\pm}$} of inception and dissolution are salvaged
by convolving the transiting linear geometry with the fixed elliptical shepherd; their initial
and final intersections will define a pair of tangent points at precisely the expected
scales \smash{$M_Y^{\odot\pm}$}.

It is useful to quantify the locations \smash{$(P_H^{x,y})_{\oslash}$} at which the
two-step line bisects the perimeter of its one-step elliptical analog in the $(\Omega \Rightarrow 0)$ limit;
this is the only scenario for which the primary geometry is materially truncated by its subordinate shepherd,
and analysis is simultaneously here simplified by linearization of the leading conic constraint.
Substituting the coordinate \smash{$(\Widehat{P}_H^{x})$} from Eq.~(\ref{eq:2stepdegen}).
into the Eq.~(\ref{eq:ellipseph}) ellipse definition,
using a single dotted accent to distinguish factors belonging to the one-step shepherd geometry,
a quadratic intersection criterion
\smash{$\{\, \alpha_\oslash (\Widehat{P}_H^{y})_{\oslash}^2 + \beta_\oslash (\Widehat{P}_H^{y})_{\oslash} + \gamma_\oslash = 0 \,\}$}
in the coordinate \smash{$(\Widehat{P}_H^{y})$} emerges, with coefficients as defined in Eqs.~(\ref{eq:yslash});
a global dimensionless factor $\Lambda^2_x$ has been multiplied through to insulate the $(\Lambda_x \Rightarrow 0)$ limit.
\begin{eqnarray}
\alpha_\oslash &\equiv&
      \dot{a\vphantom{d}}\, \Lambda_y^2
\,-\, 2\, \dot{b\vphantom{d}}\, \Lambda_y\, \Lambda_x
\,+\, \dot{c\vphantom{d}}\, \Lambda_x^2
\vphantom{{\Big(\Big)}^2} \nonumber \\[2pt]
\beta_\oslash &\equiv& 
      2\, \Big(\dot{a\vphantom{d}}\, \Lambda_y - \dot{b\vphantom{d}}\, \Lambda_x\Big)\, \Big(\Gamma - \Delta\Big)
\,-\, 2\, \dot{d}\, \Lambda_y\, \Lambda_x
\,+\, 2\, \dot{f}\, \Lambda_x^2
\vphantom{{\Big(\Big)}^2} \nonumber \\[2pt]
\gamma_\oslash &\equiv&
      \dot{a\vphantom{d}}\, {\Big(\Gamma - \Delta\Big)}^2
\,-\, 2\, \dot{d}\, \Lambda_x\, \Big(\Gamma - \Delta\Big)
\,+\, \dot{g\vphantom{d}}\, \Lambda_x^2
\label{eq:yslash}
\end{eqnarray}

\begin{figure}[tp]
	\centering
	\includegraphics[width=0.48\textwidth]{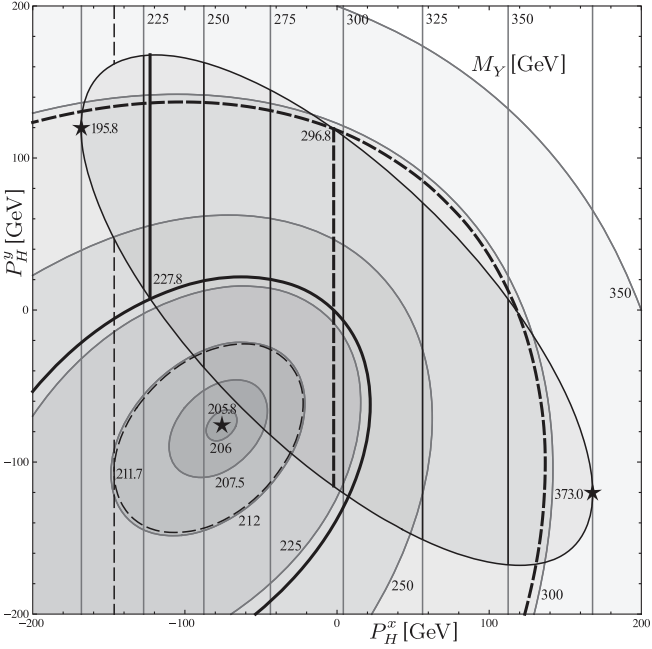}
	\caption{Intersection of elliptical and segmented-linear kinematic consistency regions
	for an example event with mixed one- and two-step decay topology,
	where the secondary visible decay product of the latter leg satisfies $(\Omega \Rightarrow 0)$.
	The lower-left (heavier, host, one-step) ellipse is selected with $P_V^\mu = (160, -80, -80, +40)$~GeV and $M_H = 100.0$~GeV.
	The left-to-right transiting (lighter, projected, two-step) vertical line is selected with $P_V^\mu = (120, +75, +75, +30)$~GeV,
	$P_S^\mu = (80, -50, +50, +20)$~GeV, $M_H = 50.0$~GeV and $M_X = 100.0$~GeV;
	the dynamic trajectory of this geometry is statically bounded by a one-step elliptical shepherd.
	The event missing energy components are selected as \smash{${\slashed{P}} {}_{\rm T}^{x,y} = (-162.5,+162.5)$}~GeV.
	The projected line materializes, in a first intersection with its elliptical shepherd, at the threshold mass \smash{$M_Y^{\odot+} = 195.8$}~GeV,
	at the position \smash{$(P_H^{x,y})_{\odot+} = (-168.0,+120.0)$}~GeV,
	and recollapses, in a final intersection with its shepherd, at the threshold mass \smash{$M_Y^{\odot-} = 373.0$}~GeV,
	at the position \smash{$(P_H^{x,y})_{\odot-} = (+168.0,-120.0)$}~GeV.
	The host ellipse materializes on the exterior of the projected elliptical shepherd, at the threshold mass \smash{$M_Y^{\odot+} = 205.8$}~GeV,
	at the position \smash{$(P_H^{x,y})_{\odot+} = (-75.6,-75.6)$}~GeV.
	A spurious preliminary intersection between the host ellipse and the projected line, exterior to the domain of its shepherd, is initiated at $M_Y = 211.7$~GeV.
	The host ellipse and projected, shepherded line segment experience a first legitimate intersection at $M_Y = 227.8$~GeV, defining the \protecting{\smash{${\Widetilde{M}}_{\rm T2}$}} event scale.
	Subsequently, the transiting host ellipse pulls away, terminating their direct intersection at $M_Y = 296.8$~GeV, but maintaining enclosure.\vspace{4.8pt}}
	\label{fig:intersection_5}
\end{figure}

When numerically rendering the coefficients from Eqs.~(\ref{eq:mt2ell}) of the one-step conic shepherd
that are referenced (with singly dotted accent) in Eqs.~(\ref{eq:yslash}), an additional computational
simplification is conferred by the $(\Omega \Rightarrow 0)$ limit; since the longitudinal rest frames of
the primary and secondary visible particle species $(V,S)$ are coincident, there is no need to perform
a supplementary longitudinal boost prior to projecting the mappings $(P_V^\mu \Rightarrow P_S^\mu)$ and
$(M_Y \Rightarrow M_X)$ onto the existing one-step formalism established in Eqs.~(\ref{eq:mt2del}).
This also implies that the existing $\Delta$ term calculated for application in the two-step geometry
may be adopted without modification for reuse in the role of $\Gamma$ for the one-step geometry.
It is emphasized that the factor $E_V$ appearing as a denominator of $\Delta$ in Eqs.~(\ref{eq:mt2del}),
and likewise as a divisor for the dimensionless ratio \smash{$\Widehat{M}_H^2$} from Eqs.~(\ref{eq:mt2ell}),
is engaged as ballast against the similar factor in \smash{$(\Widehat{P}_H^{x,y})$} of Eq.~(\ref{eq:ellipseph}),
and is thus unaffected by the substitution $(P_V^\mu \Rightarrow P_S^\mu)$; in contrast, all dimensionless
references \smash{$\Widehat{P}_V^\mu$} to the primary visible particle species $V$ in Eq.~(\ref{eq:mt2ell})
are to be exchanged for a corresponding reference to the secondary visible species $S$,
including both the four-vector orientation $P_S^\mu$ and the transverse energy scaling $E_S$.

The leading quadratic coefficient \smash{$\alpha_\oslash$} in Eqs.~(\ref{eq:yslash}) reduces to
\smash{$ \{\, ( P_\mu^{\raisebox{-1pt}{$\scriptstyle V$}}  P_S^\mu )^2 - M_V^2 M_S^2 \,\} \div (E_V E_S)^2$}, making application of the constraint $(\Omega \Rightarrow 0)$,
mimicking a positive semi-definite factor observed previously in Eq.~(\ref{eq:gammadisc}); the limit where this term vanishes, namely 
proportionality $(P_V^\mu \propto P_S^\mu)$ of the visible decay products,
shall be considered independently, subsequently.  The discriminant
\smash{$(\beta_\oslash^2 - 4\, \alpha_\oslash \gamma_\oslash)$} associated with quadratic inversion
\smash{$\{\, (\Widehat{P}_H^{y})_{\oslash\pm} \equiv ( - \beta_\oslash \pm \surd[\, \beta_\oslash^2 - 4\, \alpha_\oslash \gamma_\oslash \,] \,) /\, 2\,\alpha_\oslash \,\}$}
of the coefficients in Eqs.~(\ref{eq:yslash}) is itself a quadratic function of the factor 
$(\Gamma \propto M_Y^2)$ from Eqs.~(\ref{eq:mt2del}); the coefficients of this expression are equivalent,
modulo a global positive semi-definite rescaling $(4\, \Lambda_x^2)$, to the two-step $(\Omega \Rightarrow 0)$
coefficients defined in Eqs.~(\ref{eq:mt2min}).  The \smash{$(\Widehat{P}_H^{y})_{\oslash}$} roots
are thus indeed degenerate, signaling tangency of the associated intersection, at each of the scales \smash{$\Gamma_{\odot\pm}$}
developed from Eqs.~(\ref{eq:mt2min}) to bound materialization and recollapse of the composite shepherded geometry;
interior to these bounds, the solutions deriving from Eqs.~(\ref{eq:yslash}) are generically real.
If the \smash{$(\Widehat{P}_H^{x})_{\oslash\pm}$} coordinate partners are likewise desired, they may be
recovered from \smash{$(\Widehat{P}_H^{y})_{\oslash\pm}$} by application of Eq.~(\ref{eq:2stepdegen}), or
from Eq.~(\ref{eq:ellipseph}) if $(\Lambda_x \Rightarrow 0)$, or from a reciprocal
$(\{x,\dot{a\vphantom{d}},\dot{d},\Lambda_x\} \Leftrightarrow \{y,\dot{c\vphantom{d}},\dot{f},\Lambda_y\})$ of Eqs.~(\ref{eq:yslash})
with inversion $(\pm \Leftrightarrow \mp)$ of the root order association.
\smash{$(\Widehat{P}_H^{x,y})_{\oslash}$}, when evaluated at the scales \smash{$M_Y^{\odot\pm}$},
thereby constitute a suitable proxy for localization of \smash{$(\Widehat{P}_H^{x,y})_{\odot\pm}$}.

The intersection of a mixed one- and two-step decay topology, where the secondary
visible decay product of the latter event leg satisfies $(\Omega \Rightarrow 0)$,
is depicted in FIG.~\ref{fig:intersection_5}.  The geometric intuition of this edge scenario is continuously
connected to that of the canonical example depicted in FIG.~\ref{fig:intersection_2}, which exhibits
progressive narrowing of the transiting ellipse as $\Omega$ approaches zero from positive values;  however,
there are peculiar aspects of the practical treatment, stemming from a need
to reject spurious intersections with the two-step line that occur beyond protection of the one-step shepherd's border.
Knowledge of the previously established coordinate bounds \smash{$(\Widehat{P}_H^{x,y})_{\oslash\pm}$} on this domain
is vital, but it is also beneficial to supplement the Eqs.~(\ref{eq:yslash}) content with an analogous criterion
\smash{$\{\, \alpha_\otimes (\Widehat{P}_H^{y})_{\otimes}^2 + \beta_\otimes (\Widehat{P}_H^{y})_{\otimes} + \gamma_\otimes = 0 \,\}$}
in \smash{$(\Widehat{P}_H^{y})$} for transversal of the (bar-primed) conic perimeter cross-projected
from the conjugate event leg via Eqs.~(\ref{eq:mt2ellp}) by the linear Eq.~(\ref{eq:2stepdegen}) host geometry.
The associated quadratic coefficients are defined in Eqs.~(\ref{eq:ytimes}); the source formulae for
factors indicated by either event leg diversely refer to their own distinct leading energy scale as $E_V$,
and that of their dual as $E'_V$.
\begin{eqnarray}
\alpha_\otimes &\equiv&
      {\bar{a\vphantom{f}}}' \Lambda_y^2
\,-\, 2\, {\bar{b\vphantom{f}}}' \Lambda_y\, \Lambda_x
\,+\, {\bar{c\vphantom{f}}}' \Lambda_x^2
\vphantom{{\Big(\Big)}^2} \nonumber \\[2pt]
\beta_\otimes &\equiv& 
      2\, \Big({\bar{a\vphantom{f}}}' \Lambda_y - {\bar{b\vphantom{f}}}' \Lambda_x\Big)\, \Big(\Gamma - \Delta\Big)
\,-\, 2\, {\bar{d\vphantom{f}}}' \Lambda_y\, \Lambda_x
\,+\, 2\, {\bar{f}\vphantom{f}}' \Lambda_x^2
\vphantom{{\Big(\Big)}^2} \nonumber \\[2pt]
\gamma_\otimes &\equiv&
      {\bar{a\vphantom{f}}}' {\Big(\Gamma - \Delta\Big)}^2
\,-\, 2\, {\bar{d\vphantom{f}}}' \Lambda_x\, \Big(\Gamma - \Delta\Big)
\,+\, {\bar{g\vphantom{f}}}' \Lambda_x^2
\label{eq:ytimes}
\end{eqnarray}

{\it A priori} decorrelation of kinematic factors across opposing event legs
and active dependence of the projected event leg's geometry on the parent
particle species mass $M_Y$ distinguish the handling of quadratic inversions
\smash{$\{\, (\Widehat{P}_H^{y})_{\otimes\pm} \equiv ( - \beta_\otimes \pm \surd[\, \beta_\otimes^2 - 4\, \alpha_\otimes \gamma_\otimes \,] \,) /\, 2\,\alpha_\otimes \,\}$}
deriving from Eqs.~(\ref{eq:ytimes}) relative to that of corresponding solutions
associated with Eqs.~(\ref{eq:yslash}) for intersection with
the host line's own static shepherd geometry.  The discriminant
\smash{$(\beta_\otimes^2 - 4\, \alpha_\otimes \gamma_\otimes)$} is again a quadratic
function of $M_Y^2$, on which the factor $\Gamma$ from Eqs.~(\ref{eq:mt2del})
in turn depends linearly, although there are now competing versions of $\Gamma$ from each event
leg that must be disentangled; the roots of that discriminant again isolate the scales of degenerate
tangential intersection, but no such intersections are now guaranteed to exist; if present,
these roots again correspond to bounds on materialization or dissociation of the composite geometry,
but the interpretation may be obscure, for example, linking a single root to a terminal
scale of intersection that is apparently unbounded from below.
The leading quadratic coefficient \smash{$\alpha_\otimes$} in Eqs.~(\ref{eq:ytimes})
vanishes, if the conjugate event topology is one-step, whenever the projected geometry
is parabolic $(M_V \Rightarrow 0)$ with a symmetry axis parallel to 
to the host line, or likewise, if the conjugate event topology is two-step, whenever
the projected geometry is also linearly degenerate $(\Omega \Rightarrow 0)$ with a slope 
identical to that of the host; it is otherwise positive.  The collinearity requisite to these scenarios
is not foreclosed by a phase transition of the type experienced by the Eqs.~(\ref{eq:yslash}) inversions.

The \smash{${\Widetilde{M}}_{\rm T2}$} statistic will be defined trivially as an unbalanced (type I)
solution if the degenerate coordinate \smash{$(\Widehat{P}_H^y)_{\oslash}$} of materialization for the
shepherded host line is enveloped at the corresponding mass scale \smash{$M_Y^{\odot\pm}$}
by the projected conic geometry, and if that body is of the one-step variety.  A similar solution, albeit one
unique to the present event configuration, presents at the scale of first tangential
intersection between the host line and the projected conic geometry, if such a scale exists,
and if it corresponds to a moment of inception rather than dissolution, and if the associated
degenerate \smash{$(\Widehat{P}_H^y)_{\otimes}$} coordinate is bounded by the range of coordinate
overlap \smash{$(\Widehat{P}_H^y)_{\oslash\pm}$} between the host line and its shepherd.
A case study exhibiting the described scenarios for indigenous geometric containment
in the context of mixed one- and two-step event topologies is presented in a footnote\footnote{
	This note provides the kinematic blueprint for a triplet of example events with mixed one- and two-step decay
	topology, demonstrating the potentiality for indigenous geometric containment involving a single linearly degenerate leg.
	The two-step (heavier, host) event leg, a line, is selected with
	$P_V^\mu = (150, 0, +100, +50)$~GeV, $P_S^\mu = (75, 0, -50, +25)$~GeV, $M_H = 50.0$~GeV and $M_X = 125.0$~GeV;
	the dynamic trajectory of this geometry is statically bounded by a one-step elliptical shepherd.
	The one-step (lighter, projected) event leg, an ellipse, is selected with $P_V^\mu = (125, 0, +75, +25)$~GeV and $M_H = 125.0$~GeV.
	The event missing energy components are selected, successively, as \smash{${\slashed{P}} {}_{\rm T}^{x,y} = (0.0,\{\, +100.0, -100.0, -750.0 \,\})$}~GeV.
	The projected ellipse materializes at the threshold mass \smash{$M_Y^{\odot+} = 221.8$}~GeV,
	at the respective positions \smash{$(P_H^{x,y})_{\odot+} = (0,\{\, +3.2, -196.8, -846.8 \,\})$}~GeV.
	In the first scenario, the host line materializes, in a first intersection with its elliptical shepherd,
	on the interior of the closed projected ellipse,
	at the threshold mass \smash{$M_Y^{\odot+} = 257.6$}~GeV,
	at the position \smash{$(P_H^{x,y})_{\odot+} = (0,+26.3)$}~GeV,
	immediately defining a type I unbalanced solution for the \smash{${\Widetilde{M}}_{\rm T2}$} event scale;
	the linear geometry recollapses at the threshold mass \smash{$M_Y^{\odot-} = 415.2$}~GeV,
	at the position \smash{$(P_H^{x,y})_{\odot+} = (0,-238.8)$}~GeV.
	In the second scenario, the host line makes a first tangential intersection with the projected ellipse
	on the interior of the closed host shepherd, at the threshold mass \smash{$M_Y = 279.7$}~GeV,
	at the position \smash{$(P_H^{x,y})_{\otimes} = (0,-3.3)$}~GeV,
	defining a type II unbalanced \smash{${\Widetilde{M}}_{\rm T2}$} solution.
	In the third scenario, the projected ellipse has no spatiotemporal overlap with the host shepherded line,
	and no genuine asymmetric \smash{${\Widetilde{M}}_{\rm T2}$} event scale may be defined.}.

The status of this ``type~II\nobreak\hspace{0.6pt}\nobreak'' unbalanced containment may be established cleanly,
noting that nullification of the Eqs.~(\ref{eq:ytimes}) discriminant implies a degenerate positional root
\smash{$\{ (\Widehat{P}_H^{y})_{\otimes} \Rightarrow -\beta_\otimes /\, 2\,\alpha_\otimes \}$}, which may
be substituted into the quadratic criterion associated with Eqs.~(\ref{eq:yslash}); given
upward concavity \smash{$(\alpha_\oslash \ge 0)$}, this expression will evaluate as negative
for coordinates interior to the shepherded bounds at \smash{$(\Widehat{P}_H^y)_{\oslash\pm}$}.
Multiplying through by the positive semi-definite constant \smash{$\alpha_\otimes$}, and making a substitution
using the vanishing Eqs.~(\ref{eq:yslash}) discriminant, the quadratic (in $M_Y^2$) functional inequality
presented in Eq.~(\ref{eq:type2containment}) emerges as a gauge of root enclosure, to be evaluated at the
described scale (if any) of onsetting tangential intersection; the \smash{$(\alpha_\otimes \Rightarrow 0)$} limit is safe.
\begin{equation}
\quad \Big\{\,
\big( \alpha_\oslash \gamma_\otimes + \alpha_\otimes \gamma_\oslash \big) \,\le\, \tfrac{1}{2} \beta_\oslash \beta_\otimes
\,{{\Big\arrowvert}\hspace{-6.4pt}{\Big\arrowvert}}_{\raisebox{+3pt}{$\scriptstyle[\,\beta_\otimes^2 \,=\, 4\, \alpha_\otimes \gamma_\otimes \,]$}}
\label{eq:type2containment}
\end{equation}

Failing either of the prior compound criteria, a solution may be isolated only at a mutual triple
intersection of the host line with the outer perimeters of the host shepherd and projected conic;
quantitatively, a condition \smash{$\{(\Widehat{P}_H^y)_{\oslash} = (\Widehat{P}_H^y)_{\otimes}\}$}
must apply for one of the available $(\pm)$ sign combinations. Isolating radicals, squaring, and
repeating, the following single equivalent condition emerges, with all relative signs absorbed;
this expression is quartic in the mass-square scale $M_Y^2$, and the lightest root within the
physically viable search range may be associated with the square of \smash{${\Widetilde{M}}_{\rm T2}$}.
A global factor of \smash{$(\alpha_\oslash \alpha_\otimes)^2$} has been multiplied through; the
limit in which this constant vanishes is safe.  A global negation has also been applied,
for a purpose detailed subsequently.
\begin{eqnarray}
0 &=& -1 \times \Big\{
{\left( \alpha_\oslash \gamma_\otimes - \alpha_\otimes \gamma_\oslash \right)}^2 \nonumber \\
&+& \left( \alpha_\oslash \beta_\otimes - \alpha_\otimes \beta_\oslash \right)
    \times
    \left( \gamma_\oslash \beta_\otimes - \gamma_\otimes \beta_\oslash \right) \Big\}
\label{eq:det4}
\end{eqnarray}

If the conjugate event leg also features a linearly degenerate $(\Omega \Rightarrow 0)$ two-step event topology,
then its projection has no isolated points of degenerate tangential intersection with the host line
in the finite coordinate domain; however, as before, the deprecation of an existing tool is roundly compensated
by a dramatic algebraic simplification.
Comparing the content of Eqs.~(\ref{eq:ellipse},\ref{eq:2stepdegen}) for the two-step linear $(\Omega \Rightarrow 0)$
limit, the following alternate ``conic'' coefficient specification is suggested; rather than retracing the indicated 
line as an infinitely compacted parabola, it makes only a single traversal; correspondingly, it is linear in the square
of the trial parent particle mass $M_Y^2$, rather than quadratic.
\begin{eqnarray}
a \Rightarrow 0           \quad;\,\, &b \Rightarrow 0&           \,\,;\quad c \Rightarrow 0 \nonumber \\
d \Rightarrow \Lambda_x\,/\,2 \quad;\,\, &f \Rightarrow \Lambda_y\,/\,2& \,\,;\quad g \Rightarrow (\Gamma-\Delta)
\label{eq:linecoeffs}
\end{eqnarray}

If the surrogate Eq.~(\ref{eq:linecoeffs}) coefficient set is employed to specify the projected event leg's native geometry in Eqs.~(\ref{eq:ytimes}),
noting that the Eqs.~(\ref{eq:mt2ellp}) transformation into ``primed'' coordinates must still be enacted, then
convolution with Eqs.~(\ref{eq:yslash}), as expressed in Eq.~(\ref{eq:det4}), will yield an expression that is quadratic
in $M_Y^2$, rather than quartic, and thus readily invertible; the coefficient $\alpha_\otimes$ vanishes explicitly.
Intuitively, the intersection of two (non-parallel) lines is a point,
which will linearly traverse the $(P_H^x,P_H^y)$ coordinate plane in response to variation of the trial event scale; this roving
point may potentially intersect the host geometry's static one-step shepherd for some range of scales, the smallest of which
constitutes a candidate value for \smash{${\Widetilde{M}}_{\rm T2}$}.  However, it is necessary that any such point of
intersection be additionally bounded by the remote geometry's one-step shepherd; as such, this calculation should be performed
twice, with the roles of target and projection geometry interchanged, accepting as \smash{${\Widetilde{M}}_{\rm T2}$} the smallest
scale (if any) for which mutual overlap is achieved.  Literal simultaneous solution of each event leg's Eq.~(\ref{eq:2stepdegen}) line
for a pair of coordinates set on the perimeter of the host's static one-step shepherd,
{\it cf.} Eq.~(\ref{eq:ellipse}), yields an expression with identical root structure to that of the Eq.~(\ref{eq:det4}) quadratic reduction,
but which may differ by various positive semi-definite multiplicative constants, including $\alpha_\oslash$, $\Lambda_x^4$, and
\smash{$(\Lambda_x\Lambda'_y-\Lambda_y\Lambda'_x)^2$}; the last of these vanishes when the two lines
have a parallel orientation, a limit that remains safe for analysis, if informatically supplemented with the
mass bounds \smash{$M_Y^{\odot\pm}$} on each event leg's kinematic materialization and dissolution.
It may likewise differ by the imposition of a global sign; the phase convention
elected Eq.~(\ref{eq:linecoeffs}) serves the purpose of maintaining a positive function value in regions
of physical overlap with the shepherd, interior to the boundary scales associated with root inversion.
A case study exhibiting mutual intersection in the context of an event with
dual linearly degenerate two-step event topology is presented in a footnote\footnote{
	This note provides the kinematic blueprint for a pair of example events with dual linearly degenerate
	two-step decay topology, demonstrating the potentiality for mutual dynamic intersection interior to the
	static bounds of each geometry's one-step elliptical shepherd.
	The first (heavier, host, two-step) line is selected with
	$P_V^\mu = (100, 0, +75, +50)$~GeV, $P_S^\mu = (160, +120, 0, +80)$~GeV, $M_H = 75.0$~GeV and $M_X = 175.0$~GeV.
	The second (lighter, projected, two-step) line is selected with
	$P_V^\mu = (150, 0, +100, +50)$~GeV, $P_S^\mu = (75, -50, 0, +25)$~GeV, $M_H = 50.0$~GeV and $M_X = 125.0$~GeV.
	The event missing energy components are selected, successively, as \smash{${\slashed{P}} {}_{\rm T}^{x,y} = (-75.0,\{\, 0.0, +250.0 \,\})$}~GeV.
	The projected and host lines materialize at the threshold masses \smash{$M_Y^{\odot+} = 235.9$}~GeV and \smash{$M_Y^{\odot+} = 255.1$}~GeV,
	respectively, in a first intersection with their elliptical shepherds.
	In the first scenario, the locus of crossed linear intersection makes first contact with the projected shepherd at the threshold mass
	\smash{$M_Y = 240.6$}~GeV, at the position \smash{$(P_H^{x,y}) = (-97.5,-22.5)$}~GeV, which is exterior to the closed host shepherd;
	it passes out of contact at the threshold mass \smash{$M_Y = 313.3$}~GeV, at the position \smash{$(P_H^{x,y}) = (+137.5,-56.1)$}~GeV.
	Correspondingly, first contact with the host shepherd occurs at the threshold mass
	\smash{$M_Y = 277.7$}~GeV, at the position \smash{$(P_H^{x,y}) = (14.6,-38.5)$}~GeV, which is interior to the closed projected shepherd,
	defining the \smash{${\Widetilde{M}}_{\rm T2}$} event scale.
	it passes out of contact at the threshold mass \smash{$M_Y = 381.5$}~GeV, at the position \smash{$(P_H^{x,y}) = (+413.7,-95.5)$}~GeV.
	In the second scenario, the two elliptical shepherds have no domain of static spatial overlap,
	and a genuine asymmetric \smash{${\Widetilde{M}}_{\rm T2}$} event scale cannot be defined.}.

Returning to the task of globally classifying \smash{${\Widetilde{M}}_{\rm T2}$} critical point behavior,
conjunction of the limits $({M_S \Rightarrow 0})$ and $({\Omega \Rightarrow 0})$ proceeds cleanly,
exhibiting no uniquely diverse phenomena, but instead reframing the presentation of its constituent
ingredients.  The two-step geometry is that of a line, as enforced in Eq.~(\ref{eq:2stepdegen}), which emerges
at the mass \smash{$M_Y^{\odot+}$} from a point of degenerate tangential intersection with a
parabolic one-step shepherd. The coordinates \smash{$(\Widehat{P}_H^{x,y})_{\odot+}$} of materialization 
may again be extracted, at the appropriate scale, from the pair of degenerate roots \smash{$(\Widehat{P}_H^{x,y})_{\oslash}$}
established by Eqs.~(\ref{eq:yslash},\ref{eq:2stepdegen}); this point may exist at any location along the parabola's
perimeter, and is not generically confined to the vertex.  The initial intersection subsequently expands into a transiting
line segment with indefinitely increasing $M_Y$.

The third independent two-step boundary case to be treated consists of the limit $(P_H^\mu \propto P_S^\mu)$;
this is the first of two scenarios for which the mass thresholds \smash{$M_Y^{\odot\pm}$} deriving from Eqs.~(\ref{eq:mt2min})
become degenerate with vanishing of the associated quadratic discriminant, as expanded in Eq.~(\ref{eq:gammadisc}).
The resulting geometry is simply that of an isolated point, existing at a single consistent mass scale $M_Y$.  The primary two-step ellipse
and the secondary one-step shepherd each independently position the \smash{$(\Widehat{P}_H^{x,y})_{\odot+}$} coordinates equivalently to
direct computation of $(x_0,y_0)$ according to the Eqs.~(\ref{eq:physellcoeff}) prescription; the result
is precisely as should be expected $(P_H^{x,y} \Rightarrow \{M_H/M_S\} \times P_S^{x,y})$ from the
stipulation of four-vector proportionality between the particle species $H$ and $S$.  The conceptual ease of this result belies 
the computational taxation that may vex certain practical approaches to isolating a singular, or nearly singular, moment of solution.

Intersection of the limits $(M_S \Rightarrow 0)$ and $(P_H^\mu \propto P_S^\mu)$, which imply also $(M_H \Rightarrow 0)$,
presents an interesting logical paradox.  On the one hand, $(P_H^\mu \propto P_S^\mu)$ insinuates convergence of the scales
\smash{$M_Y^{\odot\pm}$} for elliptical materialization and recollapse, while on the other hand, $(M_S \Rightarrow 0)$
suggests that these scales must be infinitely disjoint, such that the recollapse does not actually occur at all.
This apparent contradiction is resolved, while retaining elements of intuition from both antecedent conditions,
by observing that all three quadratic coefficients from Eqs.~(\ref{eq:mt2min}) simultaneously vanish, which
implies that $(R^2 \Rightarrow 0)$ independently of the $M_Y$ scale selection; an otherwise kinematically consistent
effective solution to Eqs.~(\ref{eq:mt2min}) is given by \smash{$(\Gamma_{\odot\pm} \Rightarrow \{0,\infty\})$},
from which a corresponding lower bound on $M_Y$ may be extracted.  Consequently, the two-step elliptical geometry is indeed
contracted to a point, but it is a traveling point rather than a static one, which initially manifests
at the coordinate origin of its self-hosted $(P_H^{x,y})$ transverse hidden momentum plane, and subsequently
translates away from the origin without bound as $M_Y$ increases.  This is in keeping with the expectation 
$(P_H^{x,y} \Rightarrow \{E_H/E_S\} \times P_S^{x,y})$ for proportionality between the particle species $H$ and $S$ in the massless limit,
where $E_H$ is otherwise unconstrained, and may take any positive semi-definite value.
As $(M_S \Rightarrow 0)$, the subordinate one-step shepherd
ellipse accesses its parabolic branch; as $(P_H^\mu \propto P_S^\mu)$, which is equivalent to $(M_X \Rightarrow M_H + M_S)$,
this parabola becomes kinematically degenerate (minimally on-shell) and folds into a compact line segment; given also
$(M_H \Rightarrow 0)$, the vertex of this line segment sits at its own coordinate origin.  The shepherd
thus precisely delineates the transiting elliptical point's trajectory.
A case study featuring the described mutual limit is presented in a footnote\footnote{
	This note provides the kinematic blueprint for a pair of example events with mixed two- and one-step decay topology,
	where the secondary visible decay product of the former leg satisfies
	the criteria $(M_S \Rightarrow 0)$ and $(P_H^\mu \propto P_S^\mu)$.
	The two-step (heavier, host) event leg, a pointlike elliptical compaction, is selected with
	$P_V^\mu = (125, 0, +50, +75)$~GeV, $P_S^\mu = (75, +50, -50, +25)$~GeV, $M_H = 0.0$~GeV and $M_X = 0.0$~GeV.
	The one-step (lighter, projected) event leg, an ellipse, is selected with $P_V^\mu = (100, 0, +75, +25)$~GeV and $M_H = 75.0$~GeV.
	The event missing energy components are selected, successively, as \smash{${\slashed{P}} {}_{\rm T}^{x,y} = (+75.0,\{\, -75, +75 \,\})$}~GeV.
	The projected ellipse materializes at the threshold mass \smash{$M_Y^{\odot+} = 136.2$}~GeV, at the respective positions \smash{$(P_H^{x,y})_{\odot+} = (+75.0,\{\, -166.9, -16.9 \,\})$}~GeV.
	The host point materializes at the threshold mass \smash{$M_Y^{\odot+} = 165.8$}~GeV, at the coordinate origin,
	and linearly transits without decohering, while maintaining a perpetually collapsed geometry.
	In the first scenario, this point initially lies to the exterior of the projected ellipse,
	and first intersects the conjugate ellipse perimeter at
	$M_Y = 192.7$~GeV, defining the \protecting{\smash{${\Widetilde{M}}_{\rm T2}$}} event scale.
	In the second scenario, the host point materializes interior to the projected ellipse,
	immediately defining a type I unbalanced \protecting{\smash{${\Widetilde{M}}_{\rm T2}$}} solution.}.

Intersection of the limits $(\Omega \Rightarrow 0)$ and $(P_H^\mu \propto P_S^\mu)$ presents no independent geometric phenomenology,
directly recalling, both intuitively and technically, various elements of its distinct parent classifications.
The two-step geometry is that of a line, as enforced in Eq.~(\ref{eq:2stepdegen}),
and it is only the subordinate one-step shepherd that illuminates the consistent solution space, which is now an isolated point rather
than an elliptically shaded continuum of line segments.  Again, the coordinates \smash{$(\Widehat{P}_H^{x,y})_{\odot+}$}
are opaque to direct computation of $(x_0,y_0)$ by Eqs.~(\ref{eq:physellcoeff}), 
but transparent to the \smash{$(\Widehat{P}_H^{x,y})_{\oslash+}$} proxy of Eqs.~(\ref{eq:yslash},\ref{eq:2stepdegen}),
consistent with the kinematic proportionality expectation $(P_H^{x,y} \Rightarrow \{M_H/M_S\} \times P_S^{x,y})$.

\begin{figure}[tp]
	\centering
	\includegraphics[width=0.48\textwidth]{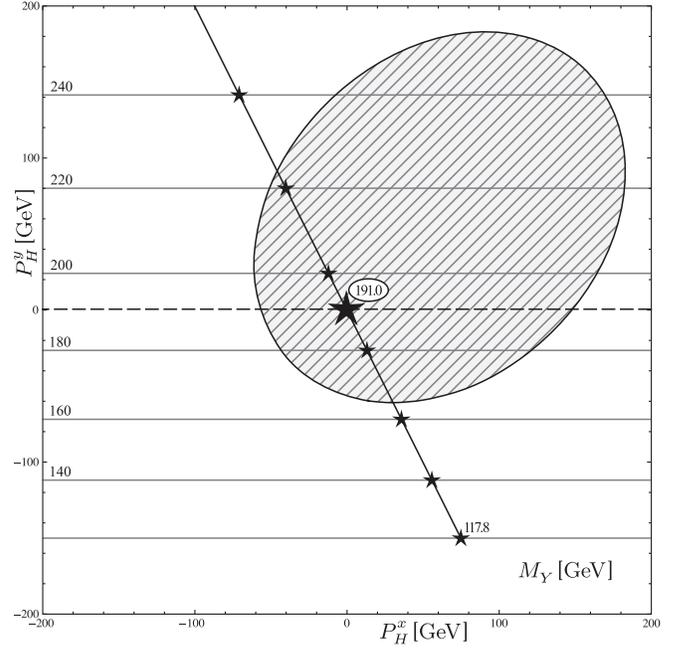}
	\caption{Intersection of instantonic-elliptical and point-like kinematic consistency regions
	for an example event with dual two-step decay topology, where the where the primary and secondary
	visible decay products of the former event leg are proportionally directed $(P_V^\mu \propto P_S^\mu)$,
	and the latter leg exhibits triple conjunction of the critical limits
	$(M_S \Rightarrow 0)$, $(\Omega \Rightarrow 0)$ and $(P_H^\mu \propto P_S^\mu)$.
	The frozen upper-right (heavier, host, two-step) ellipse, in ersatz manifestation of an effective one-step geometry,
	is selected with $P_V^\mu = (120, +40, +40, +80)$~GeV, $P_S^\mu = (75, +25, +25, +50)$~GeV, $M_H = 0.0$~GeV and $M_X = 105.0$~GeV.
	The lower-to-upper transiting (lighter, projected, two-step) horizontal line is selected with
	$P_V^\mu = (75, +25, +25, -50)$~GeV, $P_S^\mu = (120, +40, -80, -80)$~GeV, $M_H = 0.0$~GeV and $M_X = 0.0$~GeV;
	the dynamic trajectory of this geometry is statically bounded by a one-step segmented-linear shepherd.
	The event missing energy components are selected as \smash{${\slashed{P}} {}_{\rm T}^{x,y} = (+75.0,-150.0)$}~GeV.
	The projected line materializes, in a first intersection with its segmented-linear shepherd,
	at the threshold mass \smash{$M_Y^{\odot+} = 117.8$}~GeV, at the position \smash{$(P_H^{x,y})_{\odot+} = (+75.0,-150.0)$}~GeV,
	and does not decohere, although the intersected geometry remains perpetually collapsed.
	The closed host ellipse materializes non-locally, in exterior containment of the projected point momentarily
	positioned at ${(P_H^{x,y})}_{\odot} = (-0.3,+0.6)$~GeV, existing only at the degenerate threshold mass \smash{$M_Y^{\odot} = 191.0$}~GeV,
	defining a type III unbalanced solution for the \protecting{\smash{${\Widetilde{M}}_{\rm T2}$}} event scale.\vspace{10pt}}
	\label{fig:intersection_6}
\end{figure}

Triple conjunction of the limits $(M_S \Rightarrow 0)$, $(\Omega \Rightarrow 0)$
and $(P_H^\mu \propto P_S^\mu)$ yields, again,
a rather directly intuitive summation of its constituent ingredients.
The two-step geometry is a roving line, as described by Eq.~(\ref{eq:2stepdegen}), that is validated
only at its single point of intersection with the shepherding static one-step line segment, which
is directed outward from the coordinate origin, parallel to the orientation of
the secondary visible particle species $S$.  This point of intersection likewise onsets at the
origin and transits away with indefinitely increasing $M_Y$, tracing out a track 
$(P_H^{x,y} \Rightarrow \{E_H/E_S\} \times P_S^{x,y})$ consistent with massless
kinematic proportionality of the particle species $H$ and $S$.  The line and line segment
may exhibit any relative angular orientation, barring only perfect parallelism, for
which intersection would be complete but momentary; the approach to this limit
identically tracks the onset of kinematic proportionality $(P_V^\mu \propto P_S^\mu)$
in the visible decay products, which is an independent case to be subsequently resolved.
The intersection of a dual two-step decay topology, where one event leg illustrates 
the described triple limit, is depicted in FIG.~\ref{fig:intersection_6}.

The final independent two-step boundary case requiring classification consists of the limit $(P_V^\mu \propto P_S^\mu)$,
which is a subclass of the previously described limit $(\Omega \Rightarrow 0)$;
this is the second of two scenarios for which the mass thresholds \smash{$M_Y^{\odot\pm}$} deriving from Eqs.~(\ref{eq:mt2min})
become degenerate with vanishing of the Eq.~(\ref{eq:gammadisc}) quadratic discriminant.
Unique geometrical consequences engendered by this limit stem from an associated vanishing of
the factors $(\Lambda_{x,y})$ defined in Eqs.~(\ref{eq:mt2del}).  Following projection of these factors 
into Eq.~(\ref{eq:2stepdegen}), which expresses the linear $(\Omega \Rightarrow 0)$ reduction of
the two-step conic geometry, no kinematic constraint in the $(P_H^{x,y})$
transverse hidden momentum plane is retained; however, consistent nullification of the residual constant
term in Eq.~(\ref{eq:2stepdegen}) implies the rule \smash{$(\Gamma_{\odot\pm} \Rightarrow \Delta)$},
which absolutely determines $M_Y$, in keeping with degeneracy of the quadratic
$(R^2 \Rightarrow 0)$ roots.  All restriction on $(P_H^{x,y})$ thus emerges solely from the static ersatz
one-step geometry, which presents the frozen silhouette of all tracks that might otherwise be scanned with
$M_Y$ by the two-step geometry; in contrast to the limit $(P_H^\mu \propto P_S^\mu)$, which effects a condensation
of inception and dissolution masses by inducing a singularity in the geometry, approaching $(P_V^\mu \propto P_S^\mu)$
instead preserves a spatially extended event topology while accelerating the geometric transition between mass
boundaries, creating a nonlocal instantaneous union of histories in the hard limit.
Concurrently, a ``type~III\nobreak\hspace{0.6pt}\nobreak'' variant of the unbalanced \smash{${\Widetilde{M}}_{\rm T2}$}
solution may potentially be identified, if this instantonic structure bounds a projection of its conjugate
event leg's (possibly also noncompact) geometry at the isolated scale of advent.
The intersection of a dual two-step decay topology, where one event leg satisfies $(P_V^\mu \propto P_S^\mu)$,
is depicted in FIG.~\ref{fig:intersection_6}.

The baseline expectation for the ersatz one-step conic's topology is that of an ellipse,
but this may be modified according to the coincidence of additional critical phases.
A merger of the limits $(M_S \Rightarrow 0)$ and $(P_V^\mu \propto P_S^\mu)$ sends this geometry
onto its parabolic branch; although standard determination of the roots $\Gamma_{\odot\pm}$ is
here foreclosed by uniform vanishing of the quadratic coefficients in Eqs.~(\ref{eq:mt2min}),
\smash{$(\Gamma_{\odot\pm} \Rightarrow \Delta)$} remains the appropriate (degenerate) effective solution.
Conjunction of the limits $(P_H^\mu \propto P_S^\mu)$ and $(P_V^\mu \propto P_S^\mu)$, the former of which is
equivalent to $(M_X \Rightarrow M_H + M_S)$, condenses the one-step geometry to a point, again highlighting
the kinematic proportionality expectation $(P_H^{x,y} \Rightarrow \{M_H/M_S\} \times P_S^{x,y})$.
Finally, the mutual application $(M_S \Rightarrow 0)$, $(P_H^\mu \propto P_S^\mu)$ and $(P_V^\mu \propto P_S^\mu)$,
of all possible critical transitions sends the secondary conic geometry into the degenerate phase of its
parabolic branch, {\it i.e.} a line segment emanating from the coordinate origin along the
$(P_H^{x,y} \Rightarrow \{E_H/E_S\} \times P_S^{x,y})$ track; consistent with expectations, this is the
frozen image that would be traced out with advancing $M_Y$ by the roving two-step point if the
$(P_V^\mu \propto P_S^\mu)$ proportionality were relaxed;
the maintenance of $(M_X \Rightarrow M_H + M_S)$ via equivalent, non-zero, values for $(M_H,M_X)$ is
unphysical, as reflected by dispatch of the entire frozen topology to a point at infinity; 
enforcing the physical limit, the singular kinematically consistent parent mass scale is $(M_Y \Rightarrow 0)$. 
A case study featuring each itemized union of critical limits is presented in a footnote\footnote{
	This note provides the kinematic blueprint for a triplet of example events with mixed two- and one-step decay
	topology, where the primary and secondary visible decay products of the former leg are proportionally directed $(P_V^\mu \propto P_S^\mu)$,
	and the visible decay product of the latter leg is massless $(M_V \Rightarrow 0)$.
	The two-step (equivalently massless or heavier, host) event leg, in frozen ersatz manifestation of an effective one-step geometry, is successively selected with
	$P_V^\mu = (120, 0, +72, \{\, +96,0,+96 \,\})$~GeV, $P_S^\mu = (100, 0, +60, \{\, +80,0,+80 \,\})$~GeV, $M_H = 0.0$~GeV and $M_X = \{\, 80.0, 80.0, 0.0 \,\}$~GeV,
	satisfying the supplementary critical limit criteria $(M_S \Rightarrow 0)$, $(P_H^\mu \propto P_S^\mu)$, and $\{(M_S \Rightarrow 0) \;\&\; (P_H^\mu \propto P_S^\mu)\}$, accordingly.
	The one-step (massless, projected) event leg, a parabola, is selected with $P_V^\mu = (75, +50, +50, -25)$~GeV and $M_H = 0.0$~GeV.
	The event missing energy components are selected as \smash{${\slashed{P}} {}_{\rm T}^{x,y} = (+75.0,+125.0)$}~GeV.
	The closed projected parabola materializes at the threshold mass \smash{$M_Y^{\odot+} = 0.0$}~GeV, with vertex at the position ${(P_H^{x,y})}_{\odot+} = (+75.0,+125.0)$~GeV.
	The instantonic host materializes, in turn, as a non-locally extended closed parabola with vertex at the position ${(P_H^{x,y})}_{\odot} = (0.0,-26.7)$~GeV,
	as a pointlike elliptical compaction positioned at the coordinate origin, and as a segmented-linear degenerate parabola emanating from the
	same, existing, respectively, only at the degenerate threshold mass \smash{$M_Y^{\odot} = \{\, 118.7, 176.6, 0.0 \,\}$}~GeV.
	In the first scenario, the host parabola experiences a pair of disjoint intersections with the projected parabola, spanned
	by a finite region of mutually consistent kinematic phase space.  In the second scenario, the host singularity abides in a state
	of type I unbalanced containment within the conjugate parabolic projection.  In the third scenario, the host line segment makes
	a four-fold degenerate cross-cutting of the projected parabola, which is likewise momentarily in a degenerately compacted phase.
	In each case, a consistent value of the \protecting{\smash{${\Widetilde{M}}_{\rm T2}$}} event scale is concurrently defined.}.\vspace{4.8pt}

\section{\protecting{Algorithmic Treatment of \texorpdfstring{\smash{$\boldsymbol{{\Widetilde{M}}_{\rm T2}}$}}{MT2}}\label{sct:algorithm}}

An algorithmic treatment of the one- and two-step asymmetric \smash{${\Widetilde{M}}_{\rm T2}$} event scale
statistic is descriptively outlined in the present section, mirroring the composition of the
standalone \smash{${\Widetilde{M}}_{\rm T2}$} utility available for download from the author's personal website~\cite{amt2},
or as an ancillary file \mbox{``{\tt anc/amt2.pl}''} with this document's {\it ar{\raisebox{1.8pt}{$\chi$}}{\hspace{-0.8pt}}iv.org} 
electronic source materials, and also as a modular element of the generalized selection cut package \mbox{\sc AEACuS}~\cite{Walker:2012vf,aeacus}.
This software is coded in the {\sc Perl} language, employing a primarily (though not purely)
functional programming paradigm, wherein computation is realized by the sequential mapping of data through
a succession of externally stateless transformations.  Nevertheless, the presentation mode is intentionally
abstract, and may be projected onto a more procedurally imperative implementation in any suitable host language.
Pseudocode and a diagram of logical flow are provided in Appendix~\ref{sct:appendix_C}.

Philosophically, the purpose of this exercise is to construct a logically
complete algorithm that handles all \smash{${\Widetilde{M}}_{\rm T2}$} event classes
described by the main document body in a unified, numerically stable, and maximally compact manner.  Specifically,
\mbox{({\it i}\hspace{1.6pt})} a well-defined set of acceptable inputs should be strictly enforced, while the capacity
should exist to map all members of this set to their appropriate functional output,
and \mbox{({\it ii}\hspace{1.6pt})} if a viable theoretical solution exists for a given input,
the algorithm should guarantee convergence, without artificial endcaps on the
solution range, and with vanishing likelihood of overstepping a solution by fault of finite domain sampling.
Pragmatically, relaxation of the hard critical limit criteria previously described
to some epsilon-width margin may increase the likelihood of gainfully deploying the various corresponding 
edge case routines by an order of cardinality, while simultaneously softening computation in regions of phase
space that are proximal to associated numerical indeterminacies.
However, there is no implication that codes excluding certain of the described processing stages
are less than serviceably robust.

In order to quantitatively associate a numerical value for \smash{${\Widetilde{M}}_{\rm T2}$} with a given pair
of event leg topologies, it is necessary to establish a formal procedure for determining whether the respective
conic boundaries on kinematic consistency experience an intersection, for a given parent particle species mass $M_Y$,
when projected onto a common transverse hidden momentum coordinate plane $(\Widehat{P}_H^{x,y})$;
precisely such a technology has been outlined in Ref.~\cite{intersectionofellipses},
and will be reviewed and expanded upon presently.
The abstract system to be considered consists of two conic perimeters
$A$ and $B$, each satisfying an algebraic constraint of the Eq.~(\ref{eq:ellipse}) variety, which is quadratic in
each of the shared coordinates $(x,y)$; subsystem $A$ may be considered to exist natively on the plane of
analysis, playing the role of host to subsystem $B$, which is projected into this plane via some linear
coordinate transformation, as expressed by the adoption of a primed coefficient set.  It is possible to
imagine inverting each quadratic constraint in the coordinate $(x)$, and equating the two resulting functions
of an implicitly equivalent coordinate $(y)$, establishing a criterion for geometric intersection; isolating
radicals, squaring, and repeating, the following quartic expression emerges in $(y)$, with all relative $(\pm)$
signs for various root associations absorbed.
\begin{equation}
0 = u_0 + u_1 y + u_2 y^2 + u_3 y^3 + u_4 y^4
\label{eq:intersection}
\end{equation}

The five coefficients $u_i$ referenced by Eq.~(\ref{eq:intersection}) are defined  
in Eqs.~(\ref{eq:intcoeffs1}), in terms of an additional eleven coefficients $v_i$,
which are themselves defined in Eqs.~(\ref{eq:intcoeffs2}). A global positive semi-definite
factor \smash{$(\,a_A\, a'_B\,)^2$} has been divided out, and the overall phase selection is arbitrary.
\begin{eqnarray}
u_0 &\equiv& v_2 v_{10} - v_4^2 \nonumber \\
u_1 &\equiv& v_0 v_{10} + v_2 (v_7 + v_9) - 2\, v_3 v_4 \nonumber \\
u_2 &\equiv& v_0 (v_7 + v_9) + v_2 (v_6 - v_8) - v_3^2 - 2\, v_1 v_4 \nonumber \\
u_3 &\equiv& v_0 (v_6 - v_8) + v_2 v_5 - 2\, v_1 v_3 \nonumber \\
u_4 &\equiv& v_0 v_5 - v_1^2
\label{eq:intcoeffs1}
\end{eqnarray}
\begin{alignat}{10}
&v_0    \enspace&\equiv&\enspace\, 2\, &( a_A b'_B \;&-&\; a'_B b_A &)& \quad;\quad
&v_1    \enspace&\equiv&\enspace\,     &( a_A c'_B \;&-&\; a'_B c_A &)& \nonumber \\
&v_2    \enspace&\equiv&\enspace\, 2\, &( a_A d'_B \;&-&\; a'_B d_A &)& \quad;\quad
&v_3    \enspace&\equiv&\enspace\, 2\, &( a_A f'_B \;&-&\; a'_B f_A &)& \nonumber \\
&v_4    \enspace&\equiv&\enspace\,     &( a_A g'_B \;&-&\; a'_B g_A &)& \quad;\quad
&v_5    \enspace&\equiv&\enspace\, 2\, &( b_A c'_B \;&-&\; b'_B c_A &)& \nonumber \\
&v_6    \enspace&\equiv&\enspace\, 4\, &( b_A f'_B \;&-&\; b'_B f_A &)& \quad;\quad
&v_7    \enspace&\equiv&\enspace\, 2\, &( b_A g'_B \;&-&\; b'_B g_A &)& \nonumber \\
&v_8    \enspace&\equiv&\enspace\, 2\, &( c_A d'_B \;&-&\; c'_B d_A &)& \quad;\quad
&v_9    \enspace&\equiv&\enspace\, 4\, &( d_A f'_B \;&-&\; d'_B f_A &)& \nonumber \\
&v_{10} \enspace&\equiv&\enspace\, 2\, &( d_A g'_B \;&-&\; d'_B g_A &)& &&&&&&&&
\label{eq:intcoeffs2}
\end{alignat}

Yet, the root structure of Eq.~(\ref{eq:intersection}) does not directly address the most currently
relevant and interesting line of inquiry, that being isolation of the parent mass $M_Y$
at which kinematic intersection is initiated; rather, it specifies the $(y)$ coordinates, if any,
at which intersection occurs, for an individually sampled trial value of the parent mass.
Nevertheless, this expression does potentially facilitate that investigation via a bisection algorithm
trained for convergence to the mass scale of transition between intersection and non-intersection.
For this application, specific numerical roots in the coordinate $(y)$ are immaterial; the only
pertinent question is whether any real roots exist, for a given $M_Y$, at all.  The method of
Sturm sequences for counting real polynomial roots within some (possibly infinite) domain interval
is the tool of choice for this application, providing both speed and accuracy.  To summarize,
crossings of the vertical axis are inferred by comparing flips in sign at the domain boundaries
of the original function, its derivative, and a sequence of polynomials with decreasing degree
that are recursively composed by the negated remainders of long division in preceding elements.
The Sturm algorithm embedded within the present \smash{${\Widetilde{M}}_{\rm T2}$} analysis
package may be of interest for its 
\mbox{({\it i}\hspace{1.6pt})} generalized treatment of polynomials of any order,
\mbox{({\it ii}\hspace{1.6pt})} acceptance of finite and infinite (undefined) domain boundary specifications,
\mbox{({\it iii}\hspace{1.6pt})} symmetric inclusive counting of roots located at either boundary,
\mbox{({\it iv}\hspace{1.2pt})} option to individually count degenerate roots,
and \mbox{({\it v}\hspace{1.2pt})} option (by boundary omission) to return an encapsulated function with memoization of the
costly Sturm sequence computation that may be reapplied to various delayed specifications of the domain, 
although a detailed presentation of its architecture exceeds the scope of this document.

In practical application of the described procedure, there are certain circumstances
under which a spurious intersection may be indicated at a $(y)$ coordinate laying outside
the physical body of the geometric objects under consideration.  In particular, this
contingency may be anticipated when the relative alignment and global orientation of
a pair of event leg topologies conspire to facilitate a continuum of trial parent
particle species masses $M_Y$ for which a pair of pointlike intersections occur
with a common $(y)$ coordinate.  An extended effective two-fold degeneracy of this type
will generate a tangential point of contact with the Eq.~(\ref{eq:intersection})
zero-axis, which may persist (within some epsilon-width detachment) even at event
scales where the active geometries have discontinued any actual overlap.
Events with dual one-step parabolic legs are moreover acutely susceptible to the
possibility that degeneracies at infinity may pollute the ascertainment of finitely
positioned roots.  Large exponents, both explicit and implicit, attending the computation
of intersection may likewise render even ostensibly safe kinematic parameterizations
numerically vulnerable to false tagging.
An effective blockade against these hazards may be realized by collapsing the boundaries in $(y)$
of the Sturm sequence search to the physical coordinate overlap between each event
leg's geometric domain.  The $(y)$ coordinate extent of a given conic topology may be determined 
by implicitly differentiating the Eq.~(\ref{eq:ellipse}) perimeter specification
to establish a slope function $(dy/dx)$, forcing that slope to zero,
and subsequently substituting the resulting constraint on $(x)$ back into Eq.~(\ref{eq:ellipse}).
This procedure produces a quadratic criterion
\smash{$\{\, \alpha_\ominus (\Widehat{P}_H^{y})_{\ominus}^2 + \beta_\ominus (\Widehat{P}_H^{y})_{\ominus} + \gamma_\ominus = 0 \,\}$}
in \smash{$(\Widehat{P}_H^{y})$}, with coefficients defined following, by Eqs.~(\ref{eq:yminus});
a positive semi-definite factor of $(a)$ has been multiplied through.
\begin{eqnarray}
\alpha_\ominus &\equiv& b^2 - a\,c \nonumber \\[4.5pt]
\beta_\ominus &\equiv& 2\, (\,b\,d - a\,f\,) \nonumber \\[4.5pt]
\gamma_\ominus &\equiv& d^2 - a\,g
\label{eq:yminus}
\end{eqnarray}

The leading Eq.~(\ref{eq:yminus}) quadratic coefficient $\alpha_\ominus$ recalls the conic discriminant from
Eq.~(\ref{eq:conicdisc}), and will thus vanish in the case of geometric degeneracy, most
pertinently on the one-step parabolic branch $(M_V \Rightarrow 0)$.  In this case,
and if $(\Widehat{V}_y \not\Rightarrow 0)$, a single solution
\smash{$\{ (\Widehat{P}_H^{y})_{\ominus} \Rightarrow -\gamma_\ominus /\, \beta_\ominus \}$}
of the residual linear expression is expected,
which diverges, consistent with Eq.~(\ref{eq:orgparproxy}), for the infinitely displaced onset of a
massive $(M_H \not\Rightarrow 0)$ parabolic geometry.  Otherwise, given that the associated quadratic
discriminant \smash{$\{\,\beta_\ominus^2 - 4\, \alpha_\ominus \gamma_\ominus = 4\,a\,R^2\,(a\,c-b^2)\,\}$} is
positive semi-definite for physical mass scales, there are a matched pair
\smash{$\{\, (\Widehat{P}_H^{y})_{\ominus\pm} \equiv ( - \beta_\ominus \pm \surd[\, \beta_\ominus^2 - 4\, \alpha_\ominus \gamma_\ominus \,] \,) /\, 2\,\alpha_\ominus \,\}$}
of real inversions corresponding to the upper and lower coordinate reach of the associated conic.

It is additionally possible to directly establish the root structure of all tangential conic intersections as a
function of the trial event scale. Since tangential intersections represent the convergence to degeneracy of two otherwise
distinct roots, the appropriate criterion is vanishing of the quartic discriminant $(\mathbb{D}_4 \Rightarrow 0)$
associated with Eq.~(\ref{eq:intersection}), as exhibited subsequently in Eq.~(\ref{eq:qrtdiscriminant}).
However, this direct approach yields a comparatively fragile numerical function, where each displayed term is implicitly of
12\textsuperscript{th} order in the $\Gamma$ factor from Eqs.~(\ref{eq:mt2del}).
Nevertheless, its expression is useful, particularly in conjunction again with the Sturm sequence method, for
quickly establishing whether intersection occurs inside some (possibly positively infinite) physical search
bounds on the trial parent mass, without directly sampling the enclosed bulk; it is specifically
inapplicable to events featuring a linearly degenerate topology $(\Omega \Rightarrow 0)$, where
truncation of the primary two-step geometry by the static one-step shepherd exposes
a raw line-segment endcap, logically invalidating the imperative of initial tangential intersection.
\begin{eqnarray}
\mathbb{D}_4
&\equiv& 256\, u_0^3 u_4^3 - 192\, u_0^2 u_1 u_3 u_4^2 - 128\, u_0^2 u_2^2 u_4^2 \nonumber \\
&+& 144\, u_0^2 u_2 u_3^2 u_4 - 27\, u_0^2 u_3^4 + 144\, u_0 u_1^2 u_2 u_4^2 \nonumber \\
&-& 6\, u_0 u_1^2 u_3^2 u_4 - 80\, u_0 u_1 u_2^2 u_3 u_4 + 18\, u_0 u_1 u_2 u_3^3 \nonumber \\
&+& 16\, u_0 u_2^4 u_4 - 4\, u_0 u_2^3 u_3^2 - 27\, u_1^4 u_4^2 + 18\, u_1^3 u_2 u_3 u_4 \nonumber \\
&-& 4\, u_1^3 u_3^3 - 4\, u_1^2 u_2^3 u_4 + u_1^2 u_2^2 u_3^2
\label{eq:qrtdiscriminant}
\end{eqnarray}

There are certain circumstances under which the quartic coefficient $u_4$ from Eq.~(\ref{eq:intersection})
vanishes identically, or within a numerical error allowance epsilon, a principal example of which is given by
the occurrence of axial alignment between dual one-step parabolic event leg topologies; in fact, the $u_3$ coefficient
vanishes in this case as well, robustly nullifying $\mathbb{D}_4$, in a manner stable against both unified rotation
and relative lateral displacement.  The appropriate redress to such a lapse is cascaded regression 
to the discriminant $\mathbb{D}_i$ of the residual leading order, as summarized in Eqs.~(\ref{eq:cascadeqrtdiscriminant}).
A case study exhibiting loss of the intersection coefficients $(u_4,u_3)$, in addition to a degenerate string of
spurious Eq.~(\ref{eq:intersection}) roots after the fashion corralled by Eqs.~(\ref{eq:yminus}),
in the dual one-step decay topology context, is presented in a footnote\footnote{
	This note provides the kinematic blueprint for an example event with dual one-step decay topology,
	where the visible decay product of each leg is massless $(M_V \Rightarrow 0)$.
	The first (heavier, host) parabola is selected with $P_V^\mu = (80, -64, 0, +48)$~GeV and $M_H = 75.0$~GeV.
	The second (lighter, projected) parabola is selected with $P_V^\mu = (100, +80, 0, 60)$~GeV and $M_H = 0.0$~GeV.
	The event missing energy components are selected as \smash{${\slashed{P}} {}_{\rm T}^{x,y} = (+50.0,0.0)$}~GeV.
	The projected parabola materializes at the threshold mass \smash{$M_Y^{\odot+} = 0.0$}~GeV, with vertex at the position ${(P_H^{x,y})}_{\odot+} = (+50.0,0.0)$~GeV.
	The host parabola materializes on the interior of the closed projected parabola, at the threshold mass \smash{$M_Y^{\odot+} = 75.0$}~GeV,
	with vertex infinitely displaced from the coordinate origin,
	immediately defining a type I unbalanced solution for the \protecting{\smash{${\Widetilde{M}}_{\rm T2}$}} event scale.};
in actuality, the lower order coefficients $(u_2,u_1,u_0)$ vanish in kind for the indicated 
event kinematics, although this is precipitated merely by a coordinate singularity,
which may be circumvented utilizing the procedure to be outlined together with Eqs.~(\ref{eq:sincos}).
\begin{eqnarray}
\mathbb{D}_3 &\equiv& u_1^2 u_2^2 - 4\, u_0 u_2^3 - 4\, u_1^3 u_3 + 18\, u_0 u_1 u_2 u_3 - 27\, u_0^2 u_3^2 \nonumber \\
\mathbb{D}_2 &\equiv& u_1^2 - 4\, u_0 u_2 \nonumber \\
\mathbb{D}_1 &\equiv& 1 \nonumber \\
\mathbb{D}_0 &\equiv& 1 \div u_0^2 \nonumber \\
\mathbb{D}_{\slashed{0}} &\equiv& 0
\label{eq:cascadeqrtdiscriminant}
\end{eqnarray}

The \smash{${\Widetilde{M}}_{\rm T2}$} analysis, proper, begins with a reading of input kinematic assignments $(P_V^\mu,M_H,P_S^\mu,M_X)$ for each event leg;
the presence of a defined $P_S^\mu$ input for the secondary visible decay product $S$ triggers handling as a
two-step event topology, and a Boolean flag is set to indicate this logical fork in subsequent operations.
A Lorentz boost \smash{$\{\,P^{0,z} \Rightarrow (\,P^{0,z} - \Widehat{P}_V^z\,P^{z,0}\,) \div \surd[\,1-{(\Widehat{P}_V^z)}^{\raisebox{-2pt}{$\scriptstyle 2$}}\,]\,\}$}
into the longitudinal rest frame of the primary visible decay product $V$ is applied to each 
of the four-vectors $(P_V^\mu,P_S^\mu)$; if this frame does not exist, or if either of the energies
$(E_V,E_S)$ vanishes, then \smash{${\Widetilde{M}}_{\rm T2}$} is undefined. 
A unified $(x,y)$ pair of missing transverse momentum components \smash{$({\slashed{P}} {}_{\rm T}^{x,y})$}
is additionally read from input, or, if omitted, is calculated as the negation of the vector sum
of visible transverse momentum over both event legs.

Dimensionlessly scaled variables will generally be preferred in the present analysis, both because of the
formal simplifications that they tend to engender, and also the smoother numerics that attend
manipulation of commensurately scaled ({\it e.g.} all of order unity) values. As
such, the quantities \smash{$(\Widehat{P}_V^\mu,\Widehat{M}_H,\Widehat{M}_X,\Widehat{M}_V,{\slashed{\Widehat{P}}} {}_{\rm T}^{x,y})$}
are archived, after the mode of Eqs.~(\ref{eq:circumflex}); analogous treatment is made of
the variables $(P_S^\mu,M_S)$, although it proves more profitable to here divide out the
secondary visible event scale $E_S$.  The ratios $(E_V/E_V')$, where $E_V'$ is the primary
visible scale of the conjugate event leg, and $(E_S/E_V)$ are likewise computed and
cached, along with the absolute dimensionful event scale $E_V$.

These parameters are subsequently sequenced through a routine that conditionally evaluates the dimensionless
consolidation factors $(\Gamma,\Delta,\Lambda_{x,y},\Omega,\Pi)$ from Eqs.(\ref{eq:mt2del}), according
to either the one- or two-step prescription. The scale dependence of $\Gamma$ is tidily rendered, employing
techniques associated with the object-oriented programming paradigm, as an ordered pair of constant coefficients
for linear polynomial expansion in the subsequently defined dimensionless mass-square parameter $\Upsilon$;
this choice of base is symmetric under exchange of the event legs, facilitating cross-computation,
and is adopted as the exclusive internal format for all references to the parent particle species trial mass $M_Y$.
\begin{equation}
\Upsilon \,\equiv\, M_Y^2/(2\, E_V E_V')
\label{eq:massscale}
\end{equation}

A method is supplied for conversion of polynomial objects
into pure numbers via evaluation at a specified event scale; operator overloading for multiplication, addition, subtraction,
and exponentiation allows for transparent object-aware polynomial manipulation, and streamlines the propagation of
scale dependence into descendent functionals of $\Gamma$.  Two supplementary recurring factors $(\zeta,\xi)$ are also precomputed,
as defined following, and Boolean flags are triggered for the one-step $(M_V \Rightarrow 0)$ and two-step
$({M_S \Rightarrow 0},\allowbreak \, {\Omega \Rightarrow 0},\allowbreak \, {P_H^\mu \propto P_S^\mu},\allowbreak \, {P_V^\mu \propto P_S^\mu})$
critical limit criteria.
\begin{eqnarray}
\zeta &\equiv& (P_\mu^{\raisebox{-1pt}{$\scriptstyle V$}}  P_S^\mu) \div (E_V E_S) \nonumber \\
\xi &\equiv& \zeta^2 \,-\, (M_V/E_V)^2 \times (M_S/E_S)^2
\label{eq:factordefs}
\end{eqnarray}

Given that various aspects of the event analysis handle the $(x,y)$ coordinates in an
asymmetric manner, {\it e.g.}~Eqs.~(\ref{eq:yslash},\ref{eq:ytimes},\ref{eq:intersection},\ref{eq:yminus}),
it can be advantageous, and is in certain rare cases essential, to apply a global rotation
\smash{$\{\, P^{x,y} \Rightarrow (\, P^{x,y} \cos \varphi \pm P^{y,x} \sin \varphi \,)\,\}$} in the
transverse coordinate plane to all presently defined event objects bearing vector indices,
namely \smash{$(\Widehat{P}_V^\mu,\Widehat{P}_S^\mu,{ \slashed{\Widehat{P}}} {}_{\rm T}^{x,y},\Lambda_{x,y})$}.
Computation of intersections in the conic phase space boundaries on event kinematic consistency
will, by convention, isolate root degeneracies along the $\Widehat{P}_H^y$ hidden momentum axis; a
suitable protocol for optimized selection of a unified rotation angle $\varphi$ may therefore consist of
maximizing the summed projection of each event leg's elliptical major axis (or parabolic symmetry axis)
onto the $\Widehat{P}_H^y$ direction, substituting the static ersatz geometry for the primary if $(P_V^\mu \propto P_S^\mu)$.
Although the two event legs must be rotated in unison, no differential compensation is required for the assignment
of roles as target or projection geometry, as the slope of a line is symmetric under coordinate reflection.
The inclination of an individual conic may be extracted from the third member of Eqs.~(\ref{eq:physellcoeff}),
and split into finite $(\cos \theta,\sin \theta)$ components using standard trigonometric identities; the resulting
formulae couple naturally with the eccentricity, {\it cf.}~Eq.~(\ref{eq:eccentricity}), yielding the
subsequent expressions, which moreover possess the benefit of vanishing in the rotationally symmetric
circular ($\varepsilon \Rightarrow 0$) limit, while emphasizing the strongly oriented parabolic $(\varepsilon \Rightarrow 1)$ branch.
\begin{equation}
\varepsilon^2 {\left\{ \cos^2\!\theta, \sin^2\!\theta \right\}} \,=\, \frac{\surd[\, {( c - a )}^2 + 4\,b^2 \,]\pm(c-a)}{\surd[\, {( c - a )}^2 + 4\,b^2 \,]+(a+c)}
\label{eq:sincos}
\end{equation}

The associated angle may be confined to quadrants ({I~\&~IV}), {\it i.e.} $(\cos \theta \ge 0)$ may be assumed without loss of generality,
while allowing the phase of $\sin \theta$ to be inherited from $\tan \theta$ in Eqs.~(\ref{eq:physellcoeff}), which
is in turn signed oppositely to the conic coefficient $b$ from Eqs.~(\ref{eq:mt2ell}).  The orientation components
derived from each event leg may then be merged as a vector sum if the differential
angle is acute, {\it i.e.} the individual orientations have a positive inner product, or a vector difference otherwise;
no rotation is indicated if both eccentricities are null, whereas no sum is required if either eccentricity vanishes.
Taking an inverse tangent of the unified vector components yields an angle that must be subtracted
from $\pi/2$ radians to establish the target rotation $\varphi$ for optimizing mutual kinematic alignment with $P_H^y$;
this procedure ensures that the angular separation of linear two-step $(\Omega \Rightarrow 0)$ and parabolic one-step
($M_V \Rightarrow 0$) event topologies from that axis will be no greater than 45 degrees.

Next, the bounds \smash{$\Gamma_{\odot\pm}$} associated with Eqs.~(\ref{eq:mt2min}) on materialization and
dissociation of each leg's geometry are computed;  the lower limit $(\Gamma_{\odot+})$
must always exist, whereas the upper limit \smash{$(\Gamma_{\odot-})$} is unphysical
for the one-step event topology, and likewise undefined for certain critical limits of the two-step topology.
In the one-step case, \smash{$(\Gamma_{\odot+} \Rightarrow \Widehat{M}_H \Widehat{M}_V)$}; 
if $(P_V^\mu \propto P_S^\mu)$, then \smash{$(\Gamma_{\odot\pm} \Rightarrow \Delta)$};
if $\{(M_S \Rightarrow 0) \;\&\; (P_H^\mu \propto P_S^\mu)\}$, then \smash{$(\Gamma_{\odot\pm} \Rightarrow \{0,\infty\})$},
the latter of which is numerically indicated by an undefined value;
otherwise, the quadratic coefficients in Eqs.~(\ref{eq:mt2min}) may be specialized to the two-step
event topology and inverted numerically, with \smash{$(\alpha_\odot \Rightarrow - {\{M_S / E_S \}}^2)$},
noting that only the single effective \smash{$\Gamma_{\odot+}$} root will persist as $(M_S \Rightarrow 0)$.
The dimensionless scales retained for future reference are in the canonical \smash{$\Upsilon_{\odot\pm}$} 
form of Eq.~(\ref{eq:massscale}), which is easily extracted from \smash{$\Gamma_{\odot\pm}$}, and easily converted to a
physical dimensionful mass $M_Y^{\odot\pm}$.

Building on these prerequisites, a quartet of anonymous function closures
are initialized, for each event leg, to perform delayed evaluation of
\mbox{({\it i}\hspace{1.6pt})} the Eqs.~(\ref{eq:ellipseph}) conic coefficients,
\mbox{({\it ii}\hspace{1.6pt})} the Eqs.~(\ref{eq:yslash},\ref{eq:ytimes}) linearly degenerate intersection criteria,
\mbox{({\it iii}\hspace{1.6pt})} the Eqs.~(\ref{eq:physellcoeff}) elliptical center coordinates (or a suitable proxy), and
\mbox{({\it iv}\hspace{1.2pt})} the Eqs.~(\ref{eq:yminus}) conic (y) coordinate domain boundaries (or a suitable proxy).
The ``closure'' is a staple of functional programming, consisting of
a first-class reference to a segment of invocable machine code that
persistently encapsulates a lexically-scoped context of state.
In preparation, if $(P_V^\mu \propto P_S^\mu)$ and/or $(\Omega \Rightarrow 0)$, coefficients for the subordinate
conic shepherd are tabulated according to the one-step prescription of Eqs.~(\ref{eq:mt2ell},\ref{eq:mt2del}), making
the replacements $(\Gamma \Rightarrow \Delta)$ and $(\Widehat{P}_V^\mu \Rightarrow P_S^\mu/E_S)$.
Likewise, the conic discriminant $(b^2 - a\,c)$ from Eq.~(\ref{eq:conicdisc}) is precomputed; 
since forthcoming calculations may be very sensitive to this factor as a denominator, especially
near the parabolic phase transitions, and since a convenient closed form expression is available in
terms of kinematic observables, it is preferable to handle the various logical forks discreetly;
if $(P_V^\mu \propto P_S^\mu)$, then the ersatz one-step discriminant is $(M_S/E_S)^2$, unless the geometry is parabolic $(M_S \Rightarrow 0)$,
in which case the value is undefined; otherwise, for a two-step event leg, the discriminant reduces to $(\,\Omega \times \xi\,)$;
otherwise, for a natural one-step event leg, the discriminant is $\Widehat{M}_V^2$, or undefined if $(M_V \Rightarrow 0)$.

The first of the described function closures, which will be referred to symbolically as $\overline{\ocircle}$, 
is now instantiated to return a numerical evaluation of the conic coefficients referenced in Eq.~(\ref{eq:ellipseph})
at an input event scale, or if no event scale is provided, as a list of polynomial objects.
A second input flag is provided to optionally trigger projection into the conjugate event leg's coordinate plane
via the Eqs.~(\ref{eq:mt2ellp}) transverse missing momentum prescription.
The scale independent coefficients are archived (as a function of $\Upsilon$) at the outset for efficiency; if $(P_V^\mu \propto P_S^\mu)$,
the previously computed shepherd coefficients are substituted as a static ersatz one-step geometry; otherwise,
if $(\Omega \Rightarrow 0)$, the truncated linear surrogate coefficients from Eqs.~(\ref{eq:linecoeffs}) are adopted;
otherwise, the Eqs.~(\ref{eq:mt2ell}) coefficients are applied directly, referencing the precalculated
one- or two-step elements of Eqs.~(\ref{eq:mt2del}).

The next routine stored, to be labeled $\overline{\oslash}$, will be used to compute the pointlike
intersections of a two-step linearly degenerate host geometry, and is thus defined only if $(\Omega \Rightarrow 0)$, and if that
condition is not implied by comprehensive kinematic proportionality of the visible decay products, {\it i.e.} $(P_V^\mu \not\propto P_S^\mu)$.
By default, intersection is imposed with the linear geometry's own one-step shepherd, as in Eqs.~(\ref{eq:yslash}), although an
alternate set of conic coefficients may be input, {\it e.g.} a projection from the conjugate event leg's primary geometry,
facilitating code reuse for the analysis of Eqs.~(\ref{eq:ytimes}).  In either event, quadratic intersection coefficients
are accounted in the hidden momentum coordinate \smash{$\Widehat{P}_H^{y}$}; the leading term simplifies
\smash{$(\alpha_\oslash \Rightarrow \xi)$} to a precomputed factor from Eqs.~(\ref{eq:factordefs}) in the default self-intersection.
Again, inputs are provided for specifying the event scale at which intersection is to be evaluated, and whether the
point(s) of intersection should be projected into the conjugate event plane, as in Eq.~(\ref{eq:projcoords}); if
no event scale is provided, the list of quadratic coefficients is returned, without evaluation, in polynomial object form.
Otherwise, depending on the context of usage, the subroutine return value will be either an ordered pair of intersection roots
\smash{$(\Widehat{P}_H^{y})_{\pm}$} in the single $(y)$ coordinate, or a single vector coordinate \smash{$\langle\Widehat{P}_H^{x,y}\rangle$} representing
the average locus of quadratic roots in the transverse coordinate pair. 

The third function closure defined here, to be denoted $\overline{\odot}$, is tasked with establishing the central
coordinate position \smash{$(\Widehat{P}_H^{x,y})$} of the associated conic geometry at an input event scale;
a primary use case for this routine will entail analysis at the threshold parent mass $M_Y^{\odot+}$, {\it cf.}
Eqs.~(\ref{eq:mt2min}), for inception of real event kinematics, with the purpose of determining the degenerate vector
coordinate \smash{$(\Widehat{P}_H^{x,y})_{\odot+}$} at which the geometry first materializes.
The canonical method for calculating this location is evaluation of the $(x_0,y_0)$ elliptical coordinate center
from Eqs.~(\ref{eq:physellcoeff}); however, vanishing of the Eq.~(\ref{eq:conicdisc}) conic discriminant precludes
this approach for the one- and two-step parabolic geometries.
For the case of a parabolic one-step event shape $(M_V \Rightarrow 0)$, including the ersatz
$\{(P_V^\mu \propto P_S^\mu) \;\&\; (M_S \Rightarrow 0)\}$ one-step parabolic branch,
there truly is no singular coordinate of geometric onset, or finite global coordinate centroid,
although the vertex position described by Eq.~(\ref{eq:orgparproxy}) may serve as a suitable proxy,
making the standard replacements $(\Gamma \Rightarrow \Delta)$ and $(\Widehat{P}_V^\mu \Rightarrow P_S^\mu/E_S)$
in the case of an effective one-step event topology; if the vertex displacement magnitude diverges,
as is expected for the kinematic onset of a massive parabolic event leg, then a dimensionlessly large,
though numerically tame, value may be substituted, while maintaining the physical coordinate orientation.
For the case of a parabolic two-step event shape $\{(\Omega \Rightarrow 0) \;\&\; (P_V^\mu \not\propto P_S^\mu)\}$,
the tangential onset of intersection with the subordinate one-step shepherd, as regulated by Eqs.~(\ref{eq:yslash}) at $M_Y^{\odot+}$,
does constitute a legitimate point of origination; at general (physical) event scales, a suitable proxy for the
ellipse center may be generated by averaging the boundaries of linear overlap with the shepherd, which
coincides with an existing functionality provided by the closure $\overline{\oslash}$.
Failing either of these exception cases, and including $\{(P_V^\mu \propto P_S^\mu) \;\&\; (M_S \not\Rightarrow 0)\}$, which is interpreted by context as a
request for the ersatz one-step geometry's center, the determination of $(x_0,y_0)$ from Eqs.~(\ref{eq:physellcoeff}) may proceed, deferring to
the closure $\overline{\ocircle}$ for scale specialization of the Eqs.~(\ref{eq:mt2ell}) conic coefficients,
and applying the precomputed conic discriminant as a divisor.
The latter two scenarios, which make reuse of $\overline{\oslash}$ and $\overline{\ocircle}$ respectively,
will automatically inherit the capacity for projecting results into the coordinates of a conjugate target system,
whereas analogous behavior must be supplied internally for the one-step parabolic branch.

The final function closure to be instantiated, which will be referred to as $\overline{\ominus}$, serves to
establish the \smash{$\Widehat{P}_H^{y}$} coordinate extent of the corresponding conic geometry at an
input event scale, as is relevant to limiting the Sturm sequence method search domain.  The baseline
technique for determination of this interval will be inversion of the Eqs.~(\ref{eq:yminus}) quadratic
coefficients, as populated, again, by evaluation of the Eqs.~(\ref{eq:mt2ell}) conic coefficients,
including cross projection if applicable, via the closure $\overline{\ocircle}$.  This approach is
unsuitable for linearly degenerate topologies, {\it i.e.} event legs satisfying
$\{(\Omega \Rightarrow 0) \;\&\; (P_V^\mu \not\propto P_S^\mu)\}$,
in which case the closure $\overline{\oslash}$ is defined, and capable of providing a functionally
equivalent specification (with optional projection) of truncation boundaries enforced
by the coupled subordinate one-step shepherd.  Failing this, if $\alpha_\ominus$ vanishes, a
parabolic one-step (or ersatz one-step) event topology is indicated, such that only one isolated
root may be expected, the remote parabolic tail being extended to positive or negative infinity
(horizontal orientation along \smash{$\Widehat{P}_H^{x}$} is mitigated by the Eqs.~(\ref{eq:sincos}) rotation),
which is assigned the undefined position value.  If  $\beta_\ominus$ additionally vanishes
(within some epsilon-sized margin), the parabola is degenerate, and the interior \smash{$\Widehat{P}_H^{y}$} bound
may be identified with the $(y)$ component of the $\overline{\odot}$ vertex locus, which is 
either also ``infinitely'' displaced, or indeterminately positioned at the (self-hosted) coordinate origin;
the undefined value (true infinity) is a lower bound if precisely one of two criteria, those being
the application of coordinate projection (exclusively) or $(\Widehat{P}_V^y < 0)$ ersatz $(P_S^y/E_S < 0)$,
applies, or alternatively, an upper bound.  Otherwise, Eqs.~(\ref{eq:yminus}) may be inverted directly to
obtain an ordered bounding pair; if $(\alpha_\ominus \Rightarrow 0)$, then, the undefined value is
a lower bound if $(\beta_\ominus < 0)$, or alternatively, an upper bound, the physical values of the
quadratic expression associated with Eqs.~(\ref{eq:yminus}) laying above the zero-axis.

Proceeding with primary analysis, the two event legs $(A,B)$ are sorted according to their mass of initial kinematic
onset \smash{$M_Y^{\odot+}$}, and a provisional lower \smash{${\Widetilde{M}}_{\rm T2}$} search bound $\vartriangle'$
is initialized to the maximum of corresponding dimensionless scales \smash{$\{\, \vartriangle' \, \Leftarrow \max\, (\Upsilon^{A,B}_{\odot+}) \,\}$};
by convention, the $(\Widehat{P}_H^{x,y})$ hidden transverse momentum coordinates associated with this heavier event
topology will be appointed the role of host system in the hunt for a minimal consistent phase space overlap.
Similarly, an acting upper \smash{${\Widetilde{M}}_{\rm T2}$} search bound $\triangledown'$ is
provided by the minimum scale of geometric dissociation \smash{$\{\, \triangledown' \Leftarrow \min\, (\Upsilon^{A,B}_{\odot-}) \,\}$};
by contrast, only two-step event leg topologies with $(M_S \not= 0)$ manifest kinematic turn-off at a finite (defined) mass \smash{$M_Y^{\odot-}$}.
Next, various contingencies related to the described solution classifications and critical phases will be screened,
conditionally updating the interim evaluation floor $\vartriangle'$ and ceiling $\triangledown'$,
and ultimately supplying a suitable initialization of the global search bounds $(\vartriangle_0,\triangledown_0)$ and
trial event scale $\vartriangleright_0$ for continuation into the final procedure.
Following application of these filters, \smash{${\Widetilde{M}}_{\rm T2}$} will immediately be deemed undefined if
the consistency condition $(\vartriangle' \,\le \triangledown')$ is not satisfied, modulo some
epsilon-width numerical roundover allowance.

If either event leg exhibits the critical limit $(P_V^\mu \propto P_S^\mu)$, a test is undertaken to determine whether
the resulting ersatz one-step geometry is in type III unbalanced containment of its conjugate, at the degenerate 
scale $\Upsilon_\odot$ of its existence.  Temporarily considering both host versus projection assignations,
the conjugate leg's closure $\overline{\odot}$ is utilized to project a vector coordinate
\smash{$( {\Widehat{\mathcal{X}}}_{\odot'}, {\Widehat{\mathcal{Y}}}_{\odot'} )$}, {\it cf.} Eq.~(\ref{eq:projcoords}),
for its geometric center at the target analysis scale; the primed notation emphasizes that this is not the
projected geometry's intrinsic scale of materialization, but that it belongs instead to the host.
If defined, these coordinates are inserted into an appropriate
generalization of Eq.~(\ref{eq:gmmaxplus}), with $(\Widehat{P}_V^\mu \Rightarrow P_S^\mu/E_S)$, to compute the scale
$\Gamma_{\oplus+}$ at which the ersatz one-step geometry, if hypothetically allowed a dynamic deviation from the 
static replacement $(\Gamma \Rightarrow \Delta)$, would intersect the projected point.  If the required adjustment
is toward contraction $(\Gamma_{\oplus+} \!\le \Delta)$, and pending the scale inversion consistency check, then
the ersatz one-step geometry either intersects, or is in full containment of, its conjugate, thus defining
\smash{${\Widetilde{M}}_{\rm T2}$}; the global upper bound is then tentatively identified with
the standing lower bound $(\triangledown_0 \!\Leftarrow\, \vartriangle')$, condensing the viable search space to a single scale.

Failing the prior, the strategy introduced in conjunction with Eqs.~(\ref{eq:mt2max}) is applied to the identification of
standard type I unbalanced \smash{${\Widetilde{M}}_{\rm T2}$} solutions, and for the providence of a definite upper search bound
initialization $\triangledown_0$ with guaranteed containment of a legitimate intersection.  Again, temporarily considering both
role reversals, analysis proceeds only when the acting host features a one-step event topology.  The projected
event leg's $\overline{\odot}$ closure is invoked at its own scale $\Upsilon_{\odot+}$ of first kinematic materialization
to supply the corresponding vector coordinate \smash{$( {\Widehat{\mathcal{X}}}_{\odot+}, {\Widehat{\mathcal{Y}}}_{\odot+} )$},
in the Eq.~(\ref{eq:projcoords}) form.  The dimensionless scale at which the one-step host geometry passes over this point is
extracted from the associated solution $\Gamma_{\oplus+}$ to Eq.~(\ref{eq:gmmaxplus}); the probationary upper bound $\triangledown''$ is
defined as the minimum of these scales \smash{$\{\, \triangledown'' \Leftarrow \min\, (\Upsilon^{A,B}_{\oplus+}) \,\}$},
or control flow passes to the next test grouping if no such value exists.  The inversion $(\triangledown'' \!\!\le\,\, \vartriangle')$ corresponds 
to a light one-step geometry that eclipses its heavier conjugate's locus of origination prior to (or at) the scale of kinematic onset;
this unbalanced fulfillment of the \smash{${\Widetilde{M}}_{\rm T2}$} commission is respected by condensing the upper
and lower search bounds $(\triangledown_0 \!\Leftarrow\, \vartriangle')$.  Otherwise, if both event legs carry a one-step topology,
mutual intersection is certain to occur at some scale no greater than $\triangledown''$, and the
appropriate assignment is $(\triangledown_0 \Leftarrow \triangledown'')$; this appointment may be bypassed if
$\triangledown''$ is excessively large, exceeding some {\tt BIG} search domain cap, as can occur
when matching an ``infinitely'' displaced one-step parabolic vertex.

Failing the prior, the procedures anticipated in discussion ensuing from Eqs.~(\ref{eq:yslash},\ref{eq:ytimes}) are implemented
in order to cope with intersections involving linearly degenerate $(\Omega \Rightarrow 0)$ two-step event topologies.
Once again making successive permutation of projection and host roles, analysis proceeds only when the de facto host is attached
to an operational instance of the function closure $\overline{\oslash}$; this routine is utilized to render the quadratic
coefficients from Eqs.~(\ref{eq:yslash},\ref{eq:ytimes}) as implicit polynomial object functions of an undetermined event scale,
employing also the conjugate event leg's closure $\overline{\ocircle}$ for projection of the requisite Eqs.~(\ref{eq:mt2ellp})
conic coefficients in the latter case.  The Eq.~(\ref{eq:det4}) convolution of these coefficient sets is additionally precomputed
as an indicator of triple intersection among the host line, the host shepherd perimeter, and the projected conic perimeter.
If the projected conic is likewise linearly degenerate, {\it i.e.} also holding a defined $\overline{\oslash}$ closure, then
\smash{${\Widetilde{M}}_{\rm T2}$} may be identified with the largest mutually consistent lower bound on this triple
intersection, associating each event leg's one-step shepherd, in turn, with the pair of lines; otherwise the possibility
exists of identifying a type II analog of the unbalanced solution scenario, wherein the point of first tangential contact,
at a root of the Eqs.~(\ref{eq:ytimes}) quadratic discriminant \smash{$(\beta_\otimes^2 - 4\,\alpha_\otimes \gamma_\otimes)$},
between the host line and its conjugate, is enveloped by the host's static one-step shepherd.  The Eqs.~(\ref{eq:ytimes})
discriminant and the dual linear reduction of Eq.~(\ref{eq:det4}) are each implicitly a quadratic function
\smash{$(\alpha_\ovee \Upsilon^2_\ovee + \beta_\ovee \Upsilon_\ovee + \gamma_\ovee = 0)$} of the dimensionless
mass-square event scale $\Upsilon$ from Eq.~(\ref{eq:massscale}), and each takes on a positive function value
in the region of physical intersection, interior to the scale of bounding roots.  In the latter case, the
leading quadratic coefficient is negative semi-definite $(\alpha_\ovee \!\le 0)$, vanishing if the merged linear
trajectory is collinear with the symmetry axis of a parabolic one-step shepherd; the $\alpha_\ovee$ phase is
similarly negative in the former case, if the projected geometry carries a two-step event topology;
however, the sign of $\alpha_\ovee$ is undetermined, complicating the interpretation of associated quadratic inversions
\smash{$\{\, \Upsilon_{\ovee\pm} \equiv ( - \beta_\ovee \pm \surd[\, \beta_\ovee^2 - 4\, \alpha_\ovee \gamma_\ovee \,] \,) /\, 2\,\alpha_\ovee \,\}$},
for a conjugate event leg of the one-step variety.  Intuitively, this difficulty may be traced to the possibility of
a phantom tangential intersection occurring between the host line and a kinematically imaginary (negative $\Gamma$)
reflection of the projected geometry.  A suitable unified protocol for rendering roots, while evading this {\it spiegelgeist},
is as follows: if the quadratic discriminant \smash{$(\beta_\ovee^2 - 4\,\alpha_\ovee \gamma_\ovee)$} is negative,
no real intersections exist, and \smash{${\Widetilde{M}}_{\rm T2}$} is immediately undefined;
otherwise, if $(\alpha_\ovee \!< 0)$, downward parabolic concavity implies that the mass-ordered $\Upsilon_{\ovee\pm}$
roots represent a minimum and maximum of intersection, respectively, which may elevate $\vartriangle'$ and/or
diminish $\triangledown'$, as applicable --- the scale inversion filter will subsequently invalidate a dual intersection
with the imaginary geometry; otherwise, if $(\alpha_\ovee \!> 0)$, reversal of the mass-ordering implies that the
lighter root corresponds to turn-off of an imaginary intersection, which is ignored, whereas the heavier root initiates
a phase of unbounded physical intersection, which is used to buttress $\vartriangle'$ in the upward
direction --- the scale $\Upsilon_{\ovee+}$ is a lower bound in both cases, and vice versa for $\Upsilon_{\ovee-}$;
otherwise, with $(\alpha_\ovee \!\Rightarrow 0)$, attention turns to the sign of $\beta_\ovee$,
with the single linear inversion \smash{$\{\, \Upsilon_{\ovee} \Rightarrow -1 \times (\gamma_\ovee / \beta_\ovee ) \,\}$}
providing an imaginary upper bound for $(\beta_\ovee \!< 0)$, which potentially lowers $\triangledown'$,
or a physical lower bound for $(\beta_\ovee \!> 0)$, which potentially raises $\vartriangle'$;
otherwise, if $(\beta_\ovee \!\Rightarrow 0)$, there can be no intersection unless $(\gamma_\ovee \!\Rightarrow 0)$ as well,
in which case no bounds are set, and failing which \smash{${\Widetilde{M}}_{\rm T2}$} is immediately undefined.
If both event legs are linearly degenerate, or if the quadratic inversion defined a lower
scale bound satisfying the Eq.~(\ref{eq:type2containment}) criterion for enclosure of a tangential intersection 
by the host shepherd (within some epsilon-width error), then \smash{${\Widetilde{M}}_{\rm T2}$}
is considered isolated, pending the scale inversion consistency check, as reflected by consolidation of the
search bounds $(\triangledown_0 \!\Leftarrow\, \vartriangle')$.  Alternatively, for a single linearly degenerate event leg,
and again pending the scale inversion consistency check, the interim upper search bound is promoted into the global position
$(\triangledown_0 \!\Leftarrow\! \triangledown')$; although no legitimate \smash{${\Widetilde{M}}_{\rm T2}$} scale may exist
exterior to the specified containment region, the question of whether a solution exists interior to it remains open in this case,
and the existing realization of Eq.~(\ref{eq:det4}) is retained for subsequent reuse to help settle it. 

Failing the prior, the global upper search bound defaults to the least available limit
$(\triangledown_0 \!\Leftarrow\! \triangledown')$, if any, and a Boolean flag is set to
trigger use of the tangential intersection discriminant from Eqs.~(\ref{eq:qrtdiscriminant},\ref{eq:cascadeqrtdiscriminant})
for ascertaining the status of \smash{${\Widetilde{M}}_{\rm T2}$} root containment.

The last stage of preparations for the iterative confinement of a potentially latent \smash{${\Widetilde{M}}_{\rm T2}$} solution
begins by filtering against scale inversion $(\vartriangle' \,\le \triangledown')$, and
employing a very soft logarithmic type cut on numerical runaway of the dimensionless scale ratio $(\vartriangle' \,\le\! \mathtt{BIG})$,
and elsewise undefining \smash{${\Widetilde{M}}_{\rm T2}$}.  If the global upper search bound $\triangledown_0$ is
not positively guaranteed to bound a valid intersection --- this is indicated either by
the preceding flag for indeterminate root containment, or by prior retention of the quartic triple intersection
(host line, host shepherd, and generalized conic projection) criterion expressed in Eq.~(\ref{eq:det4}) ---
then it is capped at a maximal value of {\tt BIG}, or set directly to the {\tt BIG} value if undefined. 
The global trial event scale $(\vartriangleright_0 \,\Leftarrow\, \vartriangle')$ and the global
lower search bound $(\vartriangle_0 \,\Leftarrow\, \vartriangle')$ are each initialized to the standing minimum.
These three variables, along with a copy of the mass-squared scaling reciprocal $(2\,E_V E'_V)$, and a pair of new function closures
$(\, \overline{\ovee},\, \overline{\circledcirc} \,)$ to be described following, constitute the sole input to this final routine.
Both closures may potentially make use of the Eq.~(\ref{eq:intersection}) quartic coefficients $u_i$,
a polynomial object representation of which is thus precomputed via Eqs.~(\ref{eq:intcoeffs1},\ref{eq:intcoeffs2}),
calling, in turn, on each event leg's $\overline{\ocircle}$ instance to supply
respective host and projection Eq.~(\ref{eq:ellipseph}) conic coefficients.

The closure $\overline{\ovee}$ accepts a dimensionless input scale $\Upsilon'$, {\it cf.} Eq.~(\ref{eq:massscale}),
and outputs an integral count of candidate event geometry intersections occurring interior to $\Upsilon'$, interpreted
as a lower root boundary, and a static copy of the global upper search bound initialization scale $\triangledown_0$.
Internally, this routine applies the Sturm sequence method to count roots, in $\Upsilon$,
for an archived, context sensitive, polynomial object.
If either event leg exhibits a static ersatz event topology $(P_V^\mu \propto P_S^\mu)$, then legitimate intersection
may occur only at the singular physical event scale, and the null root count value $(0)$ is returned.
Otherwise, if the  Eq.~(\ref{eq:det4}) quartic triple intersection has been retained from the screening phase,
then it doubles here as the analysis function.  Otherwise, if the indeterminate
containment flag is set, then the appropriate function for analysis is the 12\textsuperscript{th} order discriminant $\mathbb{D}_i$
of tangential conic intersection from Eqs.~(\ref{eq:qrtdiscriminant},\ref{eq:cascadeqrtdiscriminant}),
as assembled from the archived Eqs.~(\ref{eq:intcoeffs1},\ref{eq:intcoeffs2}) polynomial object coefficients $u_i$; an undefined intersection count,
as may occur if $\mathbb{D}_i$ is identically null, is interpreted numerically as $(0)$.  Otherwise,
no such analysis functionality is essential, and $\overline{\ovee}$ is thus left undefined.
To expedite computation, the corresponding sequence of Sturm
polynomials is assembled only a single time, and reapplied to root counting for variable boundaries on demand.
As a note, a version of the Sturm sequence that counts degenerate roots separately is employed in this
cycle, in order to distinguish that circumstance from the crossing of a single root. 

The closure $\overline{\circledcirc}$ accepts a dimensionless input scale $\Upsilon'$, {\it cf.} Eq.~(\ref{eq:massscale}),
and outputs a Boolean assertion of whether the conic perimeters associated with each event leg
occupy a state of mutual intersection in the $(\Widehat{P}_H^{x,y})$ transverse hidden momentum plane.  Internally,
this routine applies the Sturm sequence method to count roots, at $\Upsilon'$, in $\Widehat{P}_H^y$, of the polynomial $u_i$.
The $\Widehat{P}_H^y$ search boundary is truncated to the intersection of physical coordinate domains
(possibly allowing some marginal excess width epsilon) for the host and projected conic topologies, each of which is supplied, in turn,
by a linked closure $\overline{\ominus}$.  The zeroth order $(y)$-intercept of this expression may be varied by some positive and negative margin
of size epsilon, and optionally scaled in proportion to the $u_i$ coefficient norm, in order to
tune for enhanced detection of roots occurring in (near) degenerate pairs.
A true value is returned if the enclosed count of intersections is at least one, or if all of the $u_i$
from Eq.~(\ref{eq:intersection}) vanish simultaneously (the Sturm algorithm returns an undefined
value), indicating degeneracy of the event leg geometries.

The convergence toward a minimal kinematically consistent parent particle species mass $M_Y$ for generalized one- and
two-step event topologies now culminates with an iteratively sampled probe of scales, which persists until 
the upper $\triangledown$ and lower $\vartriangle$ search bounds become sufficiently mutually nestled about
a verified moment of conic intersection, or the search space becomes exhausted.
The closure $\overline{\ovee}$ is prized for an ability to count bounded intersection roots directly in the mass domain $\Upsilon$ (as opposed to
operating in the coordinate domain at fixed scales), but it can be numerically fragile; in deference to unrivaled stability
of root isolation in $\Widehat{P}_H^y$ associated with the closure $\overline{\circledcirc}$, $\overline{\ovee}$ is used,
when defined, as a beacon to home in on scales of interest, which are subsequently reviewed in $\mathbb{N}$ steps by $\overline{\circledcirc}$.
During each process cycle, the prevailing trial event scale $\vartriangleright$ will be conditionally assigned to one of the two search bounds $(\vartriangle,\triangledown)$:
if the closure $\overline{\ovee}$ is defined, while the step count $\mathbb{N}$ is not, then
\mbox{({\it i}\hspace{1.6pt})} it is evaluated to establish a count $\eta$ of bounded roots superior to $\vartriangleright$, and for primary loop entry
\mbox{({\it ii}\hspace{1.6pt})} an initial root count is retained $(\eta_0 \!\Leftarrow\! \eta)$ for comparison while
	the lower bound $(\vartriangle \Leftarrow\, \vartriangleright)$ is elected for assignment (a null operation),
or for loop reentry, if the sequestration of roots $\eta$ is undiminished relative to $\eta_0$, then
\mbox{({\it iii}\hspace{1.6pt})} the lower bound is extended $(\vartriangle \Leftarrow\, \vartriangleright)$,
or \mbox{({\it iv}\hspace{1.2pt})} the lightest root has elseways been passed over and the upper bound is clipped $(\triangledown \!\Leftarrow\, \vartriangleright)$; otherwise, 
\mbox{({\it i}\hspace{1.6pt})} the closure $\overline{\circledcirc}$ is evaluated for topological intersection at $\vartriangleright$, and if found, then 
\mbox{({\it ii}\hspace{1.6pt})} $\overline{\ovee}$ and $\mathbb{N}$ (being redundant) are undefined while the upper scale bound is diminished $(\triangledown \!\Leftarrow\, \vartriangleright)$, or otherwise
\mbox{({\it iii}\hspace{1.6pt})} the lower scale bound is elevated $(\vartriangle \Leftarrow\, \vartriangleright)$.
Subsequently, the trial event scale is itself updated, by a step-wise increment of the scanning interval $(\triangledown \,-\! \vartriangle_0)$ if $\mathbb{N}$ is defined,
or alternatively to the bisection of event bounds \smash{$\{\, \vartriangleright \,\Leftarrow (\vartriangle \!+\, \triangledown) \div 2 \,\}$};
in the former case, a map that power-compresses sampling at lighter scales while diluting the sample density at heavier scales is desirable, {\it cf.} Appendix~\ref{sct:appendix_C}. 
The enclosing loop terminates when
\mbox{({\it i}\hspace{1.6pt})} $\eta$ is defined and either has relinquished just a single contained root to $\eta_0$ or $\eta_0$ is null, or
\mbox{({\it ii}\hspace{1.6pt})} the scale gap $(\triangledown -\! \vartriangle)$ approaches zero; for a specific proximity criterion that optimizes 
uniform mass resolution in $M_Y$, {\it cf.} Appendix~\ref{sct:appendix_C}.
If the closure $\overline{\ovee}$ is residually defined, then hunting with $\overline{\ovee}$ and scanning with $\overline{\circledcirc}$ alternate in a tic-toc cadence
with revisions to the initialization state providing the opportunity to either validate and converge upon a broadly isolated root candidate or
bypass a spurious intersection to continue searching at heavier candidate scales: if $\mathbb{N}$ is not defined, then
\mbox{({\it i}\hspace{1.6pt})} a suitable scanning sample count $(\mathbb{N} \Leftarrow \mathbb{N}_0)$ is selected ({\it cf.} Appendix~\ref{sct:appendix_C}) and the acting lower scale bound reverts to the global floor
	\smash{$\{\, (\vartriangle,\vartriangleright) \Leftarrow (\vartriangle_0,\vartriangle_0) \,\}$} while $(\eta,\eta_0)$ are undefined and the primary loop is reentered,
or, if the acting and global upper scale bounds are equivalent $(\triangledown \equiv \triangledown_0)$, then
\mbox{({\it ii}\hspace{1.6pt})} scanning has broached the global bound with no intersections found and \smash{${\Widetilde{M}}_{\rm T2}$} is reported as undefined, or
\mbox{({\it iii}\hspace{1.6pt})} the active hunting interval is pushed above the failed scanning interval
	\smash{$\{\, (\vartriangle,\vartriangleright,\triangledown) \Leftarrow (\triangledown,\triangledown,\triangledown_0) \,\}$}
	while $(\mathbb{N},\mathbb{N}_0,\eta,\eta_0)$ are undefined and the primary loop is reentered;
otherwise, at last, and optionally enforcing a final scale filter \mbox{$(\vartriangleright \,\le \mathtt{BIG})$}, a physical asymmetric s-transverse mass statistic value is assigned
\smash{$\{\, {\Widetilde{M}}_{\rm T2} \,\Leftarrow \surd[\, (2\,E_V E'_V) \hspace{0.6pt}\times\hspace{-0.6pt} \vartriangleright \,] \,\}$}.


\begin{center}
\bf{\small Conclusions}
\end{center}

The $M_{\rm T2}$, or ``s-transverse mass'', statistic was developed to cope with the difficulty of
associating a parent mass scale with a missing transverse energy signature, given that models of new physics
generally predict production of escaping particles in pairs, while collider experiments are sensitive
to just a single vector sum over all sources of missing transverse momentum.
This document focused on the generalized \smash{${\Widetilde{M}}_{\rm T2}$} extension of that statistic to
asymmetric one- and two-step decay chains, with arbitrary child particle masses and upstream missing transverse momentum.  A
\mbox{({\it i}\hspace{1.6pt})} unified theoretical formulation,
\mbox{({\it ii}\hspace{1.6pt})} complete solution classification,
\mbox{({\it iii}\hspace{1.6pt})} taxonomy of critical points,
and \mbox{({\it iv}\hspace{1.2pt})} technical algorithmic prescription were
provided for treatment of the \smash{${\Widetilde{M}}_{\rm T2}$} event scale.

Potentially novel elements of this presentation have included
\mbox{({\it i}\hspace{1.6pt})} a unified symbolic and computational environment encompassing arbitrary event configurations,
\mbox{({\it ii}\hspace{1.6pt})} a graphically enhanced intuition for the organization and interrelation of solution classes, 
\mbox{({\it iii}\hspace{1.6pt})} a heightened sensitivity in the identification and handling of exception cases,
\mbox{({\it iv}\hspace{1.2pt})} an optimization of the mass search bounds without resort to an arbitrary hard upper scale,
and \mbox{({\it v}\hspace{1.2pt})} implementation of an existence test for solutions interior to some scale bounds without sampling the enclosed bulk.

An implementation of the described algorithm is now available for download~\cite{amt2},
and is also a deployable component of the author's fully-featured selection cut software
package \mbox{\sc AEACuS} ({\sc A}lgorithmic {\sc E}vent {\sc A}rbiter and {\sc Cu}t {\sc S}elector),
which was formerly developed under the name \mbox{\sc CutLHCO}~\cite{Walker:2012vf,aeacus}.
All described programs have been freely released into the public domain under the terms of the GNU General Public License~\cite{gnugpl}.\vspace{-0.4pt}

A triad of appendices are provided following, addressing
\mbox{({\it a}\hspace{1.2pt})} combinatoric assembly of event objects, 
\mbox{({\it b}\hspace{1.2pt})} validation by public codes against Monte Carlo events,
and \mbox{({\it c}\hspace{1.2pt})} the algorithm pseudocode and logical flow diagram.


\makeatletter
\close@column@grid
\onecolumngrid
\vspace{+8pt}
\begin{center}
\includegraphics[width=0.8\textwidth]{vectorian_dot_net_flourish.eps}
\end{center}
\close@column@grid
\twocolumngrid
\makeatother


\vphantom{~}
\newpage
\vphantom{~}
\newpage
\appendix

\section{\protecting{Combinatoric Assembly of \texorpdfstring{\smash{$\boldsymbol{{\Widetilde{M}}_{\rm T2}}$}}{MT2}}\label{sct:appendix_A}}

The preceding treatment of \smash{${\Widetilde{M}}_{\rm T2}$} has assumed a well-defined
particle association for each of the primary $V$ and secondary $S$ visible particle species, {\it cf.} FIGS.~\ref{fig:onestep} and~\ref{fig:twostep}.
However, in a realistic collider environment, both the topology of the parent process and the roles played by
particular descendent tracks within that topology are obscured.  The typical application of
\smash{${\Widetilde{M}}_{\rm T2}$} therefore consists of stipulating a parent event hypothesis, permuting
the logically consistent assignment of roles to each hard track, and the minimal resulting
\smash{${\Widetilde{M}}_{\rm T2}$} scale for consistency with the event hypothesis.  The present section of the
appendices details several common strategies of this type, including the associated treatment of
pre-selection event cuts and the combinatoric assembly of constituent objects, as well as providing a brief tutorial
for the tidy automation of this full analysis within the author's generalized selection cut
package \mbox{\sc AEACuS}~\cite{Walker:2012vf,aeacus}.

Failing a robust and luminous signal of new physics, the \smash{${\Widetilde{M}}_{\rm T2}$} variable
will necessarily find employ as a discovery tool, {\it i.e.} an aid in the suppression of standard model
background, rather than as the classificant of a putative new physics mass scale.  Correspondingly,
the variable specializations \smash{$M_{\rm T2}^{\rm b}$}, \smash{$M_{\rm T2}^{{\rm b}\ell}$}, \smash{$M_{\rm T2}^{\rm W}$}~\cite{Bai:2012gs}, and \smash{$M_{\rm T2}^{\tau}$}~\cite{ATLAS-CONF-2013-037}
to be detailed following will each adopt a Standard Model event hypothesis, namely that of top quark pair production $(\,t\bar{t}\,)$,
and cut events that realize an \smash{${\Widetilde{M}}_{\rm T2}$} scale beneath the top quark mass threshold at $175$~GeV. 

The variable \smash{$M_{\rm T2}^{\rm b}$}~\cite{Bai:2012gs}, like each of those to be here considered, assumes an underlying event topology
of di-leptonic top quark pair production, with one lepton track lost to the detector.  In order to facilitate variants of the
\smash{$M_{\rm T2}$} algorithm that are limited to symmetric invisible daughter species, the visible lepton is manually discarded via
assimilation into the missing transverse momentum vector \smash{${\vec{{\slashed{P}}}} {}_{\rm T}$}.  Each event leg thereby features 
a one-step decay, with the role of the visible $(V)$ particle object played by an available bottom-quark jet, and an invariant missing
mass $(M_H \Rightarrow 80.4$~GeV$)$ associated with the lost lepton and neutrino ($W$~boson) pair.

The variable \smash{$M_{\rm T2}^{{\rm b}\ell}$}~\cite{Bai:2012gs} is similarly based upon a dual one-step event topology, but is optimized
for algorithms that allow asymmetric invisible daughter species.  The visible lepton is instead merged
with its upstream jet, such that the hidden mass $(M_H \Rightarrow 0.0$~GeV$)$ on that event leg is now attributed 
to only the lost neutrino.

The variable \smash{$M_{\rm T2}^{\rm W}$}~\cite{Bai:2012gs} instead adopts a mixed one- and two-step event topology, taking advantage
of the extra information density available by reserving independent roles for the lepton, which is adopted as
the secondary $(S)$ visible species, and jet, which retains assignment as the primary $(V)$ visible species.  The hidden mass
$(M_H \Rightarrow 0.0$~GeV$)$ on that event leg is again associated with the neutrino, whereas the on-shell mediator species $(X)$
is hypothesized $(M_X \Rightarrow 80.4$~GeV$)$ to be a $W$~boson.

The variable \smash{$M_{\rm T2}^{\tau}$}~\cite{ATLAS-CONF-2013-037}, which corresponds to a dual two-step event topology,  stipulates that
the ``lost'' lepton was actually a hadronic tau, which therefore contributed an additional measure of visible momentum to the event,
in the form of a third final state jet.  The leptonic \smash{$M_{\rm T2}^{\tau}$} event leg is treated identically as for its prior appearance in \smash{$M_{\rm T2}^{\rm W}$}.
For the hadronic event leg, the pair of neutrinos (from $W \to \tau \bar{\nu}_\tau$ and $\tau \to \nu_\tau j$) are implicitly combined
into a single, ostensibly massless $(M_H \Rightarrow 0.0$~GeV$)$, invisible object, whereas the mediator species $(X)$
is again a $W$, with $(M_X \Rightarrow 80.4$~GeV$)$.

Each of the described variables, with the exception of \smash{$M_{\rm T2}^{\rm b}$}, which counterfeits the transverse missing momentum,
is available as a native processing option in the \mbox{\sc AEACuS}~\cite{Walker:2012vf,aeacus} event selection package.
The combinatoric assembly of jets made available to the \smash{${\Widetilde{M}}_{\rm T2}$} algorithm is 
handled internally, in a manner consistent with the suggestions of Ref.~\cite{Bai:2012gs}.
If there are two or more tagged jets supplied, then all pairwise combinations are considered, whereas all non-tagged jets present
are otherwise permuted into the vacancy or vacancies; both pairwise orderings for the event leg assignment are entertained.
For \smash{$M_{\rm T2}^{\tau}$}, all available jets not currently playing the role of a primary visible $(V)$ object are cycled into
the secondary $(S)$ visible position on the hadronic event leg, for each combination of leading pairs.  The smallest well-defined
value of \smash{${\Widetilde{M}}_{\rm T2}$} extracted from all combinatoric trials is associated with the event as a whole.
 
The \mbox{\sc AEACuS} control syntax for constructing a typical analysis is demonstrated in Card~\ref{mt2card}; the object reconstruction phase of
this card enforces identification of a solitary isolated lepton, in conjunction with the existence of precisely one or two heavy-flavor
tagged jets among the hardest four jet candidates, which are supplied as a group to the \smash{${\Widetilde{M}}_{\rm T2}$} algorithm.
A missing transverse energy event selection cut of 100~GeV is enforced, in addition to a lower bound of 175~GeV for each of the
variables \smash{$M_{\rm T2}^{{\rm b}\ell}$}, \smash{$M_{\rm T2}^{\rm W}$}, and \smash{$M_{\rm T2}^{\tau}$}, which are referenced as modes $(1,2,\mathrm{~and~}3)$, respectively.

\begin{card}
\centering
\ovalbox{%
\begin{minipage}{0.97\linewidth}
\vspace{+5pt}
{\scriptsize
\begin{Verbatim}[numbers=left,numbersep=6pt]
*** AEACuS 3.4 cut_card.dat ***
* Study MT2bl, MT2W, & MT2Tau with
* 1-2 b-Tags in 4 hardest Jets & 1 isolated Lepton
**** Object Reconstruction ****
OBJ_TAU = PTM:[0,0], JET:1
  # Recast Tau Leptons as Jets
OBJ_LEP = PTM:10, PRM:[0.0,2.5], EMT:-3, CUT:[1,1]
  # Enforce exactly one e/mu with PT > 10 and ETA < 2.5
OBJ_JET = PTM:25, PRM:[0.0,2.5]
  # Define Jets with PT > 25 and ETA < 2.5
OBJ_LEP_001 = SRC:+000, EMT:+1
  # Identify Electron flavor subset for future reference
OBJ_JET_002 = SRC:+000, CMP:+001, CDR:0.2
  # Enforce Jet isolation DeltaR > 0.2 from prior Electron
OBJ_LEP_003 = SRC:+000, PTM:25, CMP:+002, CDR:0.4, CUT:[1,1]
  # Enforce Lepton isolation DeltaR > 0.4 from prior Jets
  # Enforce isolated Lepton is not soft PT > 25
OBJ_JET_004 = SRC:+002, PTM:50, CUT:[4,UNDEF,-1]
  # Identify 4 hardest Jets with PT > 50
OBJ_JET_005 = SRC:+004, HFT:0.5
  # Identify heavy-flavor (via PGS4 code) Jet subset
OBJ_JET_006 = SRC:[+004,-005], CUT:[2,3]
  # Enforce exactly 2 or 3 of hardest 4 Jets not b-Tagged 
******* Event Selection *******
EVT_MET = CUT:100
  # Enforce Missing Transverse Energy > 100 GeV
EVT_ATM_002 = MET:000, JET:004, LEP:003, MOD:1, CUT:175
  # Enforce minimum MT2bl of prior objects > 175 GeV
EVT_ATM_003 = MET:000, JET:004, LEP:003, MOD:2, CUT:175
  # Enforce minimum MT2W of prior objects > 175 GeV
EVT_ATM_004 = MET:000, JET:004, LEP:003, MOD:3, CUT:175
  # Enforce minimum MT2Tau of prior objects > 175 GeV
*******************************
\end{Verbatim}
} \vspace{-5pt} \end{minipage}}
\caption{\mbox{\sc AEACuS} syntax for \smash{$M_{\rm T2}^{{\rm b}\ell}$}, \smash{$M_{\rm T2}^{\rm W}$}, and \smash{$M_{\rm T2}^{\tau}$},
selecting four hard jets, one or two of which must carry a heavy-flavor tag, and a single isolated light lepton.}
\label{mt2card}
\end{card}

\section{\protecting{Validation by Public Codes of \texorpdfstring{\smash{$\boldsymbol{{\Widetilde{M}}_{\rm T2}}$}}{MT2}}\label{sct:appendix_B}}

The present section of the appendices summarizes results for a thorough cross-comparison of the described  
\smash{${\Widetilde{M}}_{\rm T2}$} algorithm against other public codes, in the context of realistic Monte-Carlo
data, including both Standard Model background events and the simulation of various candidates for new physics.
The \smash{$M_{\rm T2}^{\rm b}$} statistic~\cite{Bai:2012gs}, corresponding to a dual one-step event topology with symmetric invisible
daughters, is precisely of the type accessible to the original algorithm~\cite{Cheng:2008hk,mt2bisection} by Cheng and Han~(CH), which
then constitutes a first natural touchstone.  That algorithm has previously demonstrated excellent
agreement with the earlier effort~\cite{Lester:1999tx,Barr:2003rg,oxbridge} of Lester and Barr {\it et al.}, which is therefore omitted from comparison.
The \smash{$M_{\rm T2}^{{\rm b}\ell}$} and \smash{$M_{\rm T2}^{\rm W}$} event hypotheses~\cite{Bai:2012gs} correspond respectively to a dual one-step
and a mixed one- and two-step event topology, both with asymmetric daughter products; extensions of the
work by Cheng and Han, with extenuation of authorship by the latter, are also publicly available for the computation of these statistics~\cite{Bai:2013ema},
and are selected as the candidates for comparison in these cases; the shorthand notation (CH) is carried over to inclusively describe
these derivative works.  The author is unaware of other publicly available codes for the analogous treatment of dual two-step event topologies,
and the \smash{$M_{\rm T2}^{\tau}$} statistic is therefore neglected in the trials.

One hundred thousand Monte Carlo event samples, including parton showering and fast detector simulation, were generated via the standard 
{\sc MadGraph}/{\sc MadEvent}~\cite{Alwall:2011uj}, {\sc Pythia}~\cite{Sjostrand:2006za}, {\sc PGS4}~\cite{PGS4} chain for
each of three subprocesses (all plus 0,1, or 2 jets), including the \mbox{({\it i}\hspace{1.6pt})} Standard Model top quark pair production  $(\,t\bar{t}\,)$ background, as well as the
SUSY stop squark pair production  $(\,\tilde{t}\bar{\tilde{t}}\,)$ signal for \mbox{({\it ii}\hspace{1.6pt})} an mSUGRA benchmark with $M_0 = 210$~GeV, $M_{1/2} = 350$~GeV, and $\tan \beta = 40$,
and \mbox{({\it iii}\hspace{1.6pt})} a No-Scale $\mathcal{F}$-$SU(5)$~\cite{Li:2010ws,Li:2011gh,Li:2013naa} benchmark with $M_{1/2} = 850$~GeV.  A modification of the
\mbox{\sc AEACuS}~\cite{Walker:2012vf,aeacus} selection cut package, enacting the instructions specified in Card~\ref{mt2card}, was used to select events
compatible with the target topology, and catalogue all viable combinatoric variants of the associated physics objects for use in the computation of
\smash{$M_{\rm T2}^{\rm b}$}, \smash{$M_{\rm T2}^{{\rm b}\ell}$}, and \smash{$M_{\rm T2}^{\rm W}$}; in the case of \smash{$M_{\rm T2}^{\rm b}$}, the visible lepton was manually
merged into the the missing transverse momentum \smash{${\vec{{\slashed{P}}}} {}_{\rm T}$}, as previously described.  In order to amplify the statistical inclusion of
background samples, the natural $(\,t\bar{t}\,)$ mono-lepton samples were enriched with counterparts fabricated from di-lepton samples where the lighter partner
was intentionally ``lost'', being likewise absorbed into \smash{${\vec{{\slashed{P}}}} {}_{\rm T}$}; in this case, only the two hardest jets were retained for consideration,
irrespective of any heavy-flavor tags.  After selection cuts, approximately 21,500 event configurations were retained for the testing
of \smash{$M_{\rm T2}^{\rm b}$}, whereas twice this tally (via breaking of the event leg symmetry) were available for application to 
each of the \smash{$M_{\rm T2}^{{\rm b}\ell}$} and \smash{$M_{\rm T2}^{\rm W}$} statistics.

\begin{table*}[htp] {
	\renewcommand{\arraystretch}{1.0}
        \centering
	\footnotesize
        \begin{tabular}{>{\centering\arraybackslash}m{\dimexpr.14\linewidth} >{\centering\arraybackslash}m{\dimexpr.24\linewidth}%
	>{\centering\arraybackslash}m{\dimexpr.24\linewidth} >{\centering\arraybackslash}m{\dimexpr.06\linewidth} >{\centering\arraybackslash}m{\dimexpr.06\linewidth} }
	\toprule
	${\slashed{P}} {}_{\rm T}^{x,y}$ & ${(P_V^\mu)}^A$ & ${(P_V^\mu)}^B$ & \protecting{${\Widetilde{M}}_{\rm T2}^{\phantom{A}}$} & ${M}_{\rm T2}^{\rm CH}$ \\
	\midrule
(-147.2, -265.1) & (186.3, -136.3, -6.5, +89.1) & (93.0, -30.4, -52.7, -69.2) & 93.1 & 148.2 \\
(-52.3, +190.1) & (199.5, -8.2, +180.3, -36.4) & (422.5, +65.3, +194.6, +369.3) & 131.7 & \tt{undef} \\
(-425.7, -174.8) & (157.0, +132.9, -11.2, +51.2) & (139.5, -74.7, -28.2, +111.6) & 107.9 & 225.1 \\
(-586.3, -52.4) & (390.4, -23.7, +166.1, +342.5) & (106.3, -53.7, -5.8, +91.1) & 90.3 & 157.2 \\
(-315.5, -175.1) & (159.4, -36.6, -92.4, +74.6) & (139.6, -137.1, -13.9, +11.3) & 119.3 & 209.2 \\
(-513.1, +1128.8) & (198.6, +115.6, -84.7, -69.8) & (577.1, -130.0, +161.4, -529.8) & 267.9 & 671.4 \\
(-668.7, +389.7) & (496.8, +37.6, -235.0, -432.5) & (124.1, -37.6, +65.2, -97.8) & 105.4 & 404.5 \\
(-78.1, +244.7) & (268.8, +192.3, -100.6, +120.0) & (481.2, -74.0, +169.2, +442.4) & 122.0 & 137.8 \\
(-199.5, -170.6) & (169.9, -8.4, -101.0, +77.4) & (297.9, -177.4, -105.9, +212.1) & 121.0 & 179.0 \\
(-199.5, -170.6) & (112.4, +17.2, -46.8, +7.2) & (297.9, -177.4, -105.9, +212.1) & 118.7 & 190.0 \\
(+73.8, +102.8) & (118.5, -0.8, -40.3, -54.2) & (205.5, +74.3, +64.6, -178.9) & 111.6 & 206.4 \\
(-226.3, -52.8) & (164.9, -60.5, +55.7, +108.5) & (120.6, -113.6, -23.0, +30.8) & 100.7 & 141.1 \\
(-733.3, -286.9) & (332.5, +16.1, +243.9, +123.5) & (465.2, -386.3, -172.7, +154.3) & 219.0 & 252.7 \\
(-438.9, +115.4) & (124.8, +39.7, +69.8, +28.3) & (226.7, -93.1, +37.8, +202.8) & 97.6 & 180.0 \\
(+230.5, +37.3) & (179.3, +46.0, -130.4, +58.8) & (127.7, +107.5, +33.0, +51.1) & 113.2 & 134.5 \\
(+1053.3, +831.7) & (244.4, -17.6, +148.6, -172.3) & (96.6, +72.8, +31.1, +52.2) & 122.8 & 261.3 \\
(+6.0, +424.1) & (351.6, -261.3, -104.5, -193.1) & (82.6, +16.1, +68.4, -40.7) & 96.2 & 138.8 \\
(+393.2, +170.0) & (712.8, -518.6, +186.0, -442.4) & (197.2, +159.9, -16.2, +109.3) & 126.0 & 344.1 \\
(-248.4, -409.4) & (196.0, -34.3, -167.8, -56.5) & (200.2, -99.5, -105.2, -137.7) & 98.9 & 130.8 \\
(-248.4, -409.4) & (231.8, -66.7, -159.0, -133.3) & (200.2, -99.5, -105.2, -137.7) & 101.3 & 143.8 \\
(+1093.0, -402.4) & (216.5, +112.3, -132.6, +97.9) & (192.3, +160.3, -55.7, -87.1) & 109.3 & 164.3 \\
(-139.0, +285.7) & (303.5, -155.6, -88.1, +231.5) & (209.4, -22.4, +70.9, +195.6) & 90.5 & 122.4 \\
(+144.5, -42.3) & (160.1, +86.5, +74.8, -98.5) & (259.0, -228.9, +40.6, -114.1) & 155.9 & \tt{undef} \\
(+207.7, +263.7) & (181.9, -85.8, +106.3, +11.9) & (319.2, +242.8, +172.2, -107.6) & 135.0 & 192.4 \\
(+466.3, -78.6) & (158.5, +51.9, -60.1, -111.4) & (98.1, +58.9, +1.8, -78.4) & 89.3 & 222.8 \\
(+53.7, -223.1) & (231.8, +36.4, +111.2, +191.2) & (52.7, +15.7, -47.7, +15.9) & 87.4 & 264.8 \\
(+204.6, +229.1) & (110.5, -6.2, -27.8, -41.3) & (447.2, +182.2, +162.4, -374.1) & 113.5 & 174.7 \\
(-148.8, +169.7) & (99.0, +59.8, -29.6, +1.4) & (171.3, -75.4, +97.5, +118.5) & 98.0 & 182.4 \\
(+225.8, -82.9) & (168.3, -104.1, +85.5, +86.5) & (103.3, +83.2, -32.5, +51.8) & 88.7 & 263.5 \\
(-103.2, -88.5) & (196.3, +153.9, -72.3, +36.1) & (129.2, -105.2, -40.7, -61.7) & 104.6 & 153.3 \\
(-103.2, -88.5) & (107.8, +44.9, -44.7, +1.2) & (129.2, -105.2, -40.7, -61.7) & 106.8 & 169.7 \\
(-58.1, +84.2) & (301.2, -21.2, +165.3, +246.1) & (203.9, +15.8, -126.1, -159.4) & 94.3 & \tt{undef} \\
(+243.7, -700.0) & (339.2, -90.0, -67.3, -302.6) & (141.0, +78.0, -100.6, +52.7) & 132.7 & 290.6 \\
(+195.6, -30.3) & (427.6, -136.6, +177.6, +355.2) & (100.5, +65.8, -11.4, +74.9) & 91.7 & 129.2 \\
(-337.1, +291.8) & (358.2, +21.6, -116.6, +332.3) & (160.7, -72.0, +75.5, +121.8) & 91.2 & 158.8 \\
(-134.8, +104.3) & (152.8, +3.1, -124.6, +8.3) & (151.6, -89.3, +115.3, +38.1) & 106.5 & 181.8 \\
(+126.7, +415.1) & (229.9, -182.7, -67.9, -74.0) & (198.2, +127.4, +125.7, -61.3) & 154.1 & 289.0 \\
(+213.6, +20.3) & (194.1, -57.4, +157.3, -69.7) & (68.1, +58.4, -31.8, -13.8) & 91.7 & 181.3 \\
(-177.5, -59.0) & (124.6, +17.3, -15.4, -84.7) & (415.9, -78.4, -46.8, -405.4) & 102.5 & 205.8 \\
(+68.7, +50.5) & (330.2, -19.0, -193.8, -251.2) & (355.4, +77.0, +174.8, -299.5) & 113.6 & 223.2 \\
(+132.4, -20.3) & (115.1, -42.5, -40.5, -56.0) & (140.5, +59.0, -19.8, +125.9) & 96.2 & 202.1 \\
(+159.5, -66.2) & (263.0, +38.9, -120.6, +215.7) & (50.4, +48.2, +13.5, +4.3) & 97.0 & 150.3 \\
(+25.9, +101.8) & (129.2, +75.7, -23.9, -53.7) & (57.0, +18.3, +52.0, +12.6) & 97.5 & 134.9 \\
	\bottomrule
        \end{tabular}}
        \caption{Monte Carlo event samples for which a discrepancy exists in computation of the
	asymmetric dual one-step event statistic \smash{$M_{\rm T2}^{{\rm b}\ell}$}~\cite{Bai:2012gs}.
	By definition, \smash{$M_H^A = 0.0$}~GeV and \smash{$M_H^B = 80.4$}~GeV.
	Tabulated masses and momenta are likewise in GeV.}
        \label{tab:mt2bl}
\end{table*}

For approximately five percent of all trials, the local algorithm reports a finite \smash{${\Widetilde{M}}_{\rm T2}$} solution that is
above the arbitrary 500~GeV upper bound at which the \smash{${M}_{\rm T2}^{\rm CH}$} family of solutions terminate their hunt.
Although that mass ceiling may be easily adjusted, the circumstance does highlight a strategic divergence taken by the present 
approach, wherein no such linear scale bounds (there is a logarithmic bulwark against numerical runaway) exist.
Excepting this trivial variety, there are no cases found
for which the local algorithm and the (CH) routine develop a discrepancy in the \smash{$M_{\rm T2}^{\rm b}$} statistic beyond the {\it per mille}
\smash{$\{\,\vert {\Widetilde{M}}_{\rm T2} - {M}_{\rm T2}^{\rm CH} \vert\div({\Widetilde{M}}_{\rm T2} + {M}_{\rm T2}^{\rm CH}) \le .001\,\}$} level.
Similarly, no substantive discrepancies are present in the pairwise computation of \smash{$M_{\rm T2}^{\rm W}$}; two events are found to diverge
by more than a part {\it per mille}, but still less than a part {\it per centum}, and the offset is attributed to process noise.
By contrast, the evaluation of \smash{$M_{\rm T2}^{{\rm b}\ell}$} presents 43 event configurations, approximately $0.1\%$ of the relevant trials, where a
non-trivial discrepancy emerges between \smash{${\Widetilde{M}}_{\rm T2}$} and \smash{${M}_{\rm T2}^{\rm CH}$}, as documented in Table~\ref{tab:mt2bl},
with entries rounded to the nearest tenth GeV.  Validity of the \smash{${\Widetilde{M}}_{\rm T2}$} solution has been directly visually confirmed with
{\sc Mathematica~9} for each case.  In the majority (40) of these scenarios, the step-wise sampling engine employed by the (CH) algorithms has paced over a
lighter intersection, finding a secondary root, whereas a minority (3) of events exist where no defined solution is reported at all.
For fairness' sake, it must be stated that the uniformly excellent performance of the (CH) routines is singly responsible for
instigating a coda of the present algorithm's development, wherein several crucial procedural refinements have been realized.

Unsurprisingly, the (CH) algorithms may fair rather less well against test cases contrived to
probe onset of the various geometric degeneracies and critical phases described.  There is no inherent obstacle
to inclusive treatment of the parabolic branches $(M_V \Rightarrow 0)$, {\it cf.}~FIG.~\ref{fig:intersection_3}, and
$(M_S \Rightarrow 0)$, {\it cf.}~FIG.~\ref{fig:intersection_4}, and both computations of these scenarios agree.
The topologies depicted in FIGS.~\ref{fig:intersection_1} and~\ref{fig:intersection_2} are elementary, and again present no difficulty.
However, the segmented-linear $(\Omega \Rightarrow 0)$, {\it cf.}~FIG.~\ref{fig:intersection_5}, and pointlike $(P_H^\mu \propto P_S^\mu)$
or instantonic $(P_V^\mu \propto P_S^\mu)$, {\it cf.}~FIG.~\ref{fig:intersection_6}, limits are 
more challenging, and the (CH) algorithm finds no solution for these examples.  It fails similarly
for the scenarios in footnote 2 (reporting the spurious intersection), 3A, 3B, 4A, 6A, 6B, 6C, and 7 (reporting undefined),
and also 5B (reporting the finite value of 156.7~GeV, which is in the unphysical mass-gap).  It agrees
that the events in footnotes 3C and 4B have no defined solution.
It reports a somewhat large, but entirely passable, value of 194.6~GeV in association with footnote 5A.
The scenario in footnote 1 is elementary, and there is agreement.

The massless phases (with some epsilon-width acceptance) may be quite common in real data, and are sensitive to particle
species and the handling of jet kinematic reconstruction.  In the present trials, $(M_V \Rightarrow 0)$ is realized by about $0.7\%$ of the raw
dual one-step event candidates, and $(M_S \Rightarrow 0)$ by approximately half (leptons are light) of the mixed one- and two-step topologies.
The onset of $(\Omega \Rightarrow 0)$ is inherently statistical, representing a common longitudinal frame (within some tolerance) for the primary $(V)$ and secondary $(S)$ visible
species of a two-step decay chain;  in tests, this phase occurs once per about 2,700 opportunities.  The kinematic
proportionalities $(P_H^\mu \propto P_S^\mu)$ and $(P_V^\mu \propto P_S^\mu)$ are not observed; the latter,
representing an explicit collinearity, is generically precluded whenever isolation in $(\Delta R)$ is enforced on
the reconstructed physics objects, as is presently the case; the former, representing an implicit collinearity equivalent to $(M_X \Rightarrow M_H + M_S)$,
remains broadly accessible in principle, although it presently requires that the secondary visible decay product has a
mass $(M_S \Rightarrow M_W)$ consistent with that of the $W$-boson, which should not much be expected of a leptonic track assignment.

It would seem that the most immediately practical and impactful innovations offered by this work 
are \mbox{({\it i}\hspace{1.6pt})} an inherently unified structure, which holistically treats all one- and two-step asymmetric event topologies,
\mbox{({\it ii}\hspace{1.6pt})} a substantially improved logic for the initial upper and lower parent particle species $M_Y$ scale boundaries (which
is intrinsically dependent upon a clear and comprehensive understanding of the accessible critical phases), and
\mbox{({\it iii}\hspace{1.6pt})} deployment of the 12\textsuperscript{th} order polynomial quartic discriminant $\mathbb{D}_4[\Upsilon]$, {\it cf.} Eq.~(\ref{eq:qrtdiscriminant}),
as a lighthouse for the illumination of conic intersection roots.  Together, these technologies (the second is particularly relevant to \smash{$M_{\rm T2}^{{\rm b}\ell}$},
and the third to \smash{$M_{\rm T2}^{{\rm W}}$}) serve to reduce the likelihood of passing over, and to streamline the convergence toward,
a kinematically viable parent mass scale.  In principle, the described \smash{${\Widetilde{M}}_{\rm T2}$} algorithm should fare no worse for any
particular example than the (CH) varietals, given that it reverts to the lower level scanning and stepping method as
a safety net underlying the more sophisticated (theoretically complete or near so) analysis framework; in practice, differential selection of the
sample point distribution may still engender disparities for certain numerically slender cases; in fact, there remains
a non-vanishing likelihood that both routines will simultaneously fail to affirm the provenance of some particularly delicate intersections,
as may well be the circumstance, unheeded, even for a subset of those limited trials here undertaken.

The algorithms in the (CH) family do fare quite substantially better in execution time for the described tests, by about a magnitude order
for \smash{$M_{\rm T2}^{{\rm b}}$} and \smash{$M_{\rm T2}^{{\rm b}\ell}$}, and two for \smash{$M_{\rm T2}^{{\rm W}}$}.
However, it must be emphasized that a large fraction, if not the outright majority, of this time differential
may be ultimately be attributable to the fact that the author's current working implementation is realized in the interpreted programming
language {\sc Perl}, which provides a highly flexible environment to developers, although one that is well-known to execute substantially
more slowly than compiled languages such as {\sc C++} or {\sc Fortran} in head-to-head tests.  Notably, many of the
advances outlined herein are intended to provide a stable alternative to the method of finely-grained incremental review 
of potential intersection scales, thereby providing a sizeable negative offset to the net instruction count; however, certain numerical
frailties argue still for the implementation of incremented sampling as a fallback, and the default tuning here has favored accuracy (and,
to a lesser extent, precision) at the cost of speed.  Scientists and members of Physics Working Groups who might benefit from
projection of the described \smash{${\Widetilde{M}}_{\rm T2}$} algorithm into a compiled language are encouraged
to contact the author directly; an additional advantage of this refactoring could be realized in the selective application of {\tt{quad~precision}}
floating point arithmetic to compensate for comparative fragility of the quartic discriminant, thereby rebalancing the relative weights of
computational time and computational safety away from the universal application of step-wise sampling in revision.

\section{\protecting{Pseudocode and Logical Flow of \texorpdfstring{\smash{$\boldsymbol{{\Widetilde{M}}_{\rm T2}}$}}{MT2}}\label{sct:appendix_C}}

The present section of the appendices supplies a thorough pseudocode and diagram of logical flow for the
\smash{${\Widetilde{M}}_{\rm T2}$} algorithm, which symbolically supplements the detailed textual commentary
of the main document body.  Further clarification of technicalities may be realized by cross-referencing
the available~\cite{amt2} exemplar of functioning source code, as projected into the {\sc Perl} programming language.  Liberal
use is made in the following of the ternary \,{$(\mathtt{If})$\tt{~?~}$(\mathtt{Then})$\tt{~:~}$(\mathtt{Else})$}\, flow control operator.


\makeatletter
\close@column@grid
\onecolumngrid

~\vspace{0pt} \\
\begin{center}
\includegraphics[scale=0.85]{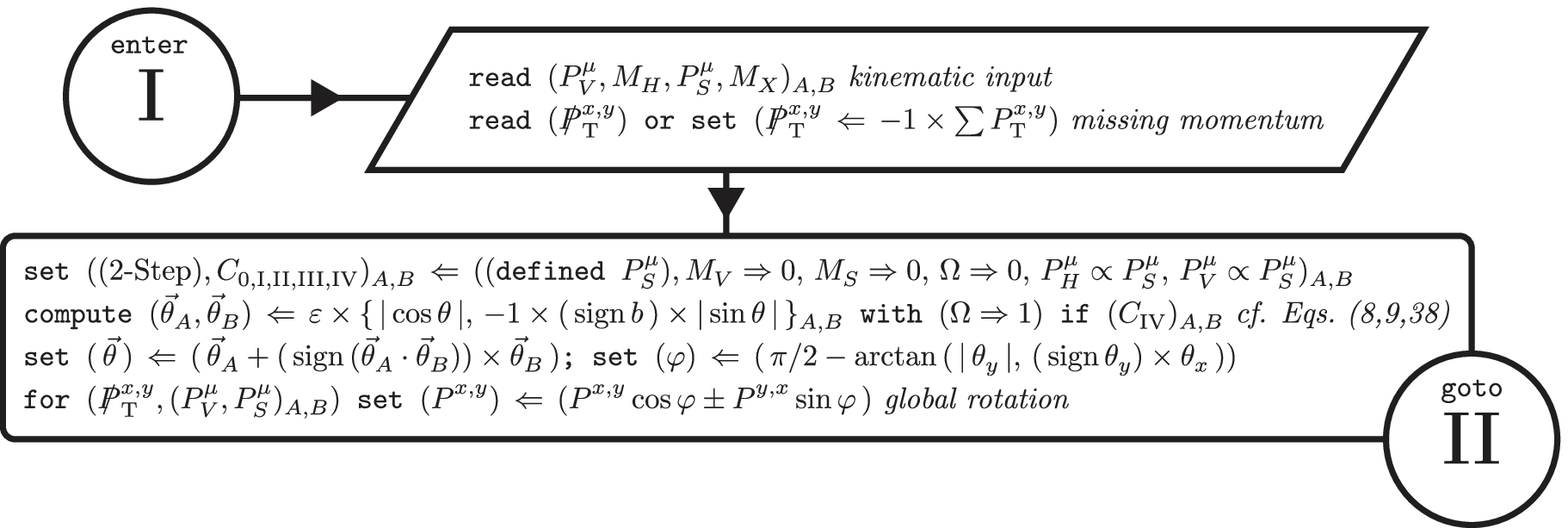}
\end{center}


\newpage
~\vspace{-8pt} \\
\begin{center}
\includegraphics[scale=0.85]{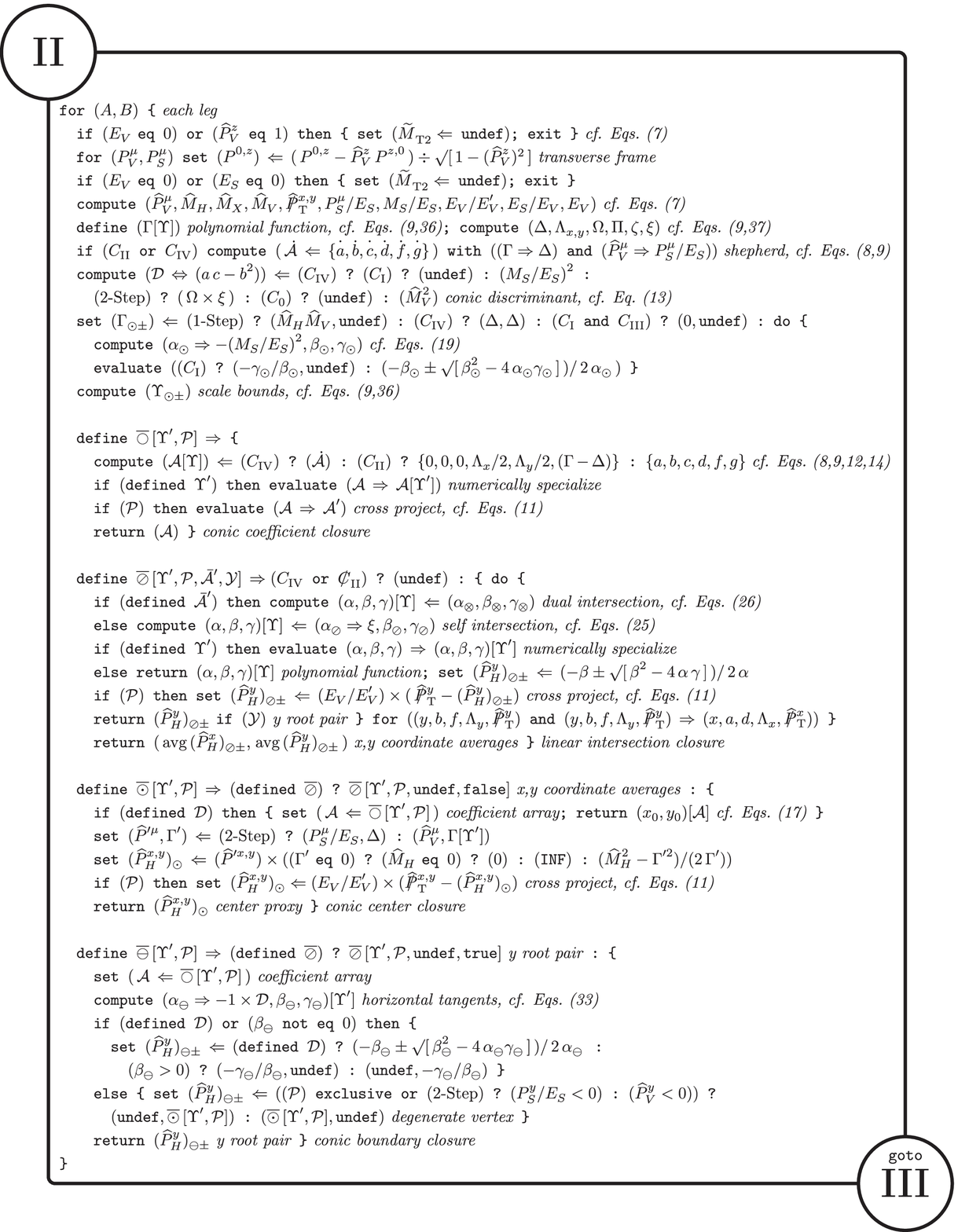}
\end{center}


\newpage
~\vspace{6pt} \\
\begin{center}
\includegraphics[scale=0.85]{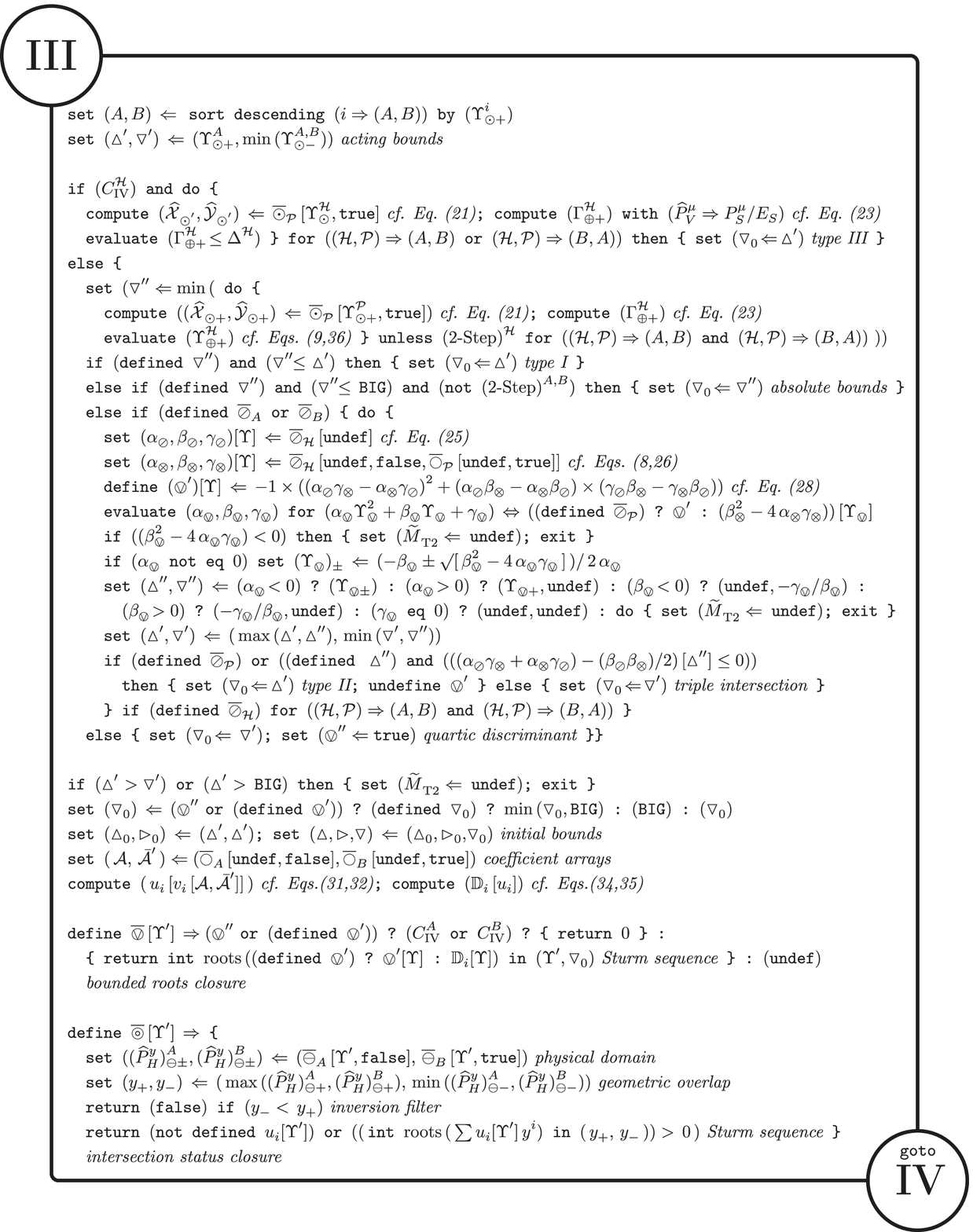}
\end{center}


\newpage
~\vspace{30pt} \\
\begin{center}
\includegraphics[scale=0.85]{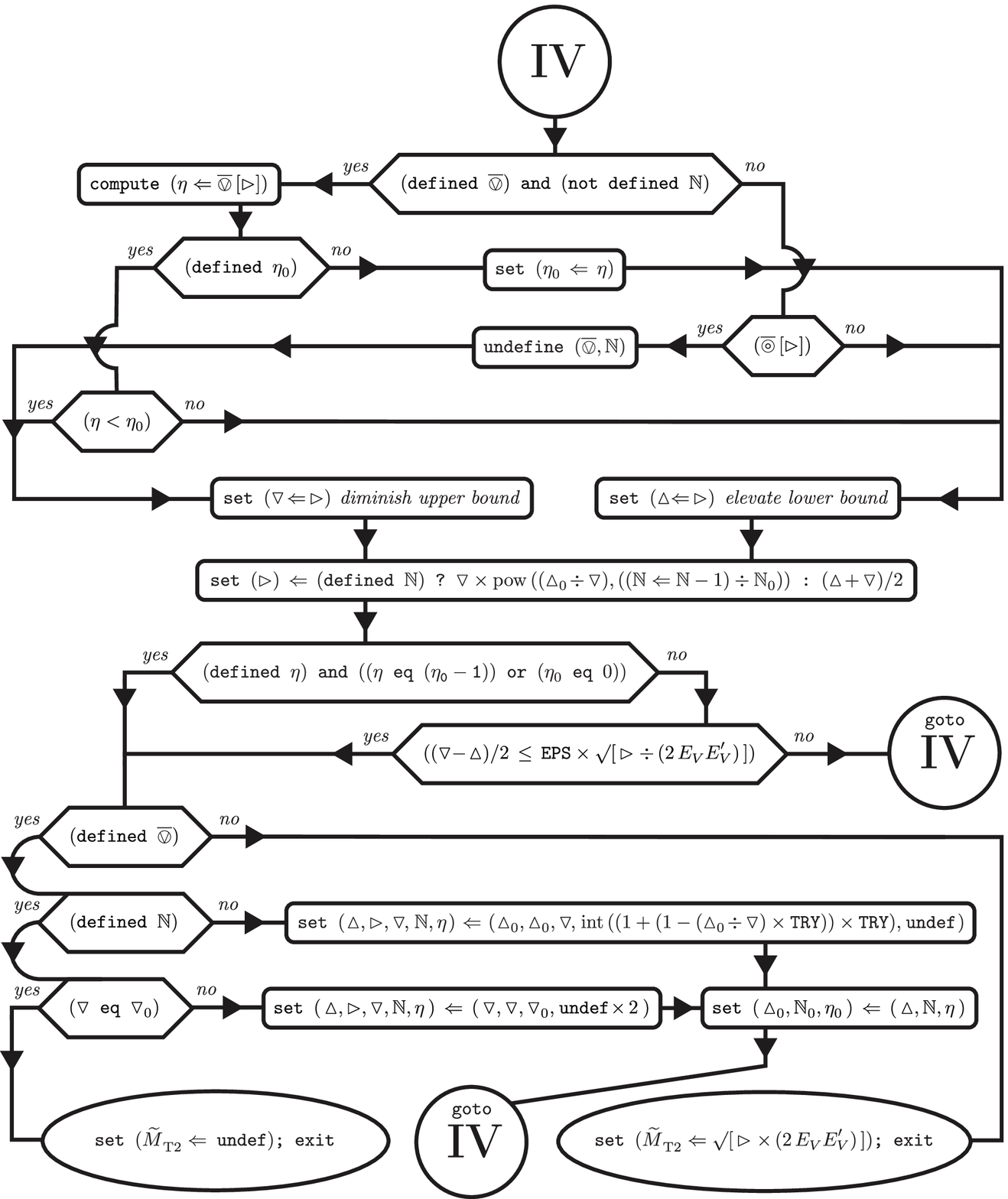}
\end{center}

\newpage
\close@column@grid
\twocolumngrid
\makeatother


\begin{acknowledgments}
This research was supported in part by the Sam Houston State University
2013 Faculty Research Grant program, and in part by the
National Science Foundation (NSF) under Grant No. NSF PHY11-25915.
I express gratitude to Dimitri V. Nanopoulos, Tianjun Li, and James A. Maxin,
colleagues in an ongoing phenomenological particle physics study that occasioned
the technical analysis and software development described in this article. 
I thank Sam Houston State University for providing the high performance computing
resources used to develop, refine and employ the associated software.
I also very much appreciate warm hospitality extended by the Kavli Institute for
Theoretical Physics, in cooperation with the NSF, during a portion of the
related research and initiation of the document authoring.
\end{acknowledgments}


\makeatletter
\close@column@grid
\onecolumngrid
\vspace{8pt}
\begin{center}
\includegraphics[width=0.8\textwidth]{vectorian_dot_net_flourish.eps}
\end{center}
\close@column@grid
\twocolumngrid
\makeatother


\def\bibsection{\section*{References}} 


\bibliography{bibliography}


\end{document}